\documentclass[twocolumn,tighten,twocolappendix]{aastex63}
\pdfoutput=1 %for arXiv submission
\usepackage{amsmath,amstext}
\usepackage[T1]{fontenc}
\usepackage{ae,aecompl}
\usepackage[utf8]{inputenc}
\usepackage{apjfonts} 
\usepackage[figure,figure*]{hypcap}
\usepackage{enumitem}
\usepackage{bm}
\usepackage{graphicx}
\input{hyperlink-year-only-natbib-patch}

\newcommand*{\http}[1]{\href{http://#1}{#1}}
\newcommand*{\https}[1]{\href{https://#1}{#1}}

\newcommand{\pc}{\ensuremath{{\rm pc}}}

\newcommand{\mpc}{\ensuremath{{\rm Mpc}}}

\newcommand{\K}{\ensuremath{{\rm K}}}
\newcommand{\s}{\ensuremath{{\rm s}}}

\newcommand{\msun}{\ensuremath{{M_{\odot}}}}

\shorttitle{Symphony Simulations}
\shortauthors{Nadler et al.}

\begin{document}

\title{Symphony: Cosmological Zoom-in Simulation Suites over Four Decades of Host Halo Mass}

\author[0000-0002-1182-3825]{Ethan O.~Nadler}
\affiliation{Carnegie Observatories, 813 Santa Barbara Street, Pasadena, CA 91101, USA}
\affiliation{Department of Physics $\&$ Astronomy, University of Southern California, Los Angeles, CA, 90007, USA}

\author[0000-0001-9863-5394]{Philip Mansfield}
\affiliation{Kavli Institute for Particle Astrophysics \& Cosmology, P. O. Box 2450, Stanford University, Stanford, CA 94305, USA}
\affiliation{SLAC National Accelerator Laboratory, Menlo Park, CA 94025, USA}
\affiliation{Department of Physics, Stanford University, 382 Via Pueblo Mall, Stanford, CA 94305, USA}

\author[0000-0001-8913-626X]{Yunchong Wang}

\affiliation{Kavli Institute for Particle Astrophysics \& Cosmology, P. O. Box 2450, Stanford University, Stanford, CA 94305, USA}
\affiliation{SLAC National Accelerator Laboratory, Menlo Park, CA 94025, USA}
\affiliation{Department of Physics, Stanford University, 382 Via Pueblo Mall, Stanford, CA 94305, USA}

\author[0000-0003-0728-2533]{Xiaolong Du}
\affiliation{Carnegie Observatories, 813 Santa Barbara Street, Pasadena, CA 91101, USA}

\author[0000-0002-0298-4432]{Susmita~Adhikari}
\affiliation{Department of Physics, Indian Institute of Science Education and Research, Homi Bhabha Road, Pashan, Pune 411008, India}

\author[0000-0002-5209-1173]{Arka~Banerjee}
\affiliation{Department of Physics, Indian Institute of Science Education and Research, Homi Bhabha Road, Pashan, Pune 411008, India}

\author[0000-0001-5501-6008]{Andrew Benson}
\affiliation{Carnegie Observatories, 813 Santa Barbara Street, Pasadena, CA 91101, USA}

\author[0000-0002-8800-5652]{Elise Darragh-Ford}
\affiliation{Kavli Institute for Particle Astrophysics \& Cosmology, P. O. Box 2450, Stanford University, Stanford, CA 94305, USA}
\affiliation{Department of Physics, Stanford University, 382 Via Pueblo Mall, Stanford, CA 94305, USA}
\affiliation{SLAC National Accelerator Laboratory, Menlo Park, CA 94025, USA}

\author[0000-0002-1200-0820]{Yao-Yuan~Mao}
\affiliation{Department of Physics and Astronomy, University of Utah, Salt Lake City, UT 84112, USA}

\author[0000-0001-5039-1685]{Sebastian Wagner-Carena}
\affiliation{Kavli Institute for Particle Astrophysics \& Cosmology, P. O. Box 2450, Stanford University, Stanford, CA 94305, USA}
\affiliation{Department of Physics, Stanford University, 382 Via Pueblo Mall, Stanford, CA 94305, USA}
\affiliation{SLAC National Accelerator Laboratory, Menlo Park, CA 94025, USA}

\author[0000-0003-2229-011X]{Risa H.~Wechsler}
\affiliation{Kavli Institute for Particle Astrophysics \& Cosmology, P. O. Box 2450, Stanford University, Stanford, CA 94305, USA}
\affiliation{Department of Physics, Stanford University, 382 Via Pueblo Mall, Stanford, CA 94305, USA}
\affiliation{SLAC National Accelerator Laboratory, Menlo Park, CA 94025, USA}

\author[0000-0002-7904-1707]{Hao-Yi Wu}
\affiliation{Department of Physics, Boise State University, Boise, ID 83725, USA}

\correspondingauthor{Ethan O.~Nadler}
\email{enadler@carnegiescience.edu}

\begin{abstract}
We present Symphony, a compilation of $262$ cosmological, cold-dark-matter-only zoom-in simulations spanning four decades of host halo mass, from $10^{11}$--$10^{15}~\msun$. This compilation includes three existing simulation suites at the cluster and Milky Way--mass scales, and two new suites: $39$ Large Magellanic Cloud-mass ($10^{11}~\msun$) and $49$ strong-lens-analog ($10^{13}~\msun$) group-mass hosts.
Across the entire host halo mass range, the highest-resolution regions in these simulations are resolved with a dark matter particle mass of~$\approx 3\times 10^{-7}$ times the host virial mass and a Plummer-equivalent gravitational softening length of $\approx 9\times 10^{-4}$ times the host virial radius, on average.
We measure correlations between subhalo abundance and host concentration, formation time, and maximum subhalo mass, all of which peak at the Milky Way host halo mass scale.
Subhalo abundances are $\approx 50\%$ higher in clusters than in lower-mass hosts at fixed sub-to-host halo mass ratios. Subhalo radial distributions are approximately self-similar as a function of host mass and are less concentrated than hosts' underlying dark matter distributions.
We compare our results to the semianalytic model \textsc{Galacticus}, which predicts subhalo mass functions with a higher normalization at the low-mass end and radial distributions that are slightly more concentrated than Symphony. 
We use \textsc{UniverseMachine} to model halo and subhalo star formation histories in Symphony, and we demonstrate that these predictions resolve the formation histories of the halos that host nearly all currently observable satellite galaxies in the universe. To promote open use of Symphony, data products are publicly available at \url{http://web.stanford.edu/group/gfc/symphony}.
\end{abstract}

\keywords{\href{http://astrothesaurus.org/uat/353}{Dark matter (353)};
\href{http://astrothesaurus.org/uat/574}{Galaxy abundances (574)};
\href{http://astrothesaurus.org/uat/1083}{$N$-body simulations (1083)};
\href{http://astrothesaurus.org/uat/1880}{Galaxy dark matter halos (1880)};
\href{http://astrothesaurus.org/uat/1965}{Computational methods (1965)}}

%---------------------------------------------------------------------------------------
%	SECTION 1
%---------------------------------------------------------------------------------------

\section{Introduction}
\label{Introduction}

The hierarchical formation and nonlinear evolution of self-gravitating dark matter systems, or ``halos,'' underpins our modern understanding of cosmic structure. Numerical simulations, which remain the most widely used and accurate technique to model structure formation, are particularly important for resolving ``subhalos'' that reside within larger host halos (e.g., see \citealt{Zavala190711775} for a review). Recent studies at the forefront of structure formation and galaxy evolution depend on the properties of subhalos with low masses relative to their host halos---and the relatively faint satellite galaxies that reside within them---in a wide range of cosmic environments (e.g., \citealt{Geha170506743,Kallivayalil180501448,Gilman190902573,Meneghetti200904471,Adhikari200811663}). Thus, a unified simulation suite that captures subhalo populations at high resolution across the entire observationally relevant range of host and subhalo masses is timely.

``Zoom-in'' simulations, in which a small region of a larger, lower-resolution ``parent'' simulation is resimulated at higher resolution, are useful for resolving subhalo populations in a cosmological context \citep{Katz1993,Bertschinger0103301}. Most ``zoom-ins'' focus on a specific host halo (or pair of host halos; e.g., \citealt{Garrison-Kimmel13106746}) and resimulate a region that contains all particles that eventually reside within that halo (its ``Lagrangian volume'') at higher resolution than in the parent box. This is done using initial conditions generated with nested regions of increasing refinement \citep{Jenkins09100258,Hahn11036031}. This approach ensures that the formation and evolution of the host halo's dark matter structure, including its subhalos, are captured at high resolution (e.g., \citealt{Onorbe13056923}). The zoom-in technique has facilitated an enormous range of studies within $\Lambda$ cold dark matter (CDM) in both $N$-body and hydrodynamic contexts (e.g., see \citealt{Vogelsberger190907976} for a recent review), and recent zoom-in simulations have increasingly included nonstandard dark matter or cosmological physics (e.g., see \citealt{Banerjee220307049} and references therein). However, existing zoom-in simulations leave room for improvements in several crucial areas.

\begin{figure*}[t!]
\hspace{-2.5mm}
\includegraphics[width=0.5\textwidth]{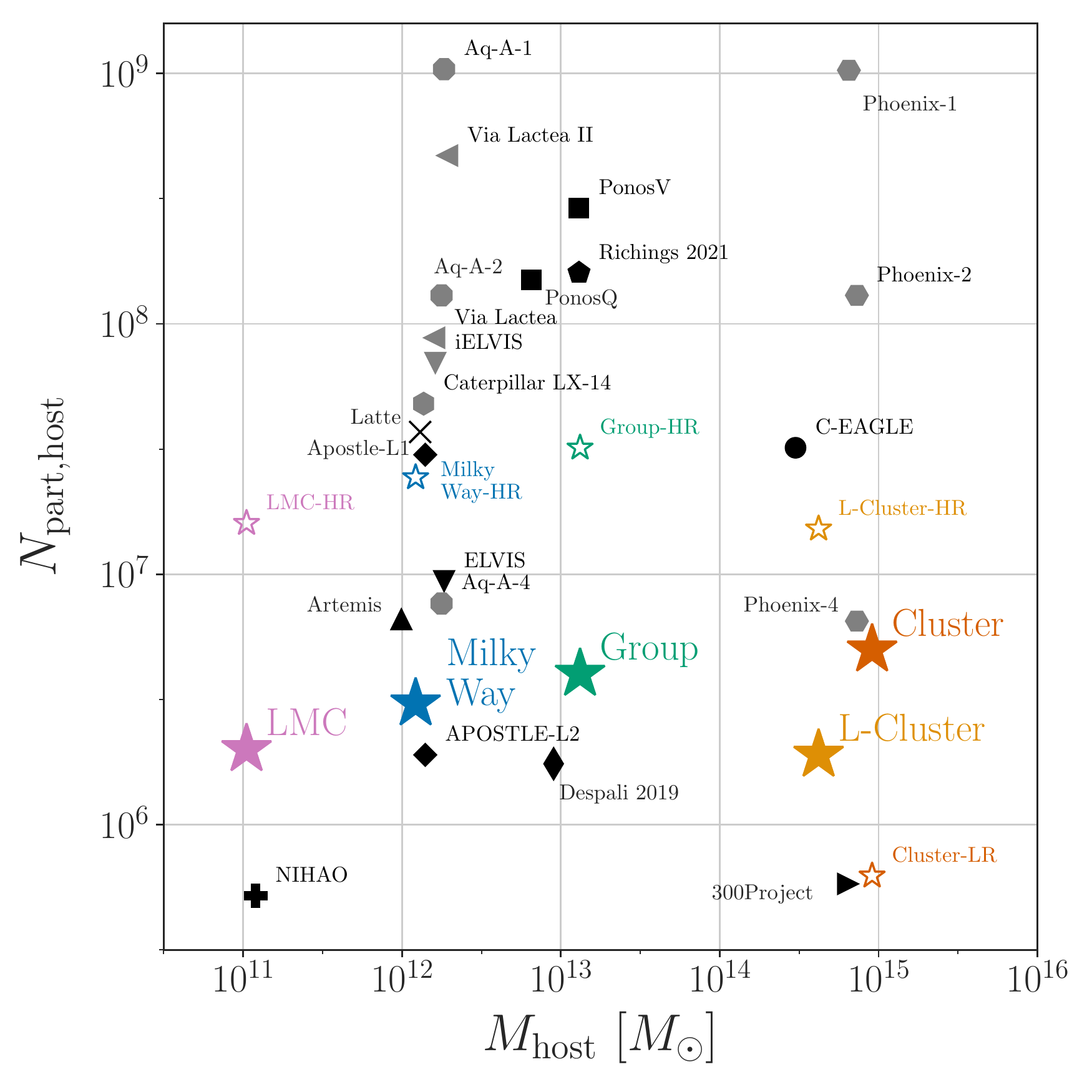}
\includegraphics[width=0.5\textwidth]{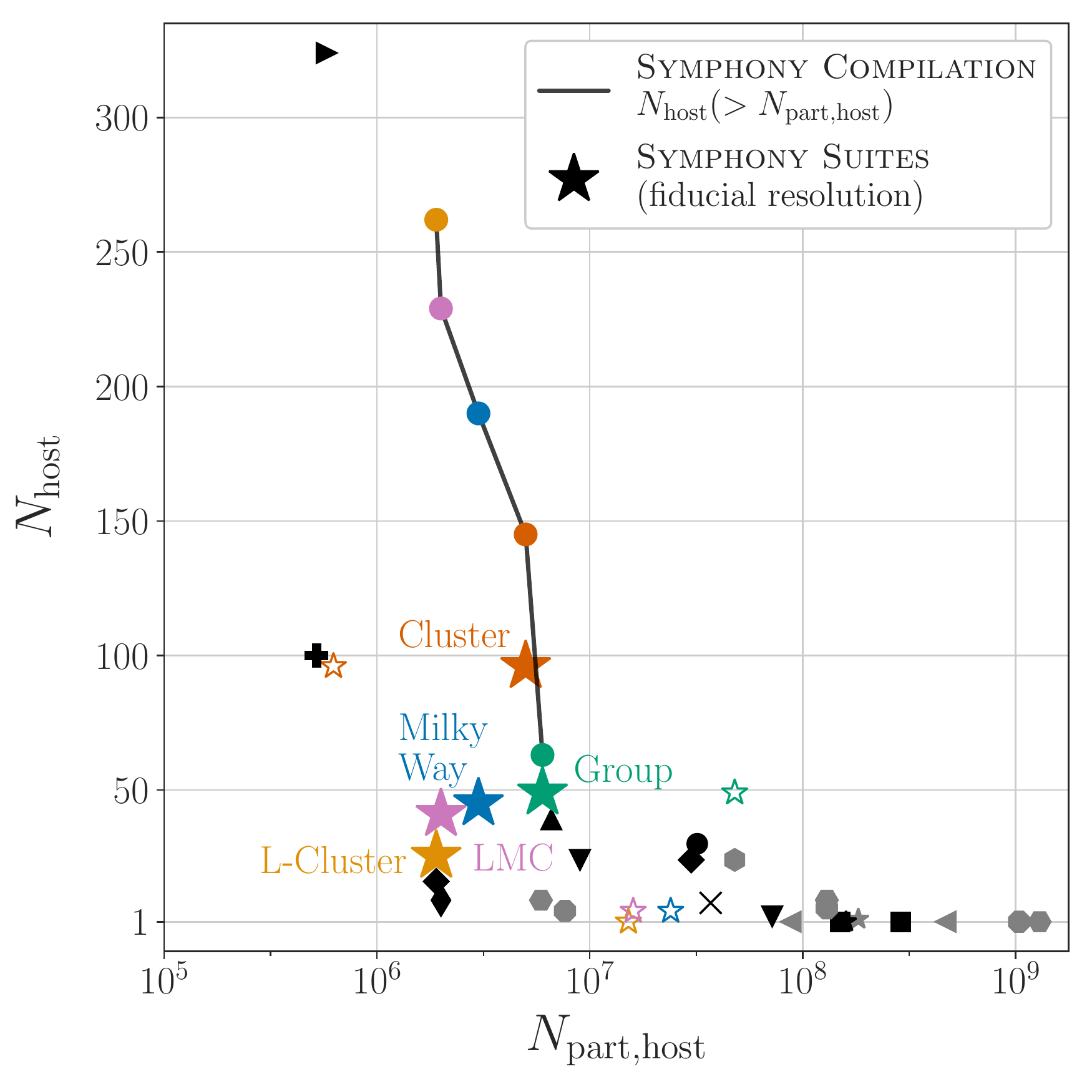}
\vspace{-1mm}
\caption{Left panel: median number of particles per host halo versus the median host mass for Symphony (stars) and a representative sample of existing cosmological zoom-in simulation suites (markers).  The five Symphony suites presented in this work are: LMC (magenta), Milky Way (blue), Group (green), L-Cluster (gold), and Cluster (red); Symphony resimulation suites are shown as small unfilled stars. Existing zoom-in suites with (without) hydrodynamic resimulations are shown in black (gray). Median host masses and particle counts are taken from each existing suite without converting to a unified virial mass definition. Host-to-host scatter for suites with multiple simulations is not shown. See Appendix \ref{sec:fig1_sims} for references. Right panel: number of host halos per zoom-in suite versus the median number of particles per host. Only target host halos are included in the $y$-axis counts; however, many zoom-ins (including those in Symphony) often contain halos of similar masses in addition to the host in the highest-resolution region. The black line shows the cumulative number of zoom-in simulations combined over all five fiducial-resolution zoom-in suites in the Symphony compilation, i.e., $N_{\mathrm{host}}(>N_{\mathrm{part,host}})$. Note that the Group point along the cumulative line is offset vertically from the Group suite for visual clarity.
}
\label{fig:simulations}
\end{figure*}

First, the majority of cosmological zoom-in simulations focus on host halos with masses similar to the Milky Way (e.g., \citealt{Diemand08051244,Springel08090898,Garrison-Kimmel13106746,Mao150302637,Griffen150901255,Sawala151101098,Wetzel160205957,Samuel190411508,Poole-McKenzie200615159}) or galaxy clusters (e.g., \citealt{Gao12011940,Barnes170310907,Cui180904622}). Although many exceptions exist (e.g., \citealt{Wang150304818,Dutton151200453,Fiacconi160203526,Fiacconi160909499,Despali181102569,Richings200514495}), zoom-in suites at other mass scales typically include only a few distinct hosts, precluding analyses of their subhalo populations that capture host-to-host scatter (see Figure \ref{fig:simulations}). Such analyses are crucial in order to characterize subhalo population statistics as a function of host halo mass, which are often simply extrapolated from Milky Way--mass hosts in semianalytic models (e.g., \citealt{Dooley170305321,Gilman190806983}). Second, existing zoom-in simulations at different mass scales often vary in resolution (e.g., in terms of how many particles comprise the host halo) and are performed or analyzed with codes that differ in detail. This limits the feasibility of unified subhalo population analyses that simultaneously cover a wide range of host mass and robustly quantify host-to-host scatter at high resolution.

To bridge these gaps, we present Symphony, the first statistical compilation of cosmological zoom-in simulations of host halos with masses from $10^{11}$--$10^{15}~\msun$. Symphony includes $39$ zoom-ins at the $10^{11}~\msun$ LMC-mass scale (the ``LMC'' suite), $45$ zoom-ins at the $10^{12}~\msun$ Milky Way--mass scale (the ``Milky Way'' suite), $49$ strong lens analog zoom-ins at the $10^{13}~\msun$ group-mass scale (the ``Group'' suite), $33$ zoom-ins at the $\approx 5\times 10^{14}$ low-mass cluster scale (the ``Low-mass Cluster,'' or ``L-Cluster'' suite), and $96$ zoom-ins at the $10^{15}~\msun$ cluster scale (the ``Cluster'' suite), for a total of $262$ distinct simulations. Each suite is run with comparable, high resolution relative to its host halo mass, such that hosts consist of greater than $10^{6}$ particles and subhalos are well resolved down to $\approx 10^{-4}$ times the host mass, on average. Thus, Symphony enables precise, self-consistent measurements of subhalo population statistics and their host-to-host scatter over four decades of host halo mass.

From an observational perspective, Symphony's new LMC and Group suites are relevant given recent advances in our ability to probe substructure in both regimes. Specifically, the combination of recent astrometric measurements \citep[e.g.][]{Gaia180409365} and photometric observations of Milky Way satellite galaxies (e.g., from the Dark Energy Survey; \citealt{DES}) indicate that several nearby ultra-faint dwarf galaxies are satellites of the LMC \citep{Kallivayalil180501448,Patel200101746}; upcoming facilities are also expected to detect satellites of LMC-mass systems throughout the Local Volume (e.g., \citealt{Mutlu-Pakdil210501658}). On the Group scale, high-resolution imaging (e.g., \citealt{Nierenberg190806344}) has enabled measurements of dark matter substructure within early-type galaxy strong lenses (e.g., \citealt{Vegetti12013643,Hezaveh160101388,Hsueh190504182,Gilman190806983,Gilman190902573}). These developments underscore the need for high-resolution simulations of dark matter substructure at the corresponding host halo mass scales. Meanwhile, Symphony's Milky Way--mass, Low-mass Cluster, and Cluster suites serve as benchmarks for analyses of subhalo evolution within the Milky Way, its analogs, and galaxy clusters throughout the universe. Zoom-ins from these suites have been resimulated with hydrodynamic or nonstandard dark matter physics (e.g., \citealt{Martizzi151000718,Nadler200108754,Nadler210912120,Bhattacharyya210608292,Mau220111740}), and the Symphony data release will facilitate further work along these lines.

From a theoretical perspective, Symphony's large dynamic range is desirable for calibrating semianalytic structure formation models, which are usually constrained using large-volume cosmological simulations (e.g., \citealt{Benson161001057}) or Milky Way--mass zoom-ins (e.g., \citealt{Pullen14078189,Yang200310646}). Furthermore, empirical and semianalytic models of galaxy evolution have been applied to zoom-in simulations (e.g., \citealt{Starkenburg12060020,Lu160502075,Newton170804247,Wang210211876,Chen220201220,Kravtsov210609724}), and this technique can be extended to other host masses using Symphony. Here, we connect to both modeling approaches by comparing Symphony to predictions from the semianalytic structure formation model \textsc{Galacticus} \citep{Benson10081786,Pullen14078189} and by using the empirical galaxy--halo connection model \textsc{UniverseMachine} \citep{Behroozi2019,Wang210211876} to predict star formation histories (SFHs) for Symphony halos and subhalos. By combining Symphony and \textsc{UniverseMachine}, we capture the relation between dark matter accretion and SFHs for nearly all currently observable satellite galaxies in the universe, with the exception of the faintest known satellites in nearby clusters. In addition, Symphony's new LMC suite provides robust predictions for the population statistics of halos below the threshold of galaxy formation (e.g., \citealt{Nadler191203303,Munshi210105822}), which will be crucial to understand in order to search for deviations from $\Lambda$CDM on small scales using forthcoming datasets (e.g., see \citealt{Bechtol220307354} and references therein).

This paper is organized as follows. Section \ref{sec:overview} describes the Symphony zoom-in simulations. Sections \ref{sec:host_halos} and \ref{sec:subhalo_populations}, respectively, study the properties of Symphony host halos and subhalo populations. Section \ref{sec:galacticus} compares Symphony results to predictions from \textsc{Galacticus}. Section~\ref{sec:universemachine} applies \textsc{UniverseMachine} to all Symphony simulations and presents the resulting central and satellite galaxy SFH predictions. Section~\ref{sec:discussion} discusses science enabled by the Symphony suites and compilation. Section~\ref{sec:conclusion} concludes. Throughout, Symphony results are presented using a color scheme of pink (LMC), blue (Milky Way), green (Group), gold (L-Cluster), and red (Cluster).

%---------------------------------------------------------------------------------------
%	SECTION 2
%---------------------------------------------------------------------------------------

\section{Symphony Simulations}
\label{sec:overview}

This section provides an overview of Symphony, including information about parent and zoom-in simulations, initial conditions, simulation parameters, zoom-in host halo selection criteria, and convergence properties.

\subsection{Parent Simulations}

Host halos for our LMC, Milky Way, and Group suites were selected from the same parent simulation, c125-1024, which has a side length of $125~ \mpc~h^{-1}$ with $1024$ particles per side; we also use a higher-resolution version of this box, c125-2048, for certain tests (see \citealt{Mao150302637} for details). c125-1024 and its corresponding zoom-in suites were run with cosmological parameters of $h = 0.7$, $\Omega_{\rm m} = 0.286$, $\Omega_{\Lambda} = 0.714$, $\sigma_8 = 0.82$, and $n_s=0.96$ \citep{Hinshaw_2013}. The \cite{Bryan_1998} virial overdensity in this cosmology corresponds to $\Delta_{\mathrm{vir}}\approx 99$ times the critical density of the universe at $z=0$.

The L-Cluster hosts were selected from a parent simulation of side length $1~ \mathrm{Gpc}~h^{-1}$ with $1024$ particles per side, which we refer to as \textsc{1000-1024a} (see \citealt{Bhattacharyya210608292} for details). This simulation uses cosmological parameters of $h = 0.7$, $\Omega_{\rm m} = 0.3$, $\Omega_{\Lambda} = 0.7$, $\sigma_8 = 0.85$, and $n_s=0.96$, corresponding to $\Delta_{\mathrm{vir}}\approx 101$ at $z=0$ \citep{Bhattacharyya210608292}.

The Cluster hosts were selected from the Carmen simulation \citep{McBrideLasDAMAS}, which has a side length of $1~ \mathrm{Gpc}~h^{-1}$ with $1120$ particles per side, hereafter referred to as 1000-1120B. This simulation uses cosmological parameters of $h = 0.7$, $\Omega_{\rm m} = 0.25$, $\Omega_{\Lambda} = 0.75$, $\sigma_8 = 0.8$, and $n_s=1$, corresponding to $\Delta_{\mathrm{vir}}\approx 94$ at $z=0$ \citep{Wu12093309,Wu12106358}. Thus, the L-Cluster and Cluster zoom-in simulations adopt different cosmological parameters compared to both each other and to the LMC, Milky Way, and Group suites. These differences have a minor impact on our results, but should be kept in mind when comparing suites in detail.

\subsection{Zoom-in Simulations}
\label{sec:symphony_overview}

For all Symphony suites, zoom-in initial conditions were generated using \textsc{MUSIC} \citep{Hahn11036031}, simulations were run with \textsc{Gadget-2} \citep{Springel0505010}, and halo catalogs and merger trees were, respectively, generated using {\sc Rockstar} and {\sc consistent-trees} \citep{Behroozi11104372,Behroozi11104370}. To promote open use of Symphony, halo catalogs and merger trees for our 262 fiducial-resolution zoom-ins are available at \url{http://web.stanford.edu/group/gfc/symphony}; furthermore, a subset of particle snapshots is publicly available.

\begin{figure*}[t!]
\hspace{-12mm}
\includegraphics[trim={0 1.25cm 0 0.625cm},width=1.15\textwidth]{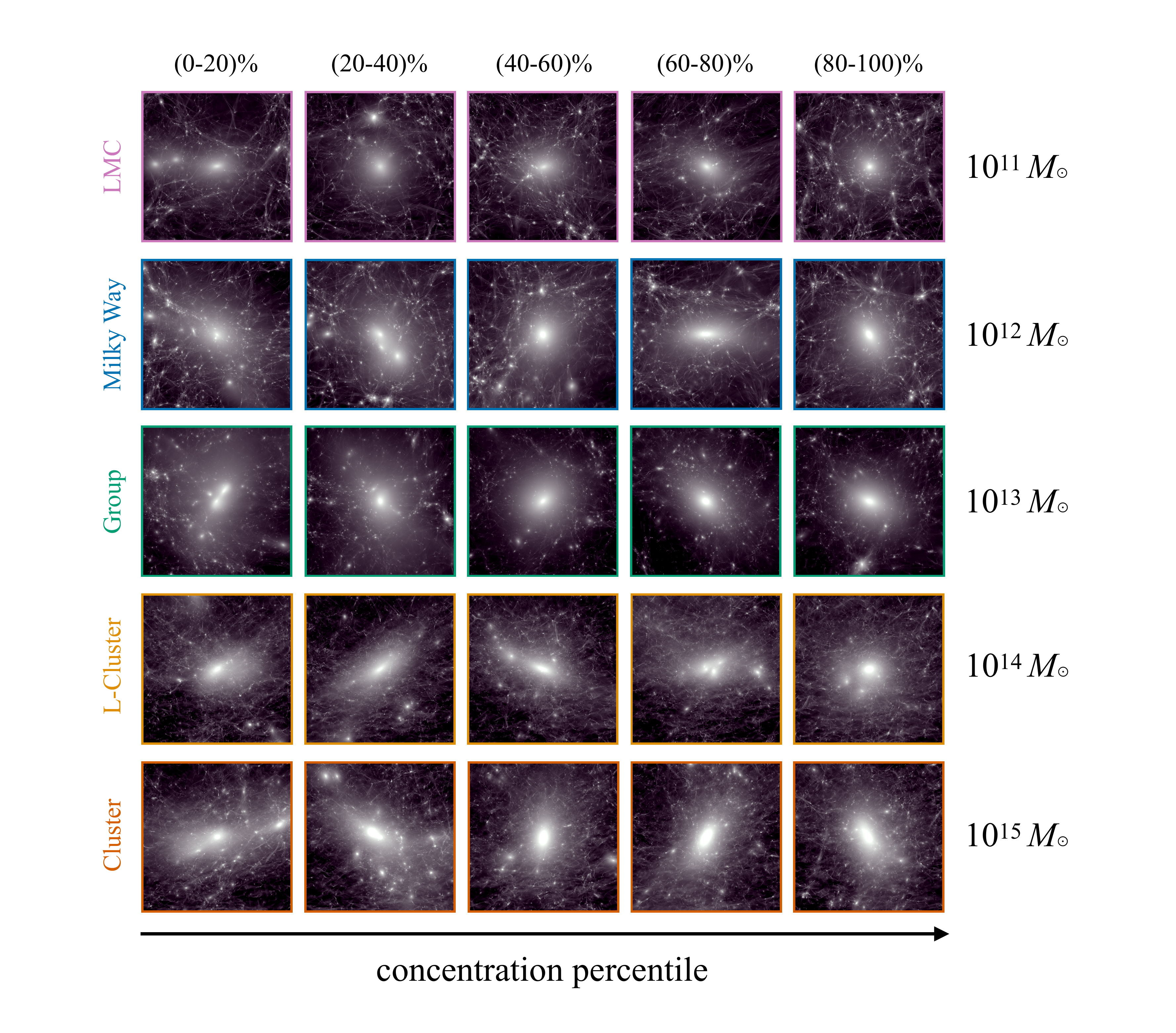}
\caption{Visualizations of the projected dark matter density at $z=0$, centered on five host halos from each of our zoom-in suites. The host halo mass scale of each zoom-in suite increases from top to bottom. Within each suite, columns show hosts randomly selected from concentration quintiles, such that concentration increases from left to right. Each visualization spans its host's virial radius in projection. Visualizations were created using the phase-space tessellation method described in \cite{KAEHLER201768} and \cite{Kaehler:2018:2470-1173:333}.}
\label{fig:vis}
\end{figure*}

Figure \ref{fig:simulations} compares the Symphony compilation to existing zoom-in suites, Figure \ref{fig:vis} visualizes five host halos in each suite at $z=0$, and Table \ref{tab:sims} lists the numerical properties of our five suites. As shown in Figure \ref{fig:simulations}, Symphony provides a unique combination of host statistics, resolution, and dynamic range compared to existing zoom-in suites (see Appendix \ref{sec:fig1_sims} for references). We note that a handful of cosmological simulations (e.g., VSMDPL from the MultiDark suite and Uchuu; \citealt{Klypin14114001,Ishiyama200714720}) contain more high-resolution Group, L-Cluster, and/or Cluster-mass hosts than Symphony; we discuss the pros and cons of zoom-ins relative to such simulations in Section \ref{sec:discussion}.

The environmental properties of Symphony's LMC, Milky Way, and Group hosts differ from typical halos of these masses to varying degrees because of the isolation criteria used to select the host halos, which are described below for each suite. For example, relative to all halos in the corresponding mass ranges from the Erebos simulations \citep{Diemer14011216,Diemer14074730}, we find that $3\%$, $87\%$, and $69\%$ of objects satisfy our isolation criteria for the LMC, Milky Way, and Group suites, respectively. We plan to characterize the impact of Symphony hosts' environments on their formation histories and subhalo populations in future work.

\subsubsection{LMC-mass Suite}
\label{sec:lmc_suite}

Host halos for the LMC suite were chosen from a $z=0$ virial mass range of $10^{11.02 \pm 0.05}~\msun$ in c125-1024, which is comparable to LMC's halo mass \citep{Erkal181208192,Shipp210713004}. Hosts in this mass range were selected subject to the constraint that no more massive halo is found within a radius of $3~ \mpc~h^{-1}$ in the parent box; thus, although we abbreviate these hosts as ``LMC''s, they are similar to the actual LMC in terms of total mass but not environment. Just 3\% of halos within this mass range meet this selection criterion.
Zoom-in initial conditions for each system were generated with five refinement regions, yielding an equivalent of $16,384$ particles per side for the most refined region. The highest-resolution region for each simulation corresponds to the Lagrangian volume containing particles within five times the virial radius of the host halo in the parent box at $z=0$. The dark matter particle mass in the highest-resolution region is $m_{\mathrm{part}}=5.0\times 10^4~ \msun$, and the comoving Plummer-equivalent gravitational softening in this region is $80~ \pc~h^{-1}$, corresponding to $0.011$ times the mean interparticle spacing.

Thirty-nine host halos were resimulated, resulting in a distribution of $m_{\mathrm{part}}/M_{\mathrm{host}}$ with a median and standard deviation of $(4.8\pm 0.8)\times 10^{-7}$ and an  $\epsilon/R_{\mathrm{vir,host}}$  distribution of $(9.3\pm 0.5)\times 10^{-4}$, where $M_{\mathrm{host}}$ and $R_{\mathrm{vir,host}}$ denote the $z=0$ host halo virial mass and radius, respectively. Over the LMC suite, zoom-in host halo masses at $z=0$ differ from the target hosts in the parent box by $1.1\%\pm 11\%$, which represents the largest amount of scatter relative to the target host masses among our c125-1024 suites. We attribute this scatter to the fact that the target hosts in the parent box are less well resolved than the target hosts in any other suite.

\subsubsection{Milky Way--mass Suite}
\label{sec:mw_suite}

The halos in the Milky Way suite were first presented in \cite{Mao150302637}; we refer the reader to this work for a complete description of the original suite. As noted in \cite{Mau220111740}, the \cite{Mao150302637} simulations used $n_s=1.0$ to generate initial conditions, while the parent box used $n_s=0.96$. We therefore resimulate this suite using $n_s=0.96$, and we exclusively present these new results here.\footnote{This change to $n_s$ alleviates the discrepancies in subhalo merger timing relative to the parent box identified in \cite{Wang210211876}.}

Host halos for the Milky Way suite were chosen from a~$z=0$ virial mass range of $10^{12.09 \pm 0.02}~\msun$ in c125-1024, and thus fall within observational estimates of the Milky Way's dark matter halo mass (e.g., see \citealt{Bland-Hawthorn160207702,Callingham180810456} and references therein). Hosts in this mass range were selected such that they were not within $4 R_{\mathrm{vir,host}}$ of any more massive halo in the parent box. Eighty-seven percent of halos in this mass range in the parent box meet this isolation criterion, and as shown in \citet{Mao150302637}, this sample has a somewhat wider concentration distribution than the mass-selected sample.
Thus, although we abbreviate these hosts ``Milky Ways,'' they are similar to the actual Milky Way in terms of total mass but not not necessarily in terms of environment. In fact, the Milky Way's large-scale environment is unusual, and constrained simulations would be required to match it in detail (e.g., see \citealt{McCall2014,Carlesi160203919,Neuzil2020,McAlpine220204099} and references therein). In addition, these hosts do not necessarily satisfy constraints on the Milky Way's formation history, such as the recent infall of the LMC and an early \emph{Gaia}--Enceladus-like merger. D.\ Buch et al.\ (2023, in preparation) presents a zoom-in suite of Milky Way--mass hosts selected from c125-1024 and using the same resolution that satisfy additional ``Milky Way--like'' constraints.

Zoom-in resimulation initial conditions for each Milky Way--mass host were generated with four refinement regions, yielding an equivalent of $8192$ particles per side in the most refined region. The highest-resolution region for each simulation corresponds to the Lagrangian volume containing particles within $10R_{\mathrm{vir,host}}$ of the host halo in the parent box at $z=0$. The dark matter particle mass in the highest-resolution regions is $m_{\mathrm{part}}=4.0\times 10^5~ \msun$ and the comoving Plummer-equivalent gravitational softening is $170~ \pc~h^{-1}$, corresponding to $0.011$ times the mean interparticle spacing.

Forty-five host halos were resimulated, resulting in a distribution of $m_{\mathrm{part}}/M_{\mathrm{host}}$  with a median and standard deviation of $(3.2\pm 0.2)\times 10^{-7}$ and an $\epsilon/R_{\mathrm{vir,host}}$ distribution of $(8.7\pm 0.1)\times 10^{-4}$. Over the Milky Way suite, zoom-in host halo masses at $z=0$ differ from the target hosts in the parent box by $1.0\pm 3\%$.

\begin{deluxetable*}{{c@{\hspace{0.07in}}c@{\hspace{0.07in}}c@{\hspace{0.07in}}c@{\hspace{0.07in}}c@{\hspace{0.07in}}c@{\hspace{0.07in}}c@{\hspace{0.07in}}c@{\hspace{0.07in}}c}}[t!]
\centering
\tablecolumns{10}
\tablecaption{Properties of the Five Symphony Simulation Suites.}
\tablehead{
\colhead{Zoom-in Suite} & \colhead{Parent Simulation} & \colhead{$N_{\mathrm{sims}}$} & \colhead{$M_{\mathrm{host}}\ (\msun)$} & $M_{\mathrm{sub,min}}\ (\msun)$ & $M_{\mathrm{sub/host,min}}$ & \colhead{$m_{\mathrm{part}}\ (\msun)$} & \colhead{$\epsilon\ (\mathrm{pc}~h^{-1})$} & \colhead{Reference}}
\startdata
LMC-mass
& c125-1024
%& WMAP9-like
& $39$
& $10^{11.02 \pm 0.05}$
& $1.5\times 10^7$
& $2.7\times 10^{-4}$
& $5.0\times 10^4$
& $80$
& This work\\
Milky Way--mass
& c125-1024
%& WMAP9-like
& $45$
& $10^{12.09 \pm 0.02}$
& $1.2\times 10^8$
& $1.1\times 10^{-4}$
& $4.0\times 10^5$
& $170$
& \cite{Mao150302637}\\
Group
& c125-1024
%& WMAP9-like
& $49$
& $10^{13.12\pm 0.11}$
& $9.9\times 10^8$
& $1.1\times 10^{-4}$
& $3.3\times 10^6$
& $360$
& This work\\
Low-mass Cluster
& 1000-1024a
%& Arka
& $33$
& $10^{14.62 \pm 0.11}$
& $6.5\times 10^{10}$
& $2.4\times 10^{-4}$
& $2.2\times 10^8$
& $1200$
& \cite{Bhattacharyya210608292}\\
Cluster
& 1000-1120B
%& Rhapsody
& $96$
& $10^{14.96 \pm 0.03}$
& $5.3\times 10^{10}$
& $6.7\times 10^{-5}$
& $1.8\times 10^8$
& $3295$
& \cite{Wu12093309,Wu12106358}\\
\enddata
{\footnotesize \tablecomments{The first column lists the name of each suite, the second column lists the name of each parent simulation, the third column lists the number of simulations per suite used in our analysis, the fourth column lists the median and standard deviation of the target host halo virial mass distribution, the fifth and sixth columns list the minimum well-resolved subhalo virial mass and sub-to-host halo mass ratio at $z=0$, the seventh column lists the dark matter particle mass in the highest-resolution zoom-in region, the eighth column lists the comoving Plummer-equivalent force softening scale in the same region, and the ninth column lists the original reference for each suite. Host halo properties are listed at $z=0$ for the Group suite; at $z=0.5$, when these hosts were selected from their parent simulation, the host mass range is $10^{12.86 \pm 0.10}~\msun$.}\vspace{-5mm}}
\label{tab:sims}
\end{deluxetable*}

\subsubsection{Group Suite}
\label{sec:group_suite}

Host halos for the Group suite were chosen from a virial mass range of $10^{12.86\pm 0.10}~\msun$ at $z=0.5$ in c125-1024. These hosts are selected in a narrow mass range at $z=0.5$ to more directly relate them to observed strong gravitational lenses studied in recent substructure analyses (e.g., \citealt{Gavazzi0701589,Auger10072880,Gilman190806983}). Hosts in this mass range were selected subject to the constraint that no halo more massive than $10^{13}~ \msun~h^{-1}$ is found within a radius of $3~ \mpc~h^{-1}$ at $z=0$ in the parent box. Sixty-nine percent of halos in this mass range meet this selection criterion. Zoom-in resimulation initial conditions for each system were generated with three refinement regions, yielding an equivalent of 4096 particles per side for the most refined region. The highest-resolution region for each simulation corresponds to the Lagrangian volume containing particles within $10R_{\mathrm{vir,host}}$ of the host halo in the parent box at $z=0$. The dark matter particle mass in the highest-resolution region is $m_{\mathrm{part}}=3.3\times 10^6~ \msun$ and the comoving Plummer-equivalent gravitational softening is $360~ \pc~h^{-1}$, or $0.012$ times the mean interparticle spacing.

Forty-nine host halos were resimulated, resulting in a distribution of $m_{\mathrm{part}}/M_{\mathrm{host}}$ with a median and standard deviation of $(2.4\pm 0.5)\times 10^{-7}$ and an $\epsilon/R_{\mathrm{vir,host}}$ distribution of $(8.3\pm 0.6)\times 10^{-4}$; these hosts are resolved with slightly higher particle counts at $z=0.5$, when they were selected from the parent box. Over the Group suite, zoom-in host halo masses at $z=0$ differ from the target hosts in the parent box by $0.0\pm 3\%$.

\subsubsection{Low-mass Cluster Suite}
\label{sec:lcluster_suite}

The L-Cluster suite was first presented in \cite{Bhattacharyya210608292}, along with a corresponding suite of self-interacting dark matter zoom-in simulations that are not included in this work. We refer the reader to \cite{Bhattacharyya210608292} for a complete description of this suite, and we summarize its properties here.

Host halos for the L-Cluster suite were chosen from a $z=0$ virial mass range of $10^{14.62\pm 0.11}~\msun$ in 1000-1024a. Fifty hosts were selected randomly from all halos in the parent box in this range with no additional isolation criteria. Zoom-in resimulation initial conditions for each system were generated with four refinement regions, yielding an equivalent of $8192$ particles per side for the most refined region. The highest-resolution region for each simulation corresponds to the Lagrangian volume containing particles within $10~\mpc\ h^{-1}$ of the host halo in the parent box at $z=0$. The dark matter particle mass in the highest-resolution region is $m_{\mathrm{part}}=2.2\times 10^8~ \msun$ and the comoving Plummer-equivalent gravitational softening is $1200~ \pc~h^{-1}$, corresponding to $0.010$ times the mean interparticle spacing

Thirty-three host halos were resimulated, resulting in a distribution of $m_{\mathrm{part}}/M_{\mathrm{host}}$ with a median and standard deviation of $(5.9\pm 1.1)\times 10^{-7}$ and an $\epsilon/R_{\mathrm{vir,host}}$ distribution of $(8.8\pm 0.7)\times 10^{-4}$. The L-Cluster zoom-in host mass distribution at $z=0$ is consistent with that in the parent box.

\subsubsection{Cluster Suite}
\label{sec:cluster_suite}

The Cluster suite (or ``Rhapsody'') was first presented in \cite{Wu12093309,Wu12106358}; we refer the reader to this work for a complete description of this suite, and we summarize its properties here.

Host halos for the Cluster suite were chosen from a $z=0$ virial mass range of $10^{14.96\pm 0.03}$ in 1000-1120B. Ninety-six hosts were selected randomly from all halos in the parent box in this range with no additional isolation criteria applied. Zoom-in resimulation initial conditions for each system were generated with four refinement regions, yielding an equivalent of $8192$ particles per side in the most refined region. The highest-resolution region for each simulation is chosen to be $40\%$ larger than the Lagrangian volume containing all friends-of-friends particles of the host halo in the parent box at $z=0$. The dark matter particle mass in the highest-resolution region is $m_{\mathrm{part}}=1.8\times 10^8~ \msun$ and the comoving Plummer-equivalent gravitational softening is $3250~ \pc~h^{-1}$, corresponding to $0.027$ times the mean interparticle spacing.

Ninety-six host halos were resimulated and presented in \cite{Wu12093309,Wu12106358}; we analyze the same systems here. These hosts have a distribution of $m_{\mathrm{part}}/M_{\mathrm{host}}$ with a median and standard deviation of $(2.0\pm 0.1)\times 10^{-7}$ and an $\epsilon/R_{\mathrm{vir,host}}$ distribution of $(18\pm 0.3)\times 10^{-4}$. The Cluster zoom-in host mass distribution at $z=0$ is consistent with that in the parent box.

\subsection{Convergence Limits}

In Appendix \ref{sec:convergence}, we use resimulations of Symphony zoom-ins to test the impact of numerical parameters on subhalo population statistics. These tests include five high-resolution resimulations for each of the LMC, Milky Way, and Group suites, one high-resolution resimulation from the L-Cluster suite \citep{Bhattacharyya210608292}, 96 low-resolution resimulations from the Cluster suite \citep{Wu12093309,Wu12106358}, and three Milky Way--mass hosts that were resimulated using a wide range of force softening scales and time-stepping criteria. From these tests, we conclude that:
\begin{itemize}
    \item Subhalo mass functions (SHMFs) evaluated using virial mass at $z=0$, $M_{\mathrm{sub}}$, are converged (i.e., differ by less than $10\%$ at varying resolution) for subhalos with greater than $300$ particles at $z=0$ in all Symphony~suites.\footnote{To calculate subhalo properties, \textsc{Rockstar} only uses particles that survive one unbinding pass after phase-space groups are determined.}
    \item SHMFs evaluated using peak virial mass, $M_{\rm peak,sub}$, are converged for subhalos that are far from the centers of their hosts at $z=0$ and have greater than $\approx 1000$ particles when $M_{\rm peak,sub}$ is achieved, and become less well converged with decreasing distance. This is consistent with previous studies \citep{VandenBosch171105276,VandenBosch180105427,MansfieldAvestruz2021}, which demonstrate the difficulty of measuring converged subhalo radial distributions at a given~$M_{\mathrm{peak,sub}}$. 
    \item Subhalo maximum circular velocity functions are highly dependent on force softening scale, as shown in previous convergence studies (e.g., \citealt{Ludlow2019,MansfieldAvestruz2021}), but our choice of $\epsilon$ for each suite avoids both suppression of inner densities due to excessive softening and runaway time integration errors due to insufficient softening. 
    \item The time-stepping criterion used in our simulations is well converged, consistent with the results of previous studies (e.g., \citealt{Ludlow2019}).
\end{itemize}

Thus, SHMFs for each Symphony suite are converged for subhalos with virial mass at $z=0$ of
\begin{equation}
M_{\mathrm{sub}}>300m_{\mathrm{part}},\label{eq:convergence}
\end{equation}
corresponding to $1.5\times 10^7~ M_\odot$, $1.2\times 10^8~ M_\odot$, $9.9\times 10^8~ M_\odot$, and $6.5\times 10^{10}~ M_\odot$, and $5.3 \times 10^{10}~ M_\odot$ for the LMC, Milky Way, Group, L-Cluster, and Cluster suites, respectively.

In turn, sub-to-host halo mass ratios within each suite are converged above a conservative limit of $300m_{\mathrm{part}}/\mathrm{min}(M_{\mathrm{host}})$, corresponding to $2.7\times10^{-4}$, $1.1\times10^{-4}$, $1.1\times10^{-4}$, $2.4\times10^{-4}$, and $6.7\times10^{-5}$ for the LMC, Milky Way, Group, L-Cluster, and Cluster suites, respectively. 
Analyses that combine all Symphony simulations across all suites should therefore focus on subhalos with
\begin{equation} \tilde{M}_{\mathrm{sub}} \equiv M_{\rm sub}/M_{\rm host} > 2.7\times 10^{-4}.\label{eq:convergence_ratio}
\end{equation}
This limit is driven by the lowest-mass host in our LMC suite and can be decreased by a factor of a few for analyses using subsets of Symphony suites and/or simulations. In particular, analyses that combine a subset of Symphony zoom-ins from one or multiple suites can be performed using the maximum $M_{\rm sub}/M_{\rm host}$ limit among the subset determined by Equation \ref{eq:convergence}. For analyses of individual zoom-ins, this reduces to an $M_{\mathrm{sub}}>300m_{\mathrm{part}}$ cut.

%---------------------------------------------------------------------------------------
%	SECTION 3
%--------------------------------------------------------------------------------------

\section{Symphony Host Halos}
\label{sec:host_halos}

\begin{figure*}[t!]
\hspace{-3mm}
\includegraphics[trim={0 0.5cm 0 0},width=0.5\textwidth]{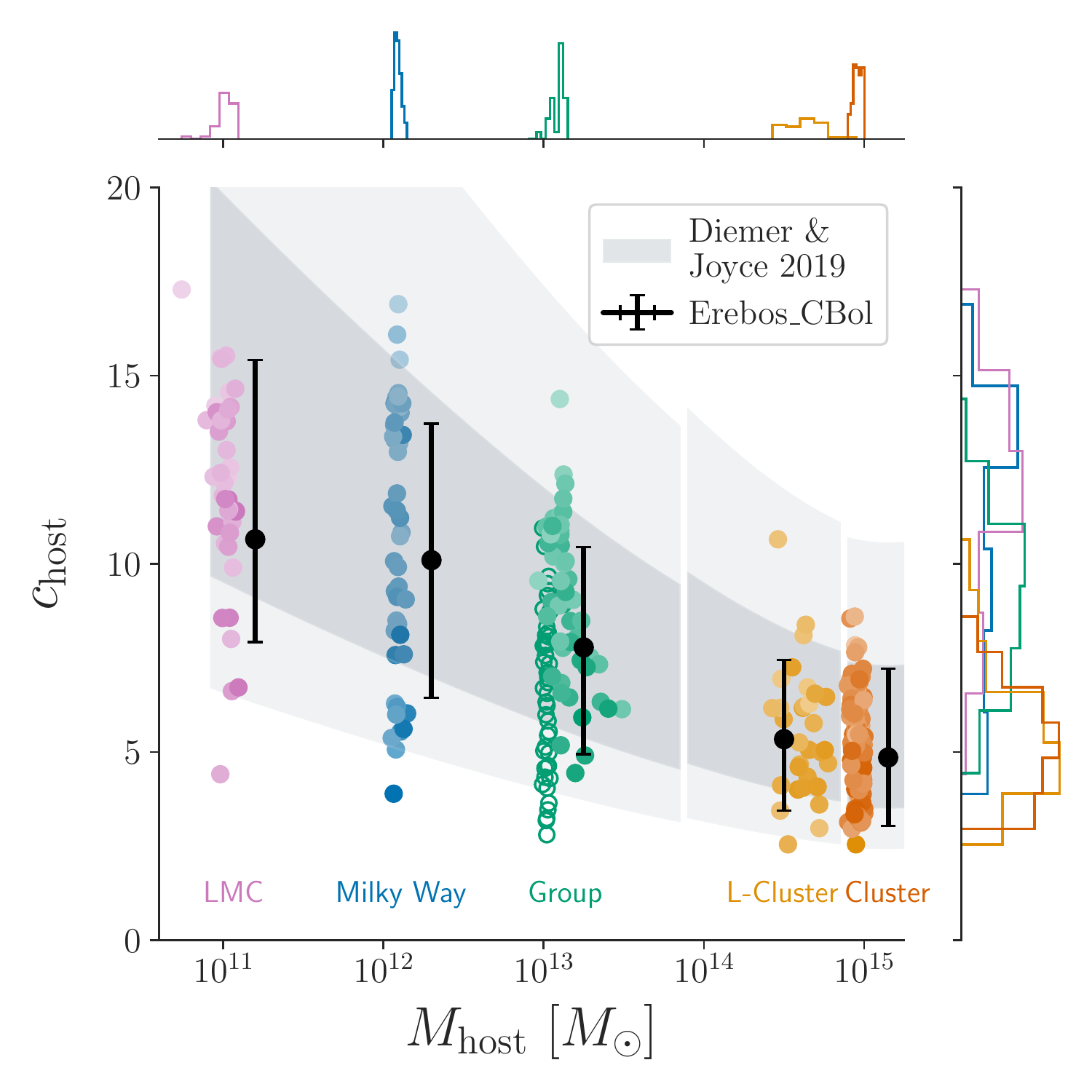}
\includegraphics[trim={0 0.5cm 0 0},width=0.5\textwidth]{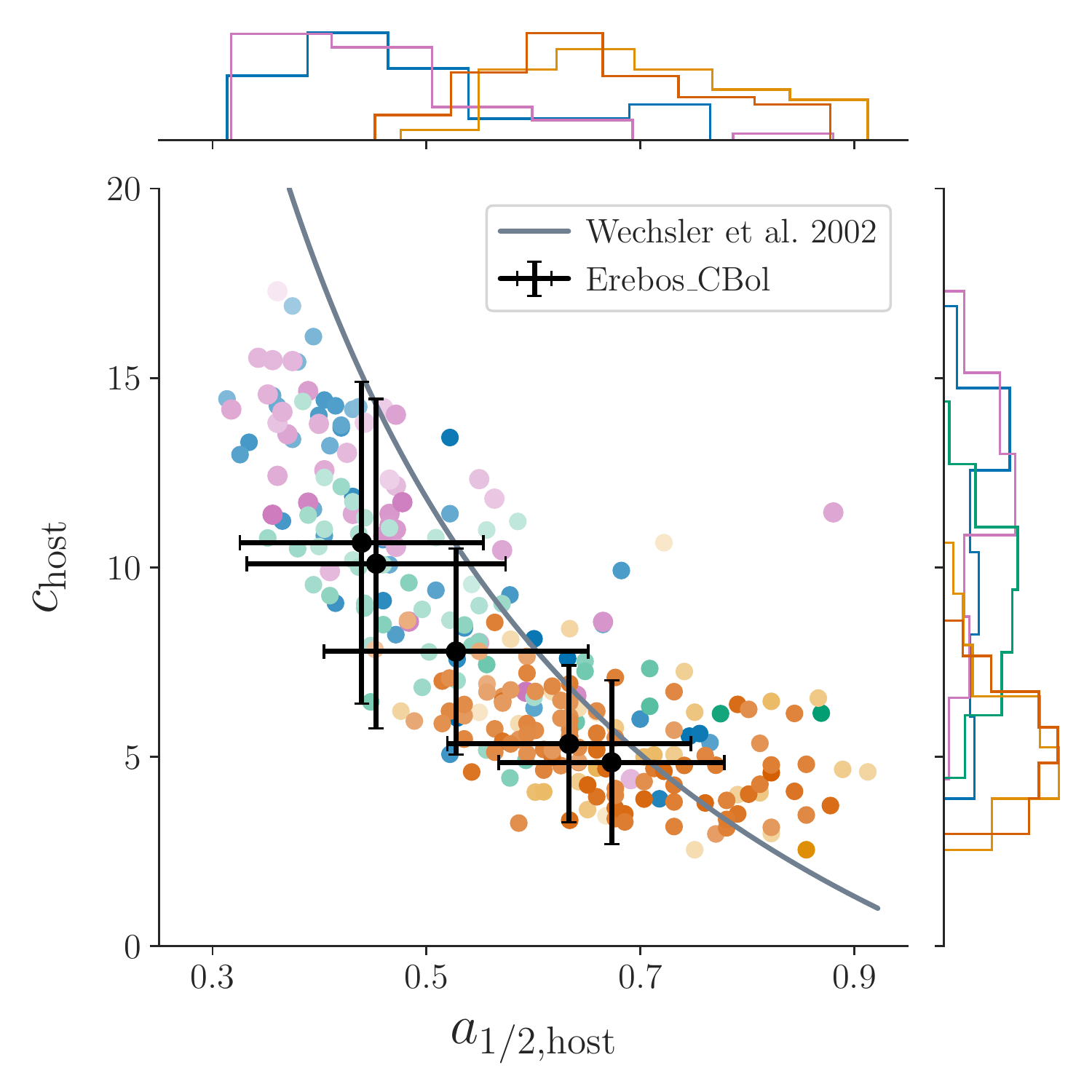}
\caption{Left panel: joint and marginal distributions of $z=0$ halo virial mass and NFW concentration  
for hosts in our zoom-in simulation suites. Each point represents a unique zoom-in simulation with particle mass of $m_{\mathrm{part}} \approx 3\times 10^{-7}M_{\mathrm{host}}$ and is shaded according to the abundance of subhalos with $M_{\mathrm{sub}}/M_{\mathrm{host}}>2.7\times 10^{-4}$; darker colors indicate hosts with higher subhalo abundance relative to the rest of the hosts in each suite. Unfilled Group points show host mass and concentration values at $z=0.5$, when these halos were mass-selected in their parent simulation. The three dark (light) gray bands show the $1\sigma$ ($2\sigma$) uncertainty on the median mass--concentration relation from \cite{Diemer180907326} for our LMC/Milky Way/Group, L-Cluster, and Cluster cosmological parameters, respectively. Black points and error bars show medians and $1\sigma$ scatter from the Erebos simulations \citep{Diemer14011216,Diemer14074730} for halos in each suite's mass range that satisfy the corresponding isolation criteria; these measurements differ from the \cite{Diemer180907326} relation for the LMC, Milky Way, and Group suites due to these cuts. Right panel: joint and marginal distributions of host half-mass scale factor and NFW concentration. The gray line shows the prediction for the $c_{\mathrm{host}}(a_{1/2,\mathrm{host}})$ relation from \cite{Wechsler0108151}, for reference; this model is calibrated using $\sigma_8=1.0$, which contributes to the normalization offset, and is not re-fit here. Manual inspection of individual mass accretion histories (MAHs) indicates that points with large $a_{1/2,\mathrm{host}}$ for a given $c_{\mathrm{host}}$ recently experienced a major merger.}
\label{fig:Mhost_chost}
\end{figure*}

We now present the properties of Symphony host halos (Section \ref{sec:host_halos_overview}), including concentrations and formation times (Section \ref{sec:mc_relation}), correlations between secondary host halo properties and the abundance of surviving subhalos at $z=0$ (Section \ref{sec:host_correlation}), mass accretion histories (Section \ref{sec:mah}), and density profiles (Section \ref{sec:density}).

We defer from providing updated fits to standard halo relations (e.g., the mass--concentration relation) due to the diverse environmental properties of our zoom-in suites. A more detailed comparison of our results to measurements from existing cosmological and zoom-in simulations, and an exploration of additional host halo properties (e.g., host halo shapes) are beyond the scope of this work.

\subsection{Overview}
\label{sec:host_halos_overview}

\begin{deluxetable}{{c@{\hspace{0.07in}}c@{\hspace{0.07in}}c@{\hspace{0.07in}}c@{\hspace{0.07in}}c}}[t!]
\centering
\tablecolumns{5}
\tablecaption{Properties of Symphony Host Halos.}
\tablehead{
\colhead{Zoom-in Suite} & $c_{\mathrm{host}}$ & $a_{1/2,\mathrm{host}}$ & $N_{\mathrm{sub,resolved}}$ & 
$\tilde{N}_{\mathrm{sub,resolved}}$}
\startdata
LMC-mass
& $12.2\pm 2.7$
& $0.46\pm 0.11$
& $57 \pm 17$
& $39 \pm 9$\\
Milky Way--mass
%& WMAP9-like
& $10.8\pm 3.5$
& $0.46\pm 0.12$
& $83 \pm 18$
& $32 \pm 9$\\
Group
%& WMAP9-like
& $9.0\pm 2.1$
& $0.50\pm 0.11$
& $136 \pm 55$
& $37 \pm 8$\\
Low-mass Cluster
%& Arka
& $5.0\pm 1.7$
& $0.70\pm 0.11$
& $104 \pm 42$
& $57 \pm 11$\\
Cluster
%& Rhapsody
& $5.3\pm 1.3$
& $0.64\pm 0.10$
& $210 \pm 31$
& $56 \pm 11$\\
\enddata
{\footnotesize \tablecomments{The first column lists the name of each suite, and the second, third, fourth, and fifth columns list the median and $1\sigma$ host-to-host scatter of the zoom-in host halo's concentration, half-mass scale factor, and the abundance of surviving subhalos at $z=0$ above resolution limits of $M_{\mathrm{sub}}>300m_{\mathrm{part}}$ and $M_{\mathrm{sub}}/M_{\mathrm{host}}>2.7\times 10^{-4}$, respectively.}\vspace{-8mm}}
\label{tab:hosts}
\end{deluxetable}

Table \ref{tab:hosts} lists the median and standard deviation of the concentration and half-mass formation time distribution for hosts in each Symphony suite, along with subhalo abundance above our absolute (Equation \ref{eq:convergence}) and normalized (Equation \ref{eq:convergence_ratio}) convergence limits.

Symphony host halos are typically resolved with several million particles, the large majority of which are the highest-resolution particle type in each suite. We verify this by measuring the distance from the center of the host within which the highest-resolution particles contribute $>90\%$ of the total mass. This ``contamination radius'' is always several times (and often about an order of magnitude) larger than the virial radius of the host halo, consistent with previous findings for the original Milky Way suite \citep{Wang210211876}.

The LMC, Milky Way, and Group hosts are resolved with $\gtrsim 10^5$ high-resolution particles at $z\lesssim 3$, while the L-Cluster and Cluster hosts are resolved with comparable particle counts at $z\lesssim 1.5$.\footnote{Note that the LMC, Milky Way, and Group halo catalogs extend to $z\approx 20$, while the L-Cluster and Cluster halo catalogs extend to $z\approx 12$.} Furthermore, hosts in all suites are typically resolved with $\gtrsim 300$ particles at $z\lesssim 7$. These properties enable the robust measurements of hosts' formation histories and dark matter structure presented below.

\subsection{Host Halo Concentrations and Formation Times}
\label{sec:mc_relation}

Figure \ref{fig:Mhost_chost} shows the host halo mass, concentration, and formation time distributions for all Symphony suites. In particular, the left panel of Figure \ref{fig:Mhost_chost} shows our hosts' mass--concentration ($M_{\rm host}$--$c_{\rm host}$) relation, and the right panel shows the relationship between $c_{\rm host}$ and $a_{1/2,\mathrm{host}}$ for our hosts. Here, $c_{\rm host}\equiv R_{\rm vir,host}/R_{s,\mathrm{host}}$ is the Navarro--Frenk--White (NFW) concentration \citep{Navarro1997}, and $a_{1/2,\mathrm{host}}$ is the scale factor at the time when a host first reached half its $z=0$ mass. For the purposes of this subsection, we assume that Symphony hosts are adequately described by NFW profiles; we explore this assumption in Section \ref{sec:density}. Both the $M_{\mathrm{host}}$--$c_{\mathrm{host}}$ and $a_{1/2,\mathrm{host}}$--$c_{\mathrm{host}}$ relations are well studied (e.g. \citealt{Navarro1997,Wechsler0108151,vandenBosch14092750,Ludlow160102624}) and our results are qualitatively consistent with earlier findings that at $z=0,$ $c_{\rm host}$ tends to decrease with increasing halo mass and increasing $a_{1/2,\mathrm{host}}.$

We quantitatively compare Symphony results against the \cite{Diemer180907326} $M_{\rm host}$--$c_{\rm host}$ relation in the left panel of Figure \ref{fig:Mhost_chost}. To account for Symphony hosts' environmental properties, we also compare against results from the Erebos\_CBol cosmological simulation suite \citep{Diemer14011216,Diemer14074730} using identical mass and isolation cuts to our zoom-in suites. Specifically, we compare each suite against the largest Erebos box that resolves halos in the corresponding host mass range.\footnote{Thus, we compare the LMC, Milky Way, Group, L-Cluster, and Cluster suites to Erebos\_CBol\_L63, L125, L250, L1000, and L2000, respectively, based on the ``0\%'' convergence criteria from \cite{MansfieldAvestruz2021}. Note that Erebos\_CBol adopts a different cosmology compared to any of our suites, with $\Omega_m = 0.27$, $h=0.7$, and $\sigma_8=0.82$; however, its cosmology is the closest to our simulations out of the suites in \cite{MansfieldAvestruz2021} that resolve LMC-mass halos.} Note that the \cite{Diemer180907326} $M_{\rm host}$--$c_{\rm host}$ relation is calibrated using the Erebos suite, meaning that the difference between the \cite{Diemer180907326} and Erebos results in Figure \ref{fig:Mhost_chost} is due to environmental selection effects.

Symphony hosts display lower median concentrations and smaller scatter than the sample of halos from cosmological simulations used to measure the mass--concentration relation in \cite{Diemer180907326}, particularly for LMC-mass halos (see \citealt{Diemer180907326} for a comparison to other mass--concentration relation models and simulation measurements). This follows because our isolation criteria require that Symphony's LMC, Milky Way, and Group hosts are not close to more massive halos. Thus, our host samples should not include ``splashback'' subhalos, which have previously passed through the virial radius of a larger host and form the entire high-concentration tail of the cosmological mass--concentration relation \citep{Mansfield190200030}. Furthermore, our isolation cuts also remove many nonsplashback halos whose growth has slowed due to tidal truncation or the high velocity dispersion of dark matter particles in dense environments. The trend toward lower concentration that we report is expected for zoom-in simulations in general: accurately simulating splashback halos at high resolution requires simulating their more massive hosts at identical mass resolution and thus much higher particle counts, which is often impractical.

With appropriate isolation criteria imposed, Symphony's LMC, Milky Way, Group, and L-Cluster concentration distributions are consistent with Erebos halos: two-sample Kolmogorov--Smirnov (K-S) tests yield $p>0.01$ in each case. Meanwhile, the Cluster distributions are statistically distinguishable ($p<10^{-5}$) because Symphony lacks the high-concentration tail measured in Erebos. No isolation criteria were imposed on our Cluster hosts; thus, selection effects cannot explain this difference. However, we note that our Cluster hosts were simulated with a cosmology that is different from the Erebos\_CBol suite, and that this is a plausible explanation for the difference.

For all suites, Symphony displays $\approx 30\%$ smaller scatter in concentration relative to Erebos, with typical values of $\sigma_{\mathrm{log}c}\approx 0.12~ \mathrm{dex}$ rather than $\sigma_{\mathrm{log}c}\approx 0.16~ \mathrm{dex}$. This discrepancy can be understood as a result of Erebos' lower resolution: the Erebos halos we match to each Symphony suite typically have $\approx 1\times 10^3$--$4\times 10^3$ particles. Assuming that any additional scatter relative to Symphony results from finite-resolution effects, \cite{Benson181206026} predicted that Symphony halos resolved with typical Erebos particle counts will display $\sigma_{\mathrm{log}c}\approx 0.18~ \mathrm{dex}$, which is similar to the scatter \cite{Diemer180907326} reported. Explicit resolution tests on Erebos halos would be needed to confirm this explanation; if it holds, our results imply that $\sigma_{\mathrm{log}c}\approx 0.12~ \mathrm{dex}$ is closer to a converged estimate of the concentration scatter in a sample without environmental constraints.

Finally, we note that recent major mergers may also play a role in setting the concentration scatter (e.g., \citealt{Wang200413732}), which is hinted at by the increased concentration in $c_{\mathrm{host}}$ at large $a_{1/2,\mathrm{host}}$ in Figure \ref{fig:Mhost_chost}. We discuss the mass--concentration relation scatter further when comparing to \textsc{Galacticus} predictions in Section \ref{sec:galacticus_host}.

\subsection{Correlations between Secondary Host Halo Properties and Subhalo Abundance}
\label{sec:host_correlation}

\begin{figure}[t!]
\hspace{-3.5mm}
\includegraphics[width=0.475\textwidth]{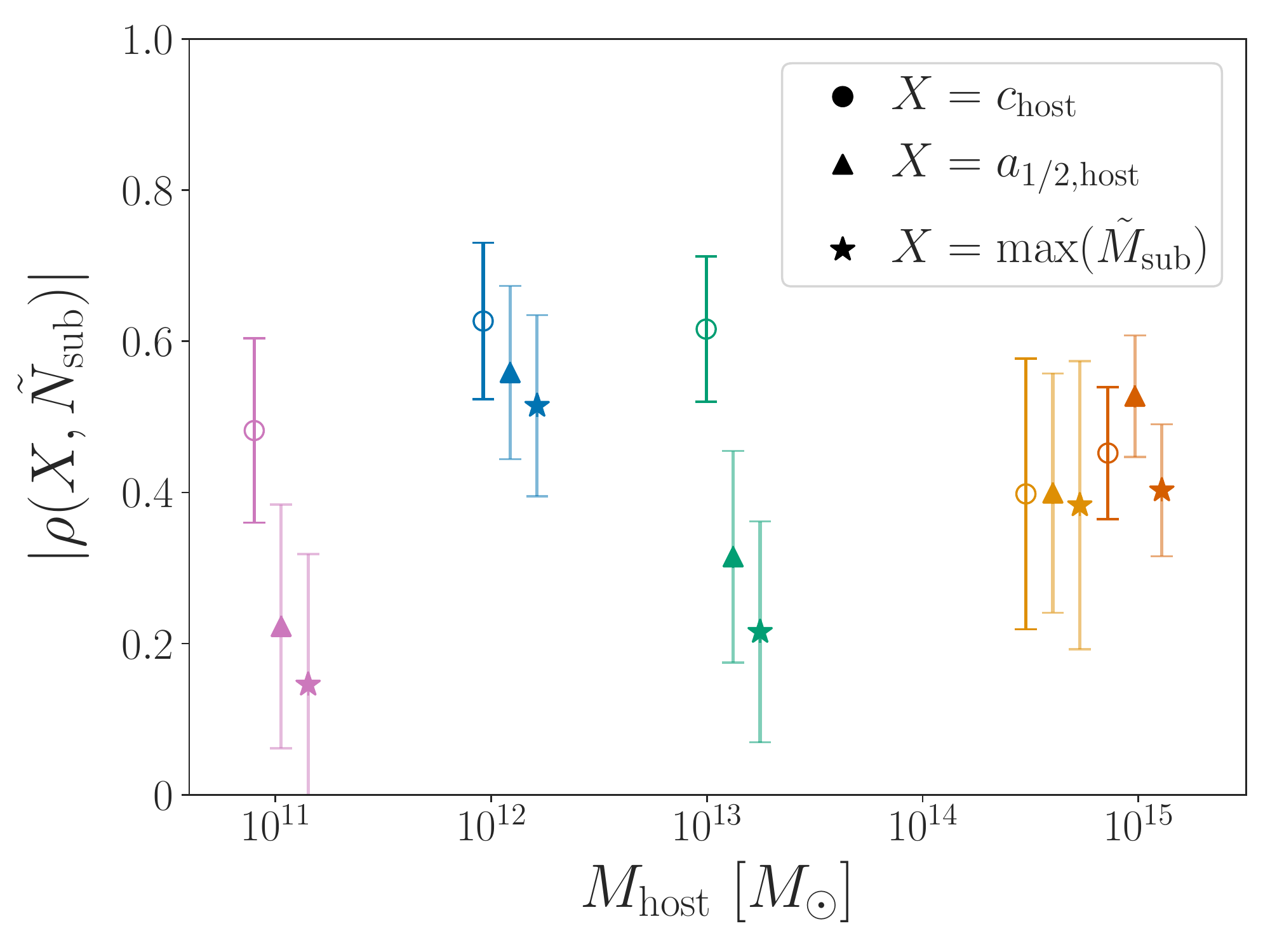}
\caption{Spearman correlation coefficients between subhalo abundance and concentration (circles), half-mass scale factor (triangles), and the sub-to-host halo mass ratio of the largest subhalo (stars) for hosts in each Symphony suite. The unfilled circles indicate that we show the absolute value of the anticorrelation with host concentration; the other correlation coefficients are positive. Error bars indicate $1\sigma$ jackknife uncertainties. Note that the horizontal error bars are smaller than the markers, and that the three error bars for each suite are offset horizontally for visual clarity. Subhalo abundance is measured above our conservative sub-to-host halo convergence limit of $M_{\mathrm{sub}}/M_{\mathrm{host}}>2.7\times 10^{-4}$ for all suites.}
\label{fig:rhoX}
\end{figure}

Next, we study the relationship between hosts' secondary properties (i.e., properties beyond host halo mass) and the abundance of surviving subhalos at $z=0$. Figure \ref{fig:rhoX} shows that the abundance of subhalos above our minimum sub-to-host halo mass ratio (Equation \ref{eq:convergence_ratio}),
\begin{equation}
    \tilde{N}_{\mathrm{sub}} \equiv N_{\mathrm{sub}}(\tilde{M}_{\mathrm{sub}}>2.7\times 10^{-4}),
\end{equation}
anticorrelates with host concentration at (roughly) fixed host mass within each suite. These anticorrelations between $\tilde{N}_{\mathrm{sub}}$ and $c_{\mathrm{host}}$ are significant (Spearman-$\rho$ tests yield $p<0.01$ within each suite) and, with the exception of the L-Cluster suite, are much stronger than the residual correlations with host mass. The strength of these anticorrelations peaks at the Milky Way host halo mass scale, with $\rho = -0.48$, $-0.63$, $-0.62$, $-0.41$, and $-0.45$ for the LMC, Milky Way, Group, L-Cluster, and Cluster suites, respectively. We caution that this trend with host mass is only significant at the $1\sigma$ level given the jackknife uncertainties on the measurement for each suite (see Figure \ref{fig:rhoX}). These findings are consistent with previous results (e.g., \citealt{Zentner0411586})---including several based on the original Milky Way suite \citep{Mao150302637,Fielder180705180,Nadler210107810}---and extend them to a wider range of host halo masses at a fixed, high-resolution sub-to-host halo mass ratio.

For the LMC and Group suites, subhalo abundance anticorrelates more strongly and significantly with host concentration than it correlates with host formation time, which we parameterize by $a_{1/2,\mathrm{host}}$. Meanwhile, the correlations between subhalo abundance and these two host properties are consistent with one another for the remaining suites. We note that Cluster subhalo abundances are slightly more sensitive to formation time than host concentration. This may be due to the prevalence of late major mergers in these systems (see the right panel of Figure \ref{fig:Mhost_chost}) and associated increases in subhalo abundance (e.g., \citealt{DSouza21041324i}). 

Several factors likely contribute to the dependence of these correlations on host halo mass. First, subhalos of early-forming hosts accrete earlier and are stripped for longer than subhalos of late-forming hosts, on average. This leads to a positive correlation between $N_{\mathrm{sub}}$ and $a_{1/2,\mathrm{host}}$; in turn, $a_{1/2,\mathrm{host}}$ anticorrelates with $c_{\rm host}$. Second, the tidal radius of a subhalo orbiting a host becomes smaller as the enclosed mass within the subhalo's orbit, $M_{\mathrm{host}}(<r),$ becomes larger and as the slope of the host's mass profile, $d\,\ln{M_{\mathrm{host}}}/d\,\ln{r}$ becomes steeper (e.g., \citealt{VandenBosch171105276}). For a fixed orbital radius and host mass, higher-concentration hosts have larger enclosed masses and steeper mass profiles than lower-concentration hosts. Thus, subhalos of higher-concentration hosts have smaller tidal radii and higher mass-loss rates on average, causing an anticorrelation between $N_{\mathrm{sub}}$ and $c_{\rm host}$; in turn, $c_{\rm host}$ anticorrelates with $a_{1/2,\mathrm{host}}$. Because $N_{\mathrm{sub}}$ anticorrelates more strongly with $c_{\rm host}$ than it correlates with $a_{1/2,\mathrm{host}}$ for our LMC, Milky Way, and Group suites, the mechanism related to the host's mass profile may be more important than the infall time effect for these host halo masses. On the other hand, $c_{\rm host}$ might simply trace accretion histories better than single-epoch measurements like $a_{1/2,\mathrm{host}}$ (e.g., \citealt{Chue2018}); a detailed follow-up study that considers hosts' entire accretion histories would be required to disentangle these effects further. 

Figure \ref{fig:rhoX} also shows that the mass of each host's largest subhalo (measured in terms of the maximum sub-to-host halo mass ratio in each zoom-in) correlates with subhalo abundance in a similar manner to formation time. Correlations between subhalo abundance and the properties of the largest surviving subhalo are particularly relevant because observations of Milky Way analogs indicate that the luminosity of the brightest surviving satellite (which is related to the ``magnitude gap'') correlates more strongly with satellite abundance than the central's stellar mass \citep{Geha170506743,Mao200812783}. We intend to study this effect by applying galaxy--halo connection models to Symphony in future work.

\subsection{Mass Accretion Histories}
\label{sec:mah}

\begin{figure*}[t!]
\centering
\includegraphics[width=\textwidth]{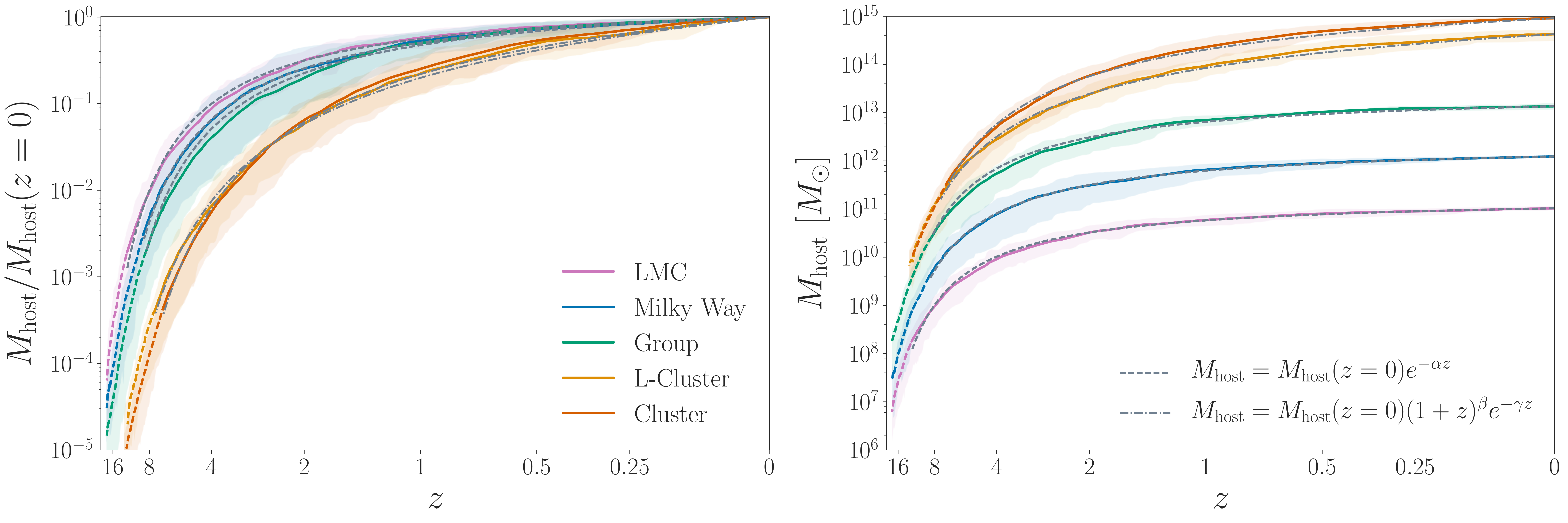}
\caption{Left panel: MAHs for the host halos in our five zoom-in simulation suites normalized to the $z=0$ virial mass for each host. Solid lines show mean MAHs for all hosts in each suite, and shaded bands show the corresponding $68\%$ host-to-host scatter about the mean. Lines transition to dashed at the redshift above which $>5\%$ of the hosts in each suite are resolved with fewer than $300$ particles. Jackknife resampling of the LMC, Milky Way, and Group suites indicates that the difference between normalized MAHs is marginally significant at $z\gtrsim 2$ and not significant at later times; the difference between L-Cluster and Cluster MAHs is not statistically significant (see Section \ref{sec:mah} for details). Dashed (dotted--dashed) gray lines show best-fit MAHs using the \cite{Wechsler0108151} exponential model and the \cite{Tasitsiomi0311062} exponential plus power-law model adopted by \cite{McBride09023659}; best-fit parameters are provided in the text. Right panel: same as the left panel, but not normalized to $z=0$ host halo virial mass.}
\label{fig:MAH}
\end{figure*}

Figure \ref{fig:MAH} shows the normalized and un-normalized mean MAHs for host halos in each Symphony suite. As suggested by the half-mass formation times in Figure \ref{fig:Mhost_chost} and demonstrated by previous studies (e.g., \citealt{Wechsler0108151}), the normalized MAHs demonstrate that lower-mass hosts build up their mass earlier, on average. Our fiducial-resolution cut conservatively requires that halos have more than 300 particles at any redshift: more than 95\% of hosts meet this cut at $z<11.8$ (LMC), $z<8.7$ (Milky Way), $z<8.5$ (Group), $z<7.4$ (L-Cluster), and $z<6.7$ (Cluster).

The dependence of accretion history on host mass is strongest for the transition from the Group to L-Cluster suites, and is not statistically significant between the L-Cluster and Cluster suites. To quantify this, we jackknife resample hosts within each suite and compare the ratios of the jackknifed MAHs among suites. The LMC vs.\ Milky Way and Milky Way vs.\ Group MAHs are distinguishable at $1\sigma$ for $z\gtrsim 2$ given these uncertainties, and are consistent at later times. Meanwhile, the L-Cluster and Cluster MAHs are statistically consistent at all redshifts we resolve. Note that our Group, Milky Way, and particularly LMC hosts occupy underdense environments relative to typical halos of these masses due to our isolation criteria (see Section \ref{sec:symphony_overview}), which biases these hosts toward later formation times than average (e.g., see \citealt{Mansfield190200030} for a review). Thus, the dependence of formation time on host halo mass without selection effects would be stronger than we report, although comparisons to MAHs in c125-1024 indicate that this systematic bias is much smaller than the host-to-host scatter in formation times within Symphony suites.

Because of the narrow mass ranges from which our hosts were selected in their parent simulations, the un-normalized MAHs display $\lesssim 0.1~ \mathrm{dex}$ host-to-host scatter near $z=0$ for all suites (see Table \ref{tab:sims} for the host halo mass distributions). Note that the Group suite displays its smallest host-to-host scatter near $z=0.5$, corresponding to the redshift at which these hosts were mass-selected in their parent box.

MAHs have been studied extensively using both empirical models calibrated to cosmological simulations (e.g., \citealt{Wechsler0108151,McBride09023659,Wu12093309,Hearin210505859}) and (semi)analytic prescriptions (e.g., \citealt{vandenBosch0105158,vandenBosch14092750,Correa14095228}). We find that the one-parameter exponential model introduced in \cite{Wechsler0108151},
\begin{equation}
    M_{\mathrm{host}}(z) = M_{\mathrm{host}}(z=0)e^{-\alpha z},
\end{equation}
fits the mean LMC, Milky Way, and Group MAHs well, with exponents of $\alpha=0.58$, $0.68$, and $0.78$, respectively. This model cannot accurately fit the L-Cluster and Cluster MAHs over the entire redshift range, consistent with the results of \cite{Wu12093309} using the Cluster suite. Instead, the two-parameter exponential plus power-law model from \cite{Tasitsiomi0311062} and adopted by \cite{McBride09023659},
\begin{equation}
    M_{\mathrm{host}}(z) = M_{\mathrm{host}}(z=0)(1+z)^{\beta}e^{-\gamma z},
\end{equation}
fits the L-Cluster and Cluster MAHs well, with best-fit parameters of $(\beta,\gamma)= (-1.34,0.69)$, and $(-0.81,0.94)$, respectively.\footnote{We find that the one-parameter exponential plus power-law model from \cite{Wu12093309} does not fit the mean L-Cluster and Cluster MAHs as well as the \cite{McBride09023659} model.} The \textsc{Galacticus} model we compare to in Section \ref{sec:galacticus} predicts MAHs that agree well with the full Symphony distributions, indicating that semianalytic models calibrated to simulations can capture hosts' MAH distributions over several decades of $z=0$ halo mass.

\pagebreak

\subsection{Host Halo Density Profiles}
\label{sec:density}

Next, we study the density profiles of Symphony host halos. We begin by reviewing standard halo profiles to contextualize our results. The NFW profile is given~by
\begin{equation}
    \rho(r) \propto x^{-1}(1+x)^{-2}
    \label{eq:NFW}
\end{equation}
for $x\equiv r/r_{s}$, where $r_s$ is the radius where the logarithmic slope is -2 \citep{Navarro1997}. The amplitude of the NFW profile is fixed for a given $M_{\rm host}$ and $r_s$, yielding a single-parameter fit in terms of $r_{s}$. The Einasto profile is given by 
\begin{equation}
    \frac{d\ln{\rho}}{d\ln{r}} \propto x^\alpha,
\end{equation}
which becomes a two-parameter fit in terms of $r_{s}$ and $\alpha$ when $M_{\rm host}$ is fixed. Einasto profiles give more accurate descriptions of inner halo profiles than NFW profiles, even if $\alpha$ is held fixed to values in the range $\alpha\approx0.17$--$0.18$ (e.g., \citealt{Navarro2004,Wang191109720}). We denote the maximum value of a halo's rotation curve, $V(<r)=\sqrt{GM(<r)/r}$, by $V_{\rm max}$, and we denote the radius at which $M(<r)=M_{\rm}/2$ by $R_{1/2}$; for hosts (subhalos), these variables are labeled by ``host'' (``sub'') subscripts. When scaled by the virial circular velocity and virial radius, respectively, both become robust, model-independent measures of halo concentration. All one-parameter halo models, including the NFW and Einasto profiles, predict a specific relationship between these halo properties.

\begin{figure*}[t!]
\centering
\includegraphics[width=0.49\textwidth]{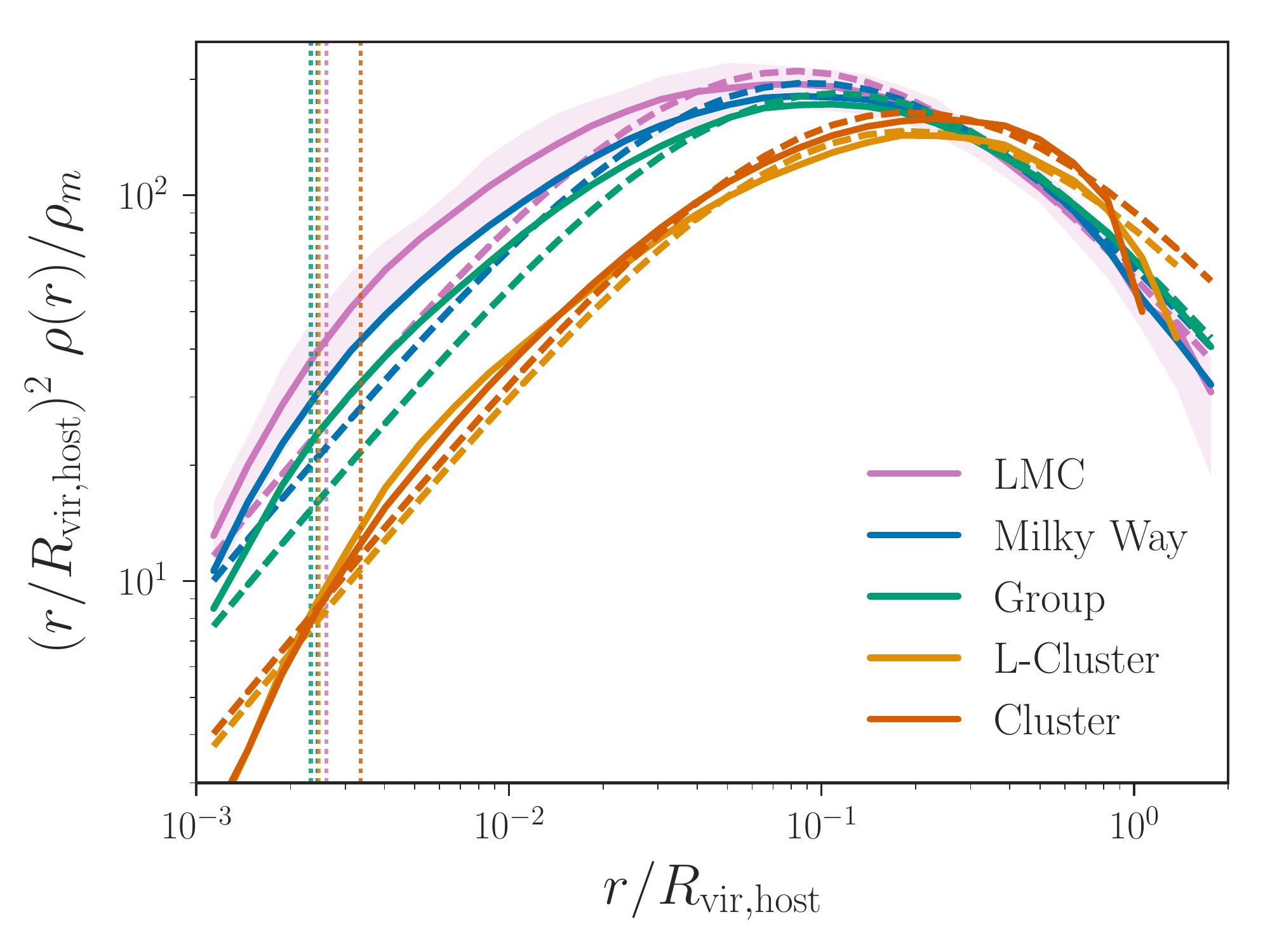}
\includegraphics[width=0.49\textwidth]{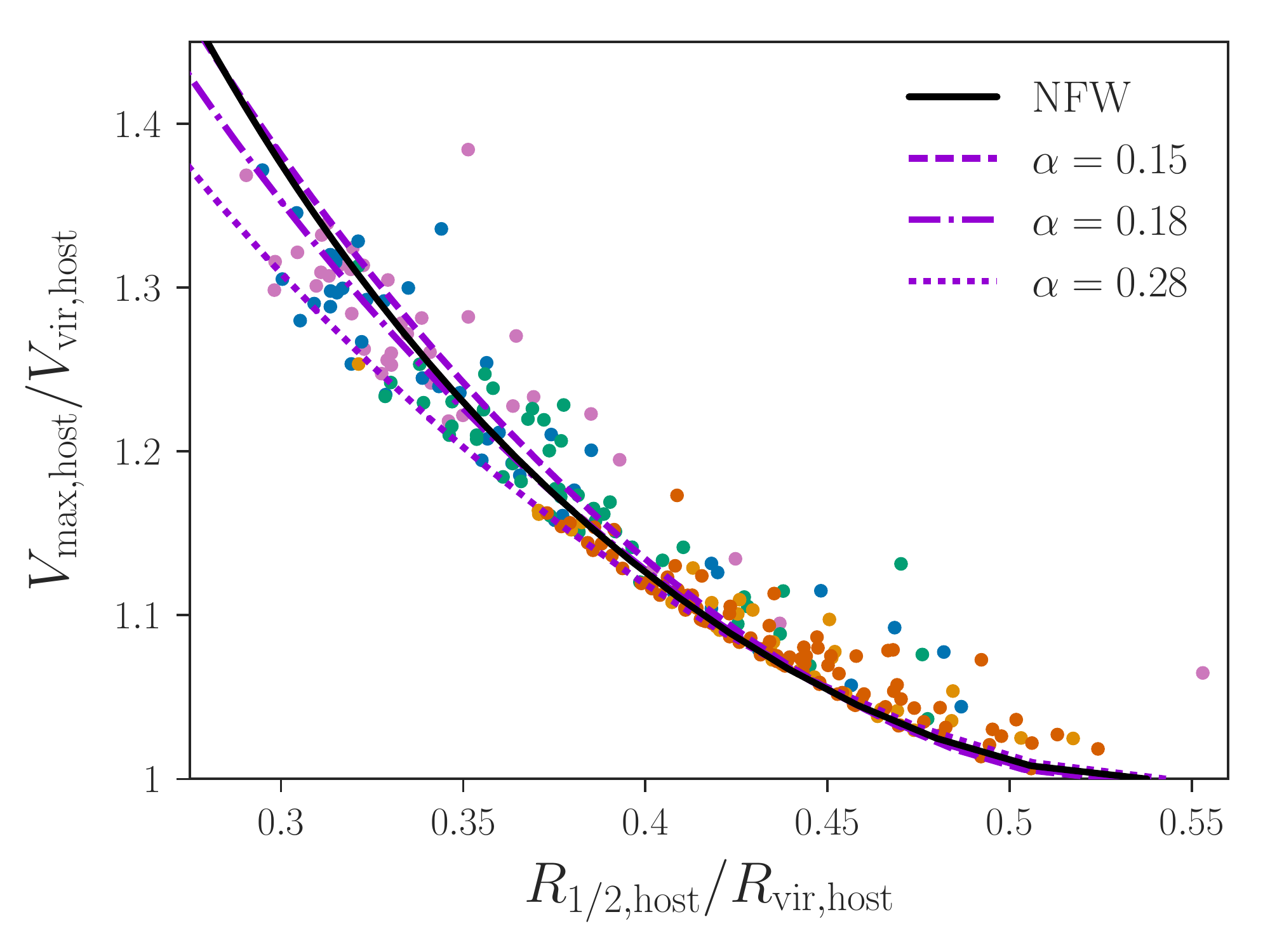}
\caption{Left panel: mean density profiles of Symphony host halos normalized to the mean cosmological matter density at $z=0$, as a function of distance from the host center in units of the virial radius, for each suite (solid lines). Dashed lines show the mean of the best-fit NFW profiles reported by \textsc{Rockstar} for each suite. Shaded bands indicate $68\%$ host-to-host scatter, which is only shown for the LMC suite for visual clarity. Dotted vertical lines show approximate convergence radii of $2.8\epsilon$ (e.g., \citealt{Ludlow2019}) for each suite, below which density profile measurements may not be converged. Right panel: maximum circular velocity normalized to the virial velocity versus half-mass radius normalized to the virial radius for all host halos in our five zoom-in suites. Points are colored by suite following the color scheme in the left panel. The black line shows the relation for an NFW profile, and purple lines show relations for Einasto profiles with $\alpha=0.15$ (solid), $\alpha=0.18$ (dotted--dashed), and $\alpha=0.28$ (dotted).}
\label{fig:density}
\end{figure*}

The left panel of Figure \ref{fig:density} shows Symphony host halo density profiles as a function of distance from the host center, in units of the virial radius, measured directly from the particle snapshots and stacked over all hosts in each suite. Dashed lines show the mean of the best-fit NFW profiles reported by \textsc{Rockstar} for each suite, which are consistent with our direct measurements at the $10\%$ level for $r/R_{\mathrm{vir,host}}\gtrsim 10^{-2}$. For distances between the ``convergence radius'' of $\approx 2.8\epsilon$ (e.g., \citealt{Ludlow2019}) and $r/R_{\mathrm{vir,host}}\approx 10^{-2}$, Symphony profiles are systematically denser than predicted by \textsc{Rockstar}'s NFW fit, consistent with previous measurements from cosmological and zoom-in simulations (e.g., \citealt{Navarro08101522,Ludlow13020288}). Furthermore, host concentrations estimated from direct profile measurements for each suite are consistent with the median values derived from \textsc{Rockstar}, in line with previous studies using the original version of our Milky Way and our Cluster suites \citep{Wu12106358,Fielder200702964}.

The right panel of Figure \ref{fig:density} compares the integrated profile properties $V_{\rm max,host}$ and $R_{1/2,\rm{host}}$ against the predictions of various halo models. The solid and dashed curves in this panel show the $V_{\rm max,host}/V_{\rm vir,host}-R_{1/2,\rm{host}}/R_{\rm vir,host}$ relation predicted for NFW profiles and for Einasto profiles across a range of $\alpha$ values. The points show Symphony host halos, color-coded by suite. At high host halo masses (i.e., mainly for L-Cluster and Cluster halos), Symphony hosts consistently lie above the NFW and Einasto curves, with $V_{\rm max,host}/V_{\rm vir,host}$ values up to $\approx 10\%$ higher than those models at fixed $R_{1/2,\mathrm{host}}/R_{\rm vir,host}$. Meanwhile, at lower host masses and typical values of $R_{1/2,\mathrm{host}}/R_{\rm vir,host}$, the simulation measurements scatter both above and below the NFW and Einasto predictions.

We hypothesize that the scatter toward larger values of $V_{\rm max,host}/V_{\rm vir,host}$ for high-mass hosts may be caused by recent major mergers, which heat the inner regions of host halos. At all masses, halo triaxiality (e.g., \citealt{Jing0202064}) and substructure (e.g., \citealt{Fielder200702964}) also contribute to the scatter about the predictions for smooth, spherically symmetric NFW and Einasto profiles. Studying these effects across our entire range of host halo mass is an interesting avenue for future work. In addition, Symphony density profile measurements will enable studies of the detailed correlation between host halos' dark matter structure and subhalo population statistics, both at (roughly) fixed host halo mass within each suite and over several decades of host mass using the entire compilation.

\section{Symphony Subhalo Populations}
\label{sec:subhalo_populations}

We now present results for subhalo populations (Section \ref{sec:subhalo_populations_overview}), including SHMFs (Section \ref{sec:shmf}) and radial distributions (Section \ref{sec:radial}); several auxiliary characteristics of the subhalo populations, including infall time and tidal stripping distributions, are presented in Appendix \ref{subhalo_appendix}.

Again, we defer from providing updated fits to standard subhalo population statistics (e.g., the SHMF), and we leave an investigation of additional subhalo population properties (e.g., subhalo orbital parameters) to future work.

\subsection{Overview}
\label{sec:subhalo_populations_overview}

Subhalos above the Symphony convergence limit contain at least 300 particles at $z=0$ and typically have fallen into their hosts several gigayears ago with $\approx 50\%$ more particles than at $z=0$ (see Appendices \ref{infall_times}--\ref{stripping_distributions}). Appendix \ref{sec:convergence} demonstrates that the abundance of Symphony subhalos above this limit is converged at the $\approx 10\%$ level; because even higher resolution is required to track the inner structure of stripped subhalos accurately (e.g., \citealt{Errani200107077}), we focus on SHMFs and radial distributions here, and we study the convergence properties of subhalo maximum circular velocity functions in Appendix \ref{sec:convergence_eps}.

For the LMC, Milky Way, Group, L-Cluster, and Cluster suites, subhalos above our sub-to-host halo convergence limit contribute $10\%$, $12\%$, $18\%$, $21\%$, and $16\%$ of the total mass within their host halo's virial radius, respectively. On average, $84\%$, $81\%$, $86\%$, $69\%$, and $73\%$ of these objects are ``first-order'' subhalos, meaning that they do not lie within the virial radius of any large halo except for the main host (see \citealt{Springel08090898} for a discussion of subsubstructure measurement subtleties).

\subsection{Subhalo Mass Functions}
\label{sec:shmf}

\begin{figure*}[t!]
\centering
\includegraphics[width=\textwidth]{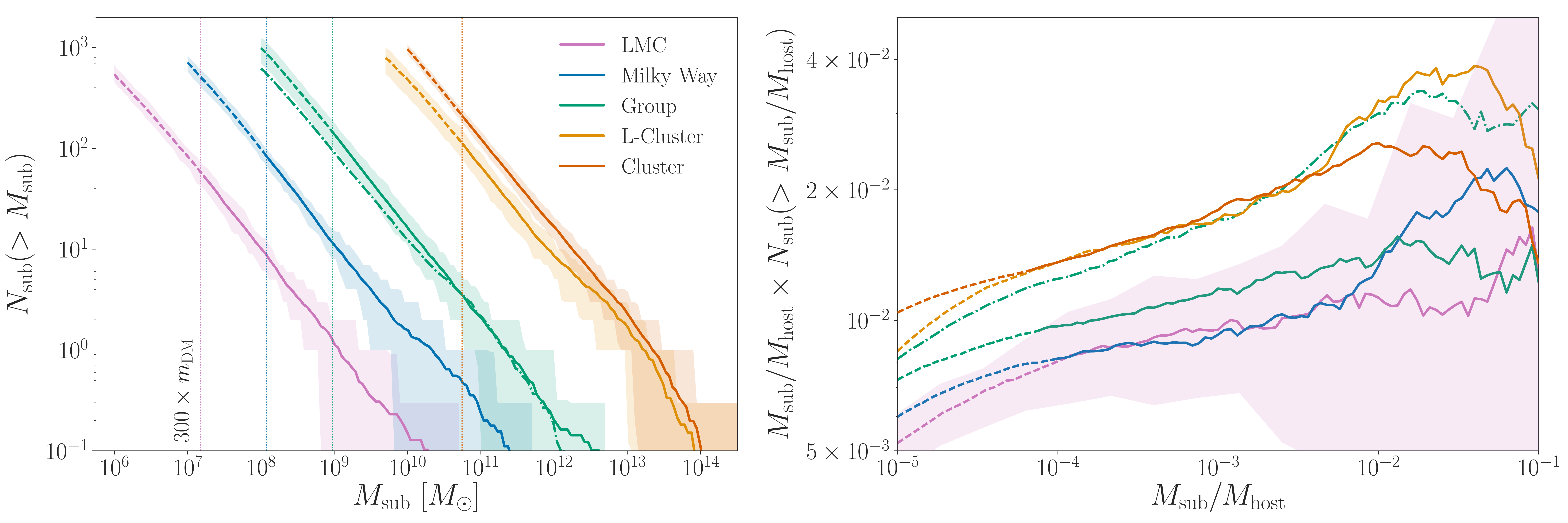}\\
\caption{Left panel: subhalo mass functions evaluated using virial mass at $z=0$ for our five zoom-in simulation suites. Solid lines show mean SHMFs stacked over each suite, and shaded bands show the corresponding $16$th--$84$th percentile of the host-to-host scatter. The dotted--dashed Group result shows the SHMF at $z=0.5$, when the corresponding host halos were mass-selected in their parent box. Vertical dotted lines show the convergence limit for each simulation, calculated as $300$ times the dark matter particle mass in the highest-resolution zoom-in region, and lines transition from solid to dashed at the median $M_{\mathrm{sub}}/M_{\mathrm{host}}$ threshold corresponding to the convergence limit for each suite. Right panel: same as the left panel, but scaled to highlight the dependence on host mass. Subhalo masses are normalized by $M_{\rm host}$ to remove most of the linear dependence of subhalo abundance on host mass, and abundances are scaled by $M_{\rm sub}/M_{\rm host}$ to reduce the dynamic range of the $y$-axis. In this panel, host-to-host scatter is only shown for the LMC suite for visual clarity.}
\label{fig:SHMF}
\end{figure*}

The left panel of Figure \ref{fig:SHMF} shows the mean cumulative subhalo virial mass function (SHMF) and host-to-host scatter for each of our five suites. Applying a conservative limit of $M_{\mathrm{sub}}/M_{\mathrm{host}}>2.7\times 10^{-4}$ to all suites, which ensures that all subhalos in all hosts contain more than 300 particles, we find that the mean cumulative SHMFs are well described by power laws with slopes of $-0.93\pm 0.01$ (LMC), $-0.92\pm 0.01$ (Milky Way), $-0.91\pm 0.01$ (Group), $-0.91\pm 0.01$ (L-Cluster), and $-0.88 \pm 0.01$ (Cluster), where errors represent $1\sigma$ uncertainties on the best-fit slope of the mean SHMF, and slopes are fit over the range $2.7\times 10^{-4}<M_{\mathrm{sub}}/M_{\mathrm{host}}<10^{-3}$ assuming Poisson uncertainties. Thus, we do not detect a systematic mass trend among the LMC/Milky Way/Group SHMF slopes. Note that the L-Cluster and Cluster slopes are difficult to interpret in detail relative to the LMC, Milky Way, and Group results due to differences in these simulations' cosmological parameters (see Section \ref{sec:overview}). Our SHMF slopes are broadly consistent with previous measurements from cosmological simulations and semianalytic models (e.g., \citealt{vandenBosch14036835,Mao150302637,Benson191104579}), accounting for both the host-to-host scatter and error of our slope measurements within each suite.

At fixed subhalo mass, the SHMF amplitude is roughly self-similar, scaling linearly with host mass. It is difficult to quantify this scaling precisely because subhalo abundance within each suite exhibits significant host-to-host scatter; meanwhile, comparisons between suites at fixed subhalo mass necessarily involve subhalos with different numbers of particles, which may suffer from numerical issues to different extents. Nonetheless, our results are broadly consistent with previous studies of the subhalo abundance--host mass scaling (e.g., \citealt{Giocoli07121563,Ishiyama08120683,Ishiyama11012020,Jiangstw439}). These studies found that hosts at different masses have nearly self-similar subhalo mass functions, but that less-massive hosts tend to have lower SHMF amplitudes, likely because they disrupt their subhalos more efficiently. To our knowledge, this is the first explicit confirmation of this scaling using LMC and Group-mass zoom-in samples (however, see, e.g., \citealt{Moline211002097} for a recent measurement from cosmological simulations).

To more directly compare subhalo abundances among Symphony suites, the right panel of Figure \ref{fig:SHMF} shows normalized SHMFs, where subhalo masses are divided by the mass of each host halo to remove most of the linear dependence of subhalo abundance on host mass, and the resulting SHMFs are scaled by $M_{\rm sub}/M_{\rm host}$ to reduce the dynamic range. The turnover at high sub-to-host halo mass ratios likely reflects the exponential cutoff in the SHMF (e.g., \citealt{Gao10062882}); however, given the rarity of high-mass subhalos in Symphony, we are not able to make decisive statistical statements about this regime. 

At fixed sub-to-host halo mass ratio, normalized subhalo abundances are consistent within the host-to-host scatter among the LMC, Milky Way, and Group suites. Meanwhile, the L-Cluster and Cluster hosts exhibit $\approx 2$ times higher normalized subhalo abundances than the lower-mass suites at fixed mass ratio, which is broadly consistent with previous studies of the ``evolved'' subhalo mass function (e.g., \citealt{Giocoli07121563}). This may result from the systematically later infall times for subhalos of the L-Cluster and Cluster hosts (see Appendix \ref{infall_times}), which leaves less time for these objects to be stripped below a fixed sub-to-host halo mass ratio. Leveraging the dynamic range of Symphony's host and subhalo populations to explore the evolution of the SHMF as a function of host halo mass is an interesting avenue for future work.

\subsection{Subhalo Radial Distributions}
\label{sec:radial}

\begin{figure*}[t!]
\centering
\includegraphics[width=\textwidth]{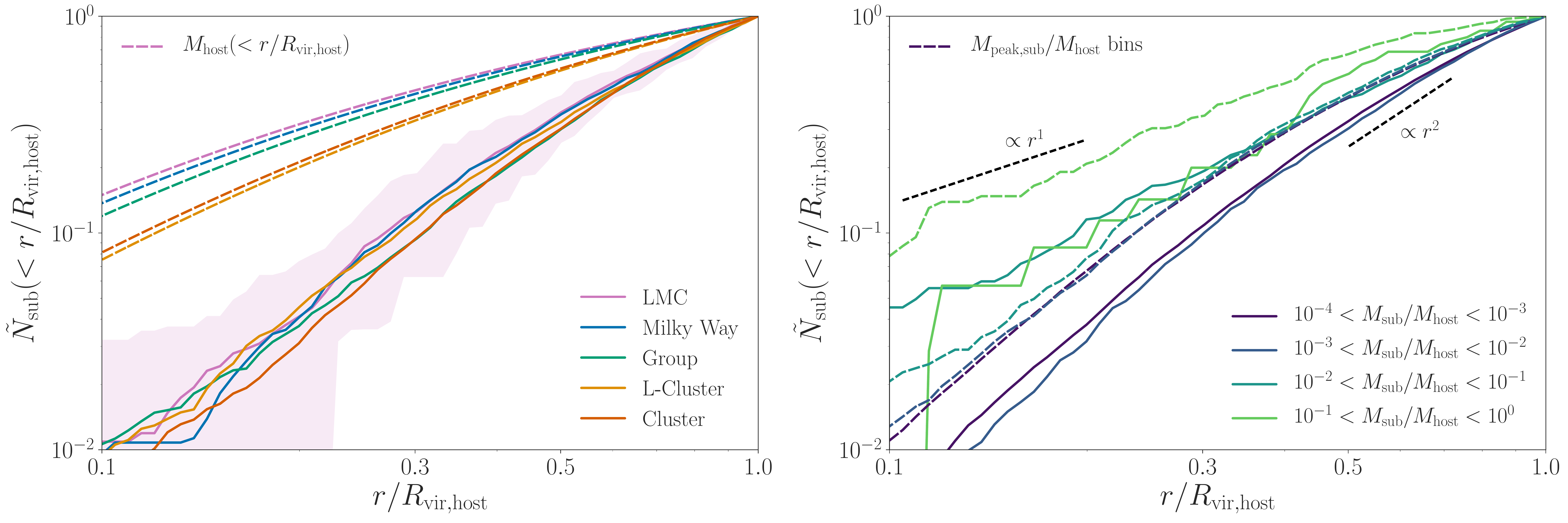}\\
\caption{Left panel: cumulative subhalo radial distributions as a function of distance from the host halo center, in units of the host halo virial radius, for our five zoom-in simulation suites. Solid lines show mean radial distributions stacked over each suite, and the shaded band shows the $16$th--$84$th percentile for the LMC suite. Only subhalos above a resolution cut of $M_{\mathrm{sub}}/M_{\mathrm{host}}>2.7\times 10^{4}$ are used to compute the radial distribution for each zoom-in, and each radial distribution is normalized to the total number of subhalos above this cut within the host halo's virial radius. Dashed lines show mean enclosed mass profiles, normalized by host halo virial mass, stacked over hosts in each suite. Right panel: normalized subhalo radial distributions, stacked over all suites, in bins of sub-to-host halo mass ratio binned by $z=0$ (solid) and peak (dashed) subhalo virial mass. Dashed black lines indicate normalized radial distributions proportional to $r$ and $r^2$ for reference.}
\label{fig:radial}
\end{figure*}

The left panel of Figure \ref{fig:radial} shows the normalized subhalo radial distributions for each Symphony suite,
\begin{equation}
    \tilde{N}_{\mathrm{sub}}(<r/R_{\mathrm{vir,host}})\equiv\frac{N_{\mathrm{sub}}(<r/R_{\mathrm{vir,host}})}{N_{\mathrm{sub}}(<R_{\mathrm{vir,host}})}
\end{equation}
using a resolution cut of $M_{\mathrm{sub}}/M_{\mathrm{host}}>2.7\times 10^{-4}$. We also measure normalized radial distributions stacked over all suites in the right panel of Figure \ref{fig:radial}. To investigate how the radial distribution depends on subhalo mass, we bin these stacked measurements according to $M_{\mathrm{sub}}/M_{\mathrm{host}}$ (solid lines) and $M_{\mathrm{peak,sub}}/M_{\mathrm{host}}$ (dashed lines). Note that the $M_{\mathrm{sub}}/M_{\mathrm{host}}$ bins extend slightly below our fiducial sub-to-host halo resolution cut; however, our qualitative results are not highly sensitive to this lower limit. Meanwhile, note that the results binned by $M_{\mathrm{peak,sub}}/M_{\mathrm{host}}$ include subhalos at small distances whose statistics are not formally converged at any mass ratio (see Appendix \ref{sec:convergence_res}), and should therefore be interpreted with caution. 

The left panel of Figure \ref{fig:radial} shows that Symphony's total radial subhalo distributions (above a $z=0$ sub-to-host halo mass ratio cut) are systematically less concentrated than their hosts' enclosed mass profiles. This is consistent with previous studies using cosmological simulations (e.g., \citealt{Zentner0411586,Springel08090898}), and may be related to the withering and artificial disruption of subhalos, even with large peak particle counts (see Appendix \ref{sec:convergence_res}, and \citealt{Green210301227}). Interestingly, the shapes of Symphony radial subhalo distributions do not noticeably depend on host mass. A host mass dependence may be expected if subhalos trace their hosts' underlying dark matter density profile; this would predict that lower-mass hosts have more centrally concentrated subhalo distributions given their more concentrated dark matter density profiles, as indicated by the dashed lines in the left panel of Figure \ref{fig:radial} (also see Figure \ref{fig:density}). However, lower-mass hosts also tidally strip their subhalos more efficiently and over longer timescales (see Appendices \ref{infall_times}--\ref{stripping_distributions}), which potentially counteracts the trend due to underlying density profiles. Disentangling these effects will require a dedicated follow-up study.

The right panel of Figure \ref{fig:radial} demonstrates that subhalos with low masses relative to their hosts exhibit the least centrally concentrated radial distributions; at higher mass ratios, dynamical friction causes subhalos to sink toward the host center more efficiently. Near hosts' outer regions, the normalized radial profiles for low-mass subhalos scale as $r^2$. Note that radial distributions are generally shallower (and therefore the radial density profiles are steeper) when binned in terms of peak rather than $z=0$ subhalo virial mass, consistent with previous studies (e.g., \citealt{Nagai0408273,Kravtsov09063295}). However, because our simulations' convergence properties are poorer when using peak rather than $z=0$ subhalo masses (see Appendix \ref{sec:convergence_res}), the radial distributions binned by $M_{\mathrm{peak,sub}}$ should be interpreted with caution, particularly at low peak particle counts.

%---------------------------------------------------------------------------------------
%	SECTION 4
%--------------------------------------------------------------------------------------

\section{Comparison with \textsc{Galacticus}}
\label{sec:galacticus}

To place our work in the context of semianalytic models, we compare our host halo properties and subhalo populations to those predicted by the \textsc{Galacticus} structure and galaxy formation model \citep{Benson10081786,Pullen14078189}.\footnote{In particular, we use the version of \textsc{Galacticus} corresponding to commit \texttt{bc4ecd2} in \href{https://github.com/galacticusorg/galacticus}{https://github.com/galacticusorg/galacticus}.} \textsc{Galacticus}' structure formation modules combine prescriptions for building merger trees using extended Press-Schechter theory with analytic, physically motivated models for halo and subhalo evolution.

We begin by briefly describing the \textsc{Galacticus} model and implementation used to generate these predictions (Section \ref{sec:galacticus_model}); we then compare host halo concentration distributions (Section \ref{sec:galacticus_host}), subhalo mass functions (Section \ref{sec:galacticus_sub}), and radial distributions (Section \ref{sec:galacticus_sub_radial}) for all suites. We comment on avenues for future work that combine Symphony with \textsc{Galacticus} and other semianalytic models (Section \ref{sec:galacticus_future}).

\subsection{\textsc{Galacticus} Realizations of Symphony Systems}
\label{sec:galacticus_model}

We generate $10$ realizations of \textsc{Galacticus} merger trees and subhalo populations for each individual Symphony zoom-in simulation, using appropriate cosmological parameters and a resolution matched to the particle mass of each suite. These merger trees have $z=0$ host halo virial masses matched to each Symphony host; thus, the \textsc{Galacticus} and Symphony host halo mass distributions match by construction. To model halo concentration, the \cite{Ludlow160102624} concentration model is applied to the formation histories of halos with progenitors in each merger tree using the best-fit parameters from \cite{Benson181206026}, and the \cite{Diemer180907326} mass--concentration relation with a scatter of $0.16~\mathrm{dex}$ is used for halos without progenitors. Thus, our comparisons of Symphony and \textsc{Galacticus} host halo concentration distributions in Section \ref{sec:galacticus_host} only rely on the \cite{Ludlow160102624} implementation in \textsc{Galacticus} because all hosts have progenitors above the resolution threshold.

We use an updated version of the \textsc{Galacticus} subhalo evolution model presented in \cite{Yang200310646}, which was calibrated to match $z=0$ SHMFs, maximum circular velocity ($V_{\mathrm{max,sub}}$) functions, and the $M_{\mathrm{sub}}$--$V_{\mathrm{max,sub}}$ relation from the ELVIS \citep{Garrison-Kimmel13106746} and Caterpillar \citep{Griffen150901255} simulations; here, we use the version of the model calibrated to Caterpillar. The subhalo evolution model we use includes a treatment of tidal stripping and is not subject to artificial subhalo disruption present in simulations. In addition, we use a new prescription for the orbital evolution of sub-subhalos (X.\ Du \& A.\ Benson 2023, in preparation), which was not included in \cite{Yang200310646}; this model leads to a slightly more concentrated radial subhalo distribution near the host center ($r/R_{\mathrm{vir,host}}\lesssim 0.1$) but does not significantly affect the SHMF or radial distribution predictions we present.

Note that, in \textsc{Galacticus}, $M_{\mathrm{sub}}$ is calculated from each subhalo's remaining bound mass at $z=0$, and $M_{\mathrm{peak,sub}}$ is identical to the virial mass at first infall onto the host halo, $M_{\mathrm{acc,sub}}$, because the evolution of subhalos before accretion (including pre-infall tidal stripping) is not modeled. However, in Symphony, $M_{\mathrm{peak,sub}}$ is typically $\approx 30\%$ higher than $M_{\mathrm{acc,sub}}$ and occurs $\approx 2~ \mathrm{Gyr}$ earlier for subhalos in all Symphony suites (consistent with, e.g., \citealt{Behroozi13102239}). Modeling pre-infall subhalo evolution in \textsc{Galacticus} is beyond the scope of our comparison, but constitutes an interesting area for future work.

\subsection{Host Halo Concentrations}
\label{sec:galacticus_host}

\begin{figure*}[t!]
\centering
\includegraphics[width=\textwidth]{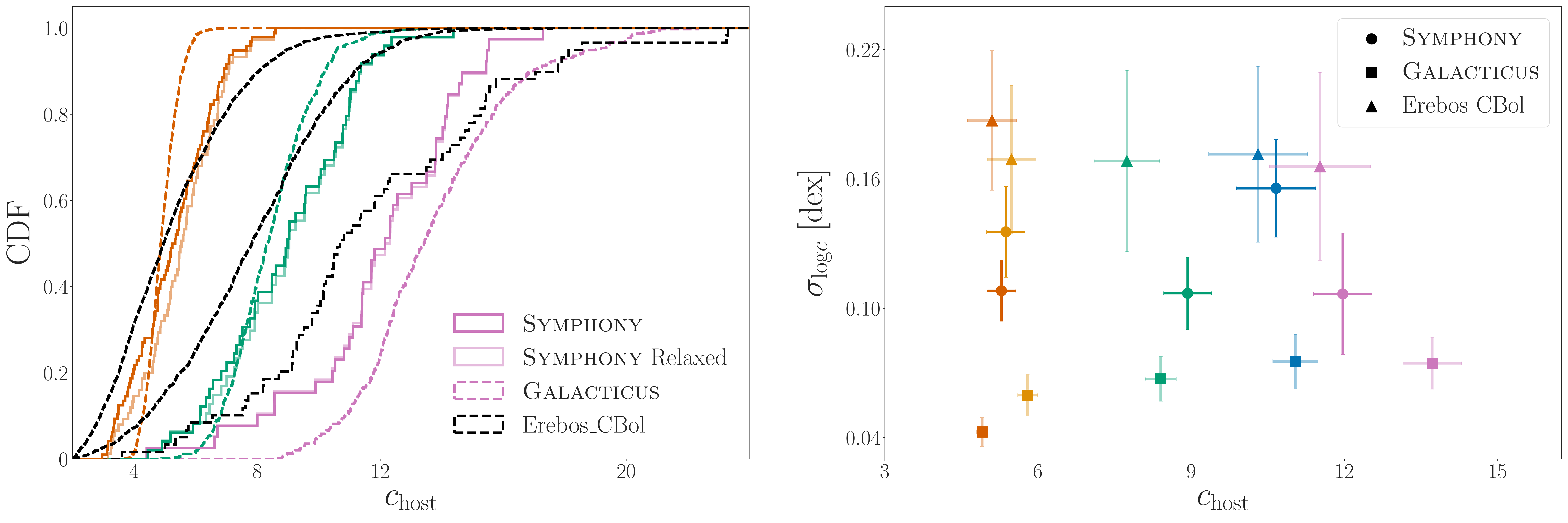}
\caption{Left panel: cumulative distributions of host halo concentration for Symphony hosts (solid), Symphony hosts that are virially relaxed according to the \cite{Ludlow160102624} criterion (faint solid), \textsc{Galacticus} realizations (dashed), and Erebos halos (black dashed) in the LMC (pink), Group (green), and Cluster (red) mass ranges. For the LMC and Group suite comparisons, we only use Erebos halos that satisfy our hosts' isolation criterion. Right panel: median host-to-host concentration scatter, measured as the standard deviation of the $\log c$ distribution in dex, versus median concentration for jackknifed samples of Symphony hosts (circles), \textsc{Galacticus} realizations (squares), and Erebos halos (triangles) in the LMC (pink), Milky Way (blue), Group (green), L-Cluster (gold), and Cluster (red) mass ranges. Error bars indicate $1\sigma$ host-to-host jackknife uncertainties.}
\label{fig:mc_comparison_galacticus}
\end{figure*}

The left panel of Figure \ref{fig:mc_comparison_galacticus} summarizes our comparisons between Symphony, \textsc{Galacticus}, and Erebos mass--concentration relations. In particular, we show cumulative distributions of host concentration for all Symphony zoom-ins, \textsc{Galacticus} realizations, and Erebos halos in the LMC, Group, and Cluster mass ranges. Because \textsc{Galacticus} assumes virial equilibrium when computing host concentrations along each merger tree, we also test the effects of applying a virial relaxation cut to the Symphony hosts. In particular, we use the relaxation criterion in \cite{Ludlow160102624}, which excludes hosts that more than double their mass within the last $3.7~ \mathrm{Gyr}$ (i.e., $a_{1/2,\mathrm{host}}\gtrsim 0.75$). This cut has a negligible effect on Symphony concentration distributions, which is expected because the LMC, Milky Way, and Group suites have very few unrelaxed hosts, while the unrelaxed L-Cluster and Cluster hosts are not significant outliers in concentration (see the right panel of Figure \ref{fig:Mhost_chost}).

Symphony and \textsc{Galacticus} concentration distributions are statistically distinguishable: two-sample K-S tests yield $p<0.01$ for all suites, with the largest discrepancy for the Cluster suite. Interestingly, \textsc{Galacticus} predicts an even lower median concentration for Cluster hosts than Symphony, which in turn predicts a concentration distribution that is shifted low compared to Erebos (see Section~\ref{sec:mc_relation}). Furthermore, as shown in the right panel of Figure \ref{fig:mc_comparison_galacticus}, \textsc{Galacticus} predicts smaller concentration scatter within each suite compared to Symphony, with typical values of~$\sigma_{\log c}\approx 0.06~ \mathrm{dex}$. The direction of this discrepancy in scatter is expected because the \cite{Ludlow160102624} model underpredicts concentration scatter when applied to semianalytic merger trees, although it predicts the expected scatter when applied to merger trees extracted from cosmological simulations \citep{Benson181206026}. 

Several effects may contribute to the discrepancies between Symphony and \textsc{Galacticus} concentration distributions, and particularly to Symphony's larger concentration scatter, including: (1) \textsc{Galacticus} merger trees do not account for the dependence of halo concentration on environment, which increases concentration scatter \citep{Benson181206026}; (2) Symphony hosts' density profiles are not spherical or smooth (i.e., they contain substructure), and they are not perfectly fit by NFW profiles, both of which increase concentration scatter \citep{Benson181206026,Fielder200702964}; (3) Symphony halos are not in perfect virial equilibrium, as assumed by the \cite{Ludlow160102624} model, even after applying the relaxation cut. Exploring the difference between the concentration distributions of Symphony and \textsc{Galacticus} hosts---for example, by applying the \citealt{Ludlow160102624} model to the Symphony merger trees to predict concentration distributions, or by removing substructure from Symphony hosts in our density profile and concentration measurements---is left for future work.

\subsection{Subhalo Mass Functions}
\label{sec:galacticus_sub}

\begin{figure*}[t!]
\includegraphics[trim={0 0.5cm 0 0},width=\textwidth]{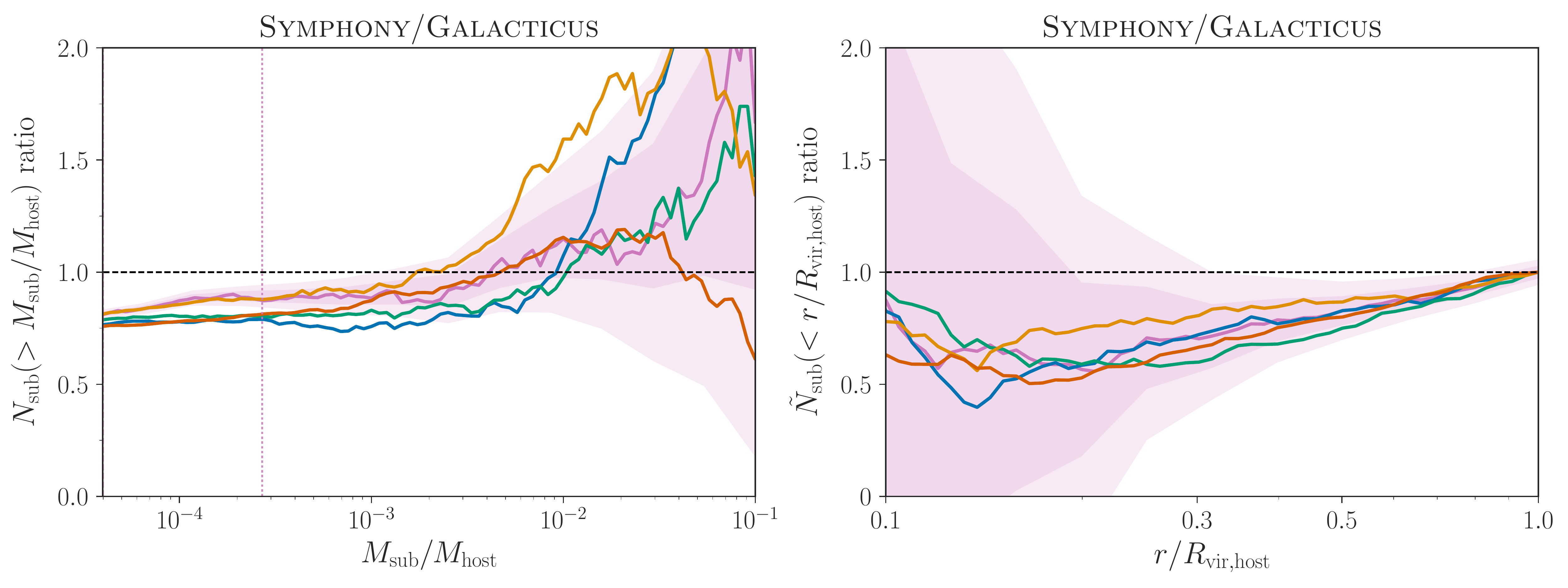}
\caption{Left panel: the ratio of cumulative subhalo mass functions in each Symphony suite to corresponding \textsc{Galacticus} predictions. Solid lines show the ratio of median cumulative SHMFs stacked over all Symphony zoom-ins and \textsc{Galacticus} realizations in our LMC (pink), Milky Way (blue), Group (green), Low-Cluster (gold), and Cluster (red) mass ranges. Dark (light) shaded bands indicate the corresponding $1\sigma$ ($2\sigma$) Poisson errors on the mean, which we only show for the LMC suite for visual clarity. The vertical dotted line indicates our conservative convergence limit of $M_{\mathrm{sub}}/M_{\mathrm{host}}>2.7\times 10^{-4}$, and the horizontal dashed line indicates perfect agreement between Symphony and \textsc{Galacticus}. Right panel: same as the left panel, for the ratio between Symphony and \textsc{Galacticus} cumulative radial distributions in units of distance from the host center divided by the host halo virial radius.}
\label{fig:SHMF_galacticus}
\end{figure*}

The left panel of Figure \ref{fig:SHMF_galacticus} compares the ratio of the mean cumulative SHMF for each Symphony suite to the corresponding \textsc{Galacticus} predictions; Appendix \ref{sec:galacticus_comparisons_additional} provides suite-by-suite SHMF comparisons. The left panel of Figure \ref{fig:SHMF_galacticus} demonstrates that Symphony and \textsc{Galacticus} SHMFs are consistent within the $2\sigma$ Poisson error on the mean for $M_{\mathrm{sub}}/M_{\mathrm{host}}\gtrsim 10^{-3}$ across all Symphony suites, but display a statistically significant $\approx 25\%$ discrepancy at lower masses. In particular, \textsc{Galacticus} predicts $\approx 25\%$ higher subhalo abundances than Symphony for $M_{\mathrm{sub}}/M_{\mathrm{host}}\lesssim 10^{-3}$ in all suites, and lower subhalo abundances than Symphony at higher sub-to-host halo mass ratios in all except the Cluster suite. Both of these discrepancies are most severe for the Milky Way suite, and we discuss them in detail below.

The cumulative SHMF slopes predicted by \textsc{Galacticus} are $-0.94\pm 0.004$ (LMC), $-0.94\pm 0.001$ (Milky Way), $-0.92\pm 0.002$ (Group), $-0.94\pm 0.002$ (L-Cluster), and $-0.92\pm 0.002$ (Cluster) over the range $2.7\times 10^{-4}<M_{\mathrm{sub}}/M_{\mathrm{host}}<10^{-3}$ assuming Poisson uncertainties. These slopes are a few percent steeper than the corresponding slopes measured in Symphony (see Section \ref{sec:shmf}).

To interpret these results, we focus on the comparison between Symphony and \textsc{Galacticus} Milky Way predictions because, as described in Section \ref{sec:galacticus_model}, the \textsc{Galacticus} model we compare to accurately reproduces the normalization and slope of SHMFs from the Milky Way--mass Caterpillar zoom-ins \citep{Yang200310646}. Thus, the Symphony--\textsc{Galacticus} Milky Way comparison can be understood by comparing Symphony and Caterpillar SHMFs, which we do systematically in Appendix \ref{sec:caterpillar_comparison}. In particular, the tests in Appendix \ref{sec:caterpillar_comparison} demonstrate that:
\begin{enumerate}
    \item At high sub-to-host halo mass ratios, $M_{\mathrm{sub}}/M_{\mathrm{host}}\gtrsim 10^{-2}$, an upward fluctuation in the mean SHMF of Symphony's target halos relative to the full sample of objects that pass the relevant mass and environmental cuts combines with a comparable downward fluctuation in the mean Caterpillar SHMF to yield a $\approx 60\%$ (or $\approx 2\sigma$) discrepancy. Both of these fluctuations are caused by the selection of the specific samples of $45$ Symphony and $35$ Caterpillar hosts, and manifest in both the zoom-in and parent cosmological simulations. Due to the low statistical strength, these fluctuations could be purely random, but they could also be related to the methods used to select target hosts after initial mass and environmental cuts are applied.
    \item At intermediate sub-to-host halo mass ratios, $10^{-3}\lesssim M_{\mathrm{sub}}/M_{\mathrm{host}}\lesssim 10^{-2}$, an overabundance in the mean Caterpillar SHMF relative to cosmological simulations---which may, in part, be a fluctuation due to the selection of the specific Caterpillar sample---yields a $\approx 25\%$ (or $\approx 2\sigma$) discrepancy.
    \item At low sub-to-host halo mass ratios, $M_{\mathrm{sub}}/M_{\mathrm{host}}\lesssim 10^{-3}$, the Symphony and Caterpillar SHMF slopes do not significantly differ. Thus, the overabundance relative to Symphony at intermediate subhalo masses propagates to very low sub-to-host halo mass ratios, resulting in the $\approx 25\%$ discrepancy at the lowest resolved masses shown in Figure \ref{fig:SHMF_galacticus}.
\end{enumerate}

To accurately calibrate semianalytic models like \textsc{Galacticus}, SHMF biases among zoom-in host samples relative to all halos in cosmological volumes that pass the relevant mass and environmental cuts must be carefully modeled. We note that the effects of zoom-in host halo selection on the SHMF at high sub-to-host halo mass ratios can be identified relatively easily through comparisons between zoom-ins and the corresponding systems in their parent boxes, or successively higher-resolution resimulations thereof. On the other hand, detailed study would be required to understand the origin of discrepancies between zoom-ins at lower sub-to-host halo mass ratios, where zoom-in procedures or analyses themselves might influence the results.

Further investigation is needed to confirm whether discrepancies between Symphony and \textsc{Galacticus} predictions for the remaining Symphony suites are mainly inherited from the Milky Way comparison, or if they are affected by (1) selection effects or random fluctuations in other Symphony suites, and/or (2) the dependence of \textsc{Galacticus} predictions on host halo mass. We expect the effects of environmental cuts within each host halo mass range to be most severe for the LMC suite, relatively weak for the Group suite, and unimportant for the L-Cluster and Cluster suites. However, the effects of zoom-in host halo selection---i.e., differences between specific zoom-in host halo samples relative to all systems that pass the relevant mass and environmental cuts---likely bias all suites' SHMFs to some extent, with the possible exception of the Cluster suite, because all hosts in its parent box above a mass threshold were resimulated. In this context, it is reassuring that the Symphony--\textsc{Galacticus} SHMF discrepancies are less severe at high sub-to-host halo mass ratios for the non--Milky Way suites, and are least severe for the Cluster suite.

\subsection{Subhalo Radial Distributions}
\label{sec:galacticus_sub_radial}

The right panel of Figure \ref{fig:SHMF_galacticus} compares the ratio of normalized distributions for each Symphony suite to the corresponding \textsc{Galacticus} predictions; Appendix \ref{sec:galacticus_comparisons_additional} provides suite-by-suite comparisons and explores the dependence of \textsc{Galacticus} radial distributions on subhalo $z=0$ and peak mass. Symphony and \textsc{Galacticus} normalized radial distributions are consistent within the $2\sigma$ Poisson error on the mean for subhalos down to our conservative sub-to-host halo resolution limit, at all at distances $r/R_{\mathrm{vir,host}}>0.1$; we limit the comparison to these distances because we cannot explicitly verify that Symphony subhalo population statistics are converged at smaller radii (see Appendix \ref{sec:convergence_res}). Above our sub-to-host halo cut, Symphony radial profiles are less centrally concentrated compared to \textsc{Galacticus}. Furthermore, \textsc{Galacticus} radial distributions are also more concentrated when binned by $M_{\mathrm{peak,sub}}/M_{\mathrm{host}}$, consistent with the comparisons in \cite{Nadler210107810} for two hosts from the original Milky Way suite (see Appendix \ref{sec:galacticus_comparisons_additional}). A dedicated study that models the effects of withering below the mass resolution limit and artificial disruption in Symphony will be needed to assess whether this discrepancy is physical or numerical.

Symphony radial distributions display a weak subhalo mass dependence for all $M_{\mathrm{sub}}/M_{\mathrm{host}}\lesssim 10^{-2}$ and $M_{\mathrm{peak,sub}}/M_{\mathrm{host}}\lesssim 10^{-1}$ (see the right panel of Figure \ref{fig:radial}), while \textsc{Galacticus} radial distributions are similar for all $M_{\mathrm{sub}}/M_{\mathrm{host}}$ but become systematically more concentrated with increasing $M_{\mathrm{peak,sub}}/M_{\mathrm{host}}$ (see Appendix \ref{sec:galacticus_comparisons_additional}). Such discrepancies may indicate the need for more thorough calibration of the \textsc{Galacticus} subhalo evolution model used to generate these predictions, which was not calibrated to Caterpillar subhalo radial distributions. On the other hand, the differences in radial distributions as a function of sub-to-host halo mass ratio that we identify might require implementing additional subhalo evolution physics in \textsc{Galacticus} (for example, related to the strength and mass dependence of dynamical friction, or the efficiency of tidal stripping). We note that zoom-in host halo selection effects do not seem to significantly affect the radial distribution comparison, since the discrepancy between Symphony and \textsc{Galacticus} predictions is similar for all suites.

\subsection{Areas for Future Work}
\label{sec:galacticus_future}

We anticipate several additional avenues for future work that combine Symphony with \textsc{Galacticus} and other semianalytic structure formation models. First, the \textsc{Galacticus} model used here does not account for our hosts' specific environments; however, Symphony's LMC, Milky Way, and Group hosts generally occupy underdense regions due to the selection criteria described in Section \ref{sec:overview}. Building on \cite{Benson181206026}, it is interesting to consider how significantly the subhalo populations of hosts in particular environments, which have biased formation histories relative to the cosmic average, differ from typical subhalo populations at fixed host halo mass.

Second, it remains challenging to robustly resolve subhalos in cosmological simulations due to numerical effects, even at Symphony's resolution. In particular, subhalos can be stripped below the mass resolution limit (or ``wither''), which affects the low-mass end of simulated subhalo populations, and they can undergo artificial disruption, which may affect subhalos even with relatively high peak particle counts (e.g., see \citealt{VandenBosch171105276,VandenBosch180105427,Green210301227}). Furthermore, the internal structure of stripped subhalos can only be accurately resolved for subhalos with thousands of particles or more (e.g., see \citealt{Errani200107077}). Semianalytic models including \textsc{Galacticus} are well suited to quantify the impact of withering and artificial disruption on subhalo population statistics because they are relatively inexpensive, which allows subhalo populations to be resolved at much higher resolution than in cosmological simulations (e.g., see \citealt{Benson220601842}). Exploiting these advantages to quantify the impact of resolution effects and the convergence of subhalo population statistics in Symphony and its convergence resimulations (see Appendix \ref{sec:resims}) represents an interesting area for future work.

Third, in addition to $z=0$ subhalo populations, we emphasize that Symphony resolves: (1) the redshift evolution of subhalo populations with high temporal resolution out to $z\gtrsim 10$, and (2) high-resolution halos out to many times the virial radius of each target host, including isolated halos, subhalos of lower-mass hosts, and splashback subhalos of the main host. Although we have not studied the redshift evolution of subhalo populations or the statistics of halos beyond the virial radius of the host in this work, Symphony data products provide opportunities to calibrate semianalytic models in these relatively under-explored regimes. The evolution of Symphony subhalo and halo populations is particularly informative for (re)calibrating semianalytic models of the conditional mass function and halo merger rates (e.g., \citealt{Benson161001057}). As mentioned above, performing such (re)calibration robustly will require modeling the impact of environment on halo and subhalo populations (e.g., using the peak background split methodology of \citealt{Sheth99001122}), as well as secondary biases imprinted by zoom-in host halo selection from a mass and environmentally selected host sample.

%---------------------------------------------------------------------------------------
%	SECTION 5
%--------------------------------------------------------------------------------------

\section{Applying \textsc{UniverseMachine} to Symphony}
\label{sec:universemachine}

In this section, we use \textsc{UniverseMachine} to model the SFHs of galaxies occupying halos and subhalos in the high-resolution regions of all Symphony simulations. We describe our \textsc{UniverseMachine} model and zoom-in technique in Section \ref{sec:um_model}, predictions for Symphony hosts' SFHs in Section \ref{sec:um_host}, predictions for the stellar mass--halo mass (SMHM) relation in Section \ref{sec:um_smhm}, the observational relevance of our predictions in Section \ref{sec:um_obs}, and areas for future work combining Symphony and \textsc{UniverseMachine} in Section \ref{sec:um_future}.

\subsection{\textsc{UniverseMachine} Model and Zoom-in Application}
\label{sec:um_model}

\textsc{UniverseMachine}~\citep{Behroozi2019} is an empirical galaxy--halo connection model that paints star formation rates (SFRs) onto dark matter halo merger trees. This modeling is performed probabilistically using halos' maximum circular velocities and accretion rates as a function of redshift. \textsc{UniverseMachine} has been calibrated to match luminosity functions, quenched fractions, and auto- and cross-correlation functions of star-forming and quenched galaxies with stellar masses $M_{*}\gtrsim 10^8\msun$ for $0<z<10$ \citep{Behroozi2019}. The public version of the \textsc{UniverseMachine} DR1 model was calibrated on the cosmological Bolshoi-Planck simulation~\citep{Klypin14114001,2016MNRAS.462..893R}, with a box size of 250 Mpc$~h^{-1}$ and dark matter particle mass of $m_{\mathrm{part}}\approx 2 \times 10^{8}$ \msun.\footnote{We use the \textsc{UniverseMachine} DR1 version dated 2020-12-19, which can be found at \url{https://bitbucket.org/pbehroozi/universemachine/src/main/}.}

\begin{figure*}[t!]
\centering
\includegraphics[width=\textwidth]{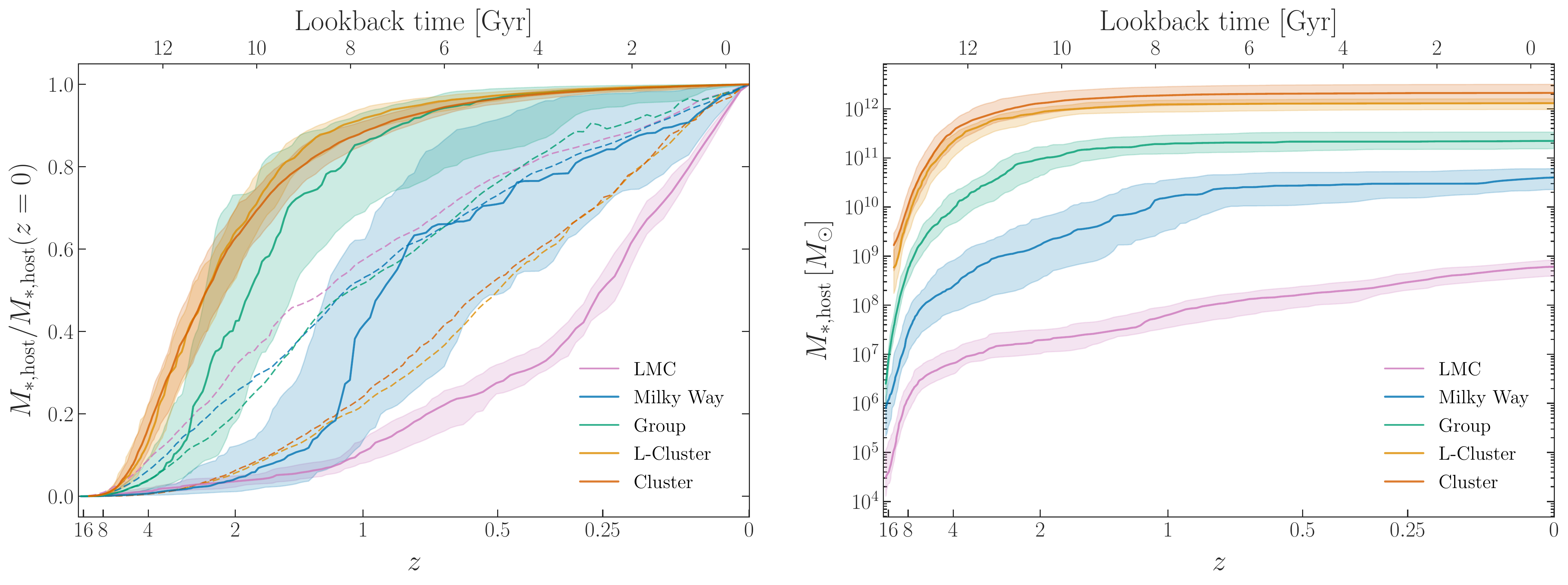}
\caption{Left panel: SFHs of central galaxies that occupy Symphony hosts, stacked over all host halos in each suite. Solid lines show the median evolution of the intrinsic stellar mass (i.e., the predicted stellar mass uncorrected for observational systematics and massive stars' death over time), normalized to each host's predicted stellar mass at $z=0$, and shaded regions denote the $68\%$ host-to-host scatter for each suite. Dashed lines show the corresponding mean normalized dark matter MAHs (see the left panel of Figure \ref{fig:MAH}). Right panel: same as the left panel, but for hosts' un-normalized SFHs.}
\label{fig:SFH_host}
\end{figure*}

\citet{Wang210211876} presented the first application of \textsc{UniverseMachine} to zoom-in simulations by using the DR1 model to model the SFHs of halos and subhalos in the original version of Symphony's Milky Way suite \citep{Mao150302637}. These authors developed a technique to join \textsc{Rockstar} halo catalogs from individual simulations to obtain accurate SFH predictions for an entire zoom-in suite. Here, we use the same methodology and DR1 model to obtain SFHs for each Symphony suite's halo and subhalo populations. In particular, we join halo lists within each suite of Symphony simulations to obtain a statistically robust halo accretion rate distribution for each suite (see \citealt{Wang210211876} for details). Each distribution is then used to generate SFHs for all simulations within the corresponding suite.

As emphasized by \cite{Wang210211876}, the \textsc{UniverseMachine} DR1 model is \emph{not} constrained for central or satellite galaxies with $M_{*} \lesssim 10^8~\msun$, which typically occupy (sub)halos with $M_{\mathrm{peak,sub}} \lesssim 10^{10}~\msun$ (e.g., see \citealt{Wechsler180403097}). For example, reionization and environmental processes are known to quench galaxies at these stellar masses, while \textsc{UniverseMachine} DR1 predicts that nearly all low-mass galaxies actively form stars down to $z=0$ \citep{Wang210211876}. This impacts our predicted SFHs for the galaxies that occupy low-mass (sub)halos in Symphony's LMC, Milky Way, and Group suites, even though their luminosity functions are consistent with extrapolations of global luminosity function measurements at higher stellar masses \citep{Wang210211876}. We also note that, although our L-Cluster and Cluster simulations have similar resolution to the simulation \textsc{UniverseMachine} DR1 was calibrated on, the details of galaxy quenching and evolution for low-mass satellites of cluster-mass hosts may not fully be captured by the DR1 model.

\subsection{Predictions for Symphony Host Halo Star Formation Histories}
\label{sec:um_host}

SFHs for the central galaxies occupying Symphony's target hosts in each suite are presented in Figure \ref{fig:SFH_host}. Lower-mass hosts form their stars systematically later than higher-mass hosts, in contrast to their dark matter MAHs (see Figure \ref{fig:MAH}). In particular, Symphony hosts form half of their stars at half-stellar-mass scale factors of $a^*_{1/2,\mathrm{host}} = 0.8\pm 0.04$, $0.56\pm 0.1$, $0.37\pm 0.09$, $0.3\pm 0.03$, and $0.3\pm 0.07$ for the LMC, Milky Way, Group, L-Cluster, and Cluster suites, respectively. These findings are consistent with many previous studies (e.g., see \citealt{Conroy08053346,Wechsler180403097}) and are expected because higher-mass galaxies are preferentially quenched at early times, while lower-mass galaxies---and particularly our LMC and Milky Way centrals---form stars continuously until low redshifts.

Interestingly, although the dependence of dark matter MAHs on host mass is strongest for the transition between our Group and L-Cluster suites (see Section \ref{sec:mah}), the most noticeable transition among normalized SFHs occurs between the LMC, Milky Way, and Group hosts. This is a robust prediction of the DR1 model, which is well constrained in the stellar mass range of Symphony's predicted central galaxies. However, note that our LMC, Milky Way, and Group hosts occupy underdense environments that may not be well represented in the \textsc{UniverseMachine} DR1 calibration.

\subsection{Predictions for the Stellar Mass--Halo Mass Relation Using Symphony}
\label{sec:um_smhm}

\begin{figure*}[t!]
\hspace{-5mm}
\includegraphics[width=\textwidth]{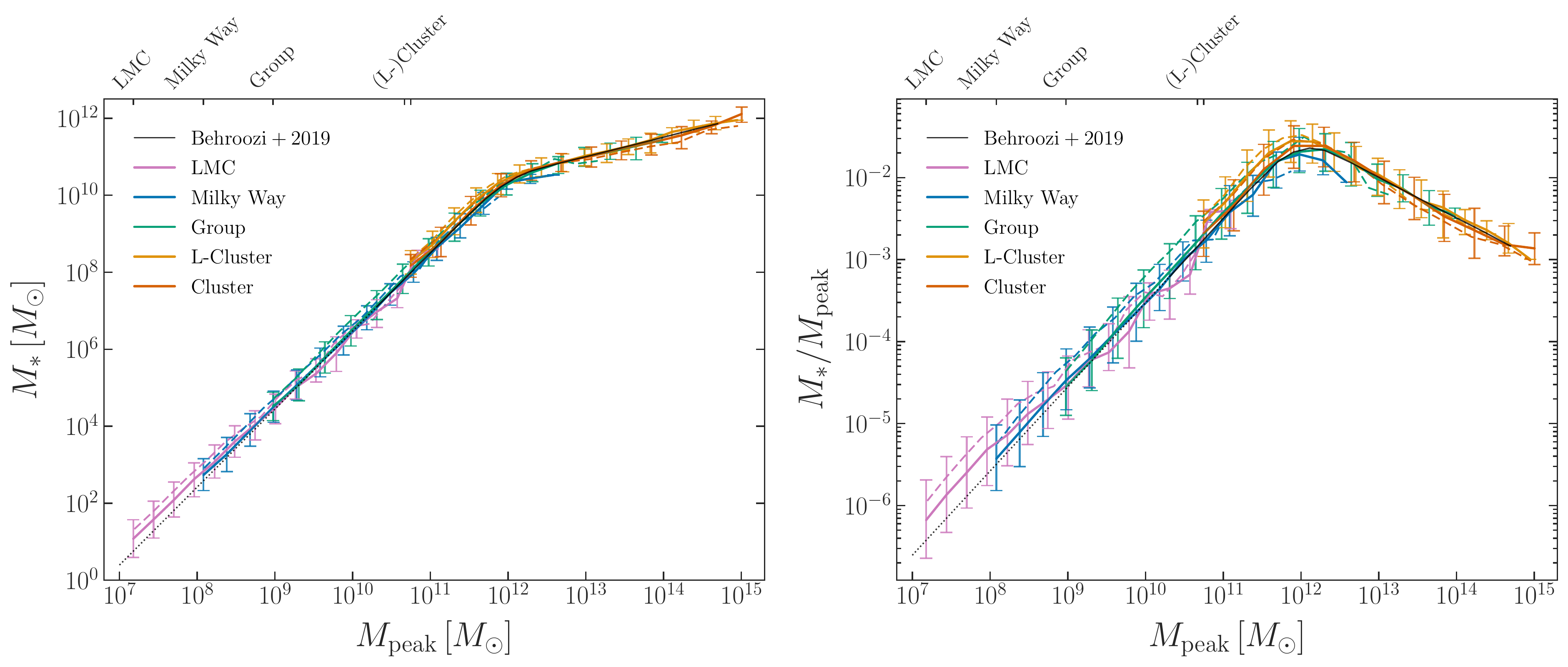}
\caption{Left panel: stellar mass--halo mass (SMHM) relations predicted by applying the \textsc{UniverseMachine} DR1 model to all Symphony suites. Solid lines represent the median SMHM relation stacked over all simulations in each Symphony suite using all halos and subhalos above our fiducial-resolution cut of $M_{\mathrm{sub}}>300m_{\mathrm{part}}$ and within $5R_{\mathrm{vir,host}}$ of each Symphony host; dashed lines only include subhalos of the target host in each simulation. Error bars denote the $68\%$ scatter, which is dominated by intrinsic scatter in the galaxy--halo connection rather than host-to-host scatter. The top axis denotes each suite's resolution limit in terms of $z=0$ (sub)halo virial mass (i.e. $M_{\mathrm{sub}}> 300 m_{\mathrm{part}}$); however, $M_{\mathrm{peak,sub}}$ is less well converged than $M_{\mathrm{sub}}$ (see Appendix \ref{sec:convergence_res}), so these ticks should only be interpreted as rough estimates of the resolution limits. The black curve shows the $z=0$ SMHM relation from \citet{Behroozi2019} for comparison, and transitions to a dotted power-law extrapolation of the DR1 prediction at stellar masses where it is not constrained. Right panel: same as the left panel, but with stellar mass normalized to each (sub)halo's peak virial mass.}
\label{fig:SMHM_UM}
\end{figure*}

The SMHM relation predicted by \textsc{UniverseMachine} at $z=0$ for halos and subhalos in the high-resolution regions of our 260 zoom-in simulations is presented in Figure \ref{fig:SMHM_UM}, using our fiducial-resolution cut of $M_{\mathrm{sub}}>300m_{\mathrm{part}}$. In particular, solid lines show the SMHM relation for all halos and subhalos within $5R_{\mathrm{vir,host}}$ of each Symphony host, and dashed lines show the SMHM only for subhalos of the target host in each zoom-in. Note that $5R_{\mathrm{vir,host}}$ corresponds to the size of the smallest high-resolution regions in Symphony, from the LMC suite, while other Symphony suites' high-resolution regions extend to $\approx 10R_{\mathrm{vir,host}}$ (see Section \ref{sec:symphony_overview}). We plan to leverage this volume and explore the impact of low-resolution contamination near the high-resolution boundaries in future work.

The SMHM relation, combined over all Symphony suites, spans an immense dynamic range of more than 10 orders of magnitude in stellar mass and 7 orders of magnitude in peak (sub)halo mass. At fixed $M_{\mathrm{peak,sub}}$, satellites of Symphony hosts have higher stellar masses than isolated systems, consistent with previous \textsc{UniverseMachine} results \citep{Behroozi2019}. Furthermore, the median and the scatter of SMHM relations across different Symphony suites all agree well within the $1\sigma$ intrinsic and host-to-host scatter where overlap exists. This suggests that our procedure for joining halo lists within each Symphony suite does not significantly bias the resulting SMHM relations. We note that the SMHM scatter is dominated by intrinsic scatter in the galaxy--halo connection, which would persist even when applying \textsc{UniverseMachine} to any individual Symphony zoom-in, rather than host-to-host scatter. Exploring the dependence of SFH predictions on (sub)halo environment is an interesting area for future study.

At stellar masses above $\approx 10^{8}~\msun$, \textsc{UniverseMachine} is calibrated to match various galaxy population statistics, and our SMHM predictions are consistent with fiducial DR1 predictions from \cite{Behroozi2019}. This agreement is nontrivial because \textsc{UniverseMachine} DR1 was calibrated and run on a cosmological simulation in which host halos at all mass scales are less well resolved than in Symphony. In addition, zooming in on a single massive host halo changes the overall satellite fractions in our simulations relative to those in cosmological volumes, even for suites that do not impose specific isolation criteria (namely, our L-Cluster and Cluster suites). In a follow-up study, we plan to quantify how these effects lead to variations in the faint-end SMHM slope for our LMC, Milky Way, and Group suites and relative to fiducial \textsc{UniverseMachine} predictions from \cite{Behroozi2019}.

At stellar masses below $\approx 10^{8}~\msun$, \textsc{UniverseMachine} is designed to smoothly extrapolate the global luminosity function even though the DR1 model is not calibrated to the observed properties of such faint galaxies. Thus, our predicted SMHM relation for Symphony's Milky Way suite is consistent with SMHM relations inferred from Milky Way satellite galaxies with $10^2 \lesssim M_{\ast} \lesssim 10^8$ \msun using abundance matching because these relations are in turn consistent with the global luminosity function \citep{Nadler191203303,Wang210211876}. The predictions at even lower stellar masses and for satellites of lower-mass hosts---particularly for galaxies occupying the lowest-mass (sub)halos in our LMC suite---have not been compared to data or other theoretical predictions. At this extreme faint end of the SMHM, the efficiency and stochasticity of galaxy formation are influenced by both reionization and environmental processes (e.g., \citealt{Rey190904664,Benitze-Llambay200406124,Manwadkar211204511,Munshi210105822}). Thus, as we discuss further in Section \ref{sec:um_future}, we plan to revisit our predictions for ultra-faint dwarf galaxy SFHs once \textsc{UniverseMachine} is appropriately recalibrated.

\subsection{Modeling Observable Galaxy Populations using \textsc{UniverseMachine} and Symphony}
\label{sec:um_obs}

The results above highlight the dynamic range of Symphony's well-resolved halo and subhalo populations, which we have exploited to statistically model the connection between dark matter accretion histories and galaxy SFHs using \textsc{UniverseMachine} over an unprecedented range of stellar mass. Here, we argue that our \textsc{UniverseMachine} predictions statistically capture the SFHs of systems with luminosities comparable to nearly all currently observable satellite galaxies in the universe, with the exception of the faintest known satellites in nearby clusters. For this to hold, resolved satellites with the lowest stellar masses from our \textsc{UniverseMachine} predictions for each suite must be fainter than the dimmest known satellites at the corresponding host halo mass scales, which are generally found in systems at the smallest distances due to selection effects.

For LMC and Milky Way--mass hosts, the faintest observed satellites are ultra-faint dwarfs orbiting the Milky Way itself \citep{Simon190105465}, some of which were originally associated with the LMC (e.g., \citealt{Kallivayalil180501448,Patel200101746}). These galaxies have stellar masses as small as $\approx 100~\msun$ (e.g., \citealt{Drlica-Wagner191203302}) and occupy halos with peak masses down to $\approx 10^8~\msun$, or smaller (e.g., \citealt{Read180707093,Nadler191203303}). Our LMC and Milky Way--mass suites resolve the abundance and formation histories of subhalos with masses down to $\approx 10^7\msun$ ($\approx 10^8\msun$), and therefore statistically capture the formation histories of subhalos that host these observed ultra-faints. The details of this comparison depend on the relationship between $z=0$ and peak subhalo mass; thus, robustly modeling the ultra-faint end of the observed Milky Way and/or LMC satellite populations still requires a model for the (potentially premature) disruption of low-mass subhalos (e.g., \citealt{Nadler180905542,Nadler191203303}).

On the Group scale, we compare our predictions to the faintest known satellites in Centaurus A (Cen A), because it is one of the closest and best-observed systems with a halo mass of $\approx 10^{13}~\msun$ (e.g., \citealt{Karachentsev0410065,Woodley0608497}). Cen A's dimmest satellites have absolute $V$-band magnitudes of $M_V\approx -8~\rm{mag}$ and luminosities of $\approx 10^5~L_{\mathrm{\odot}}$ \citep{Crnojevic180905103}, or stellar masses of $\approx 10^5~\msun$ assuming mass-to-light ratios of $\approx 1$. Our Group suite resolves the abundance and formation histories of subhalos with masses down to $\approx 10^9\msun$, which host satellites with stellar masses of a few times $10^4~\msun$; these are dimmer than the limit set by Cen A observations.

Our L-Cluster and Cluster suites have similar resolution; thus, it is sufficient to consider the faintest observed satellites at either host halo mass scale to assess whether our simulations can resolve such galaxies. We assume that the faintest observed satellites of the Virgo cluster---which have absolute $g$-band magnitudes of $M_g\approx -10~\rm{mag}$ \citep{Ferrarese160406462}---set this limit, noting that other nearby clusters have been observed to comparable depths (e.g., the Coma and Fornax clusters; \citealt{Yamanoi12055366,Venhola190408660}). The faintest observed Virgo satellites have stellar masses of $\approx 10^6~\msun$, which is below our L-Cluster and Cluster suites' capabilities given that these suites resolve subhalos down to $\approx 6\times 10^{10}~\msun$, which have stellar masses of $\approx 10^{8}~\msun$, on average. However, we note that observations of Virgo (and other nearby clusters) are much more complete at the faint end than typical observations of higher-redshift clusters. Thus, our L-Cluster and Cluster suites' resolution is sufficient for modeling the majority of observed cluster satellite populations, with the exception of the faintest satellites in several nearby clusters.

Finally, we emphasize that our SFH predictions robustly span a wide range of redshifts because they are tied to dark matter accretion histories that are well resolved at early times. For example, our LMC and MW-mass hosts are well resolved out to $z\gtrsim 9$ (see Section \ref{sec:mah}); thus, Symphony enables SFH predictions for the progenitors of LMC and MW-mass central galaxies at epochs that are highly relevant given HST observations over the last decade (e.g., \citealt{Bouwens14034295,Finkelstein14105439,Atek180309747}) and ongoing JWST studies. Thus, we anticipate that our predictions will aid the development of empirical models that rely on simulation-based SFH predictions at high redshifts (e.g., \citealt{Behroozi200704988}).

\subsection{Areas for Future Work}
\label{sec:um_future}
  
Given the dynamic range of our galaxy--halo connection predictions, we anticipate several interesting observational applications of the \textsc{UniverseMachine}--Symphony combination. For example, this combination enables:
\begin{enumerate}
    \item Predictions for the satellite populations of isolated LMC-mass hosts, which will be relevant given upcoming observations of ultra-faint dwarf satellite throughout much of the Local Volume (e.g., \citealt{Mutlu-Pakdil210501658}), using the LMC suite;
    \item Predictions for the satellite populations of Milky Way analogs, and particularly their quenched fractions, which are in moderate tension with observations of the quenched fraction for satellites of the Milky Way and Andromeda \citep{Geha170506743,Mao200812783}, using the Milky Way suite;
    \item Predictions for the luminosity functions, radial distributions, and SFHs of satellites of early-type galaxy strong lenses, to improve upon current lens modeling techniques and analyses of observed lenses, using the Group suite;
    \item Predictions for the properties of subhalos and splashback systems around cluster hosts at low redshifts, which have been measured in recent years (e.g., \citealt{Shin181106081}), using the L-Cluster suite;
    \item Predictions for the satellite populations of intermediate-redshift clusters, some of which display anomalously large quenched fractions (e.g., \citealt{2021MNRAS.507.5272N}), using the Cluster suite.
\end{enumerate}

Once again, we caution that \textsc{UniverseMachine} DR1 is not calibrated to dwarf galaxy observations; indeed, DR1 predicts unrealistic SFHs and quenched fractions for $M_{\ast} \lesssim 10^{8} \msun$ that are inconsistent with Local Group dwarf observations~\citep{Wang210211876}. Hence, an updated version of \textsc{UniverseMachine} that incorporates these constraints is required before using the faintest satellite galaxies predicted for our LMC, Milky Way, and Group suites to interpret observational results. Such updates will be presented in future work (Y.\ Wang et al.\ 2023, in preparation) that integrates more realistic satellite quenching models into the current \textsc{UniverseMachine} framework; these models will be constrained by Local Volume and SAGA satellite populations \citep{Geha170506743,Mao200812783,Carlsten220300014} and individual Local Group dwarf galaxies SFHs (e.g., \citealt{2014ApJ...796...91B,2014ApJ...789..147W,2014ApJ...789..148W,2015ApJ...804..136W,2021arXiv210804271S}). Relevant observables will be studied with the updated model, once \textsc{UniverseMachine} is calibrated to match available dwarf galaxy constraints.

%---------------------------------------------------------------------------------------
%	SECTION 6
%--------------------------------------------------------------------------------------

\section{Symphony Science Cases}
\label{sec:discussion}

In this Section, we discuss potential science cases for the individual Symphony suites (Sections \ref{sec:lmc_suite_science}--\ref{sec:lcluster_suite_science}) and of the compilation as a whole (Section \ref{sec:compilation_science}). The applications we discuss are simply examples of the science Symphony enables, and are not meant to be comprehensive. Instead, our aim is to highlight the unique characteristics of Symphony that benefit particular science cases at each mass scale, bearing in mind that existing zoom-in suites are also useful for addressing many of these questions.

\subsection{LMC-mass Suite}
\label{sec:lmc_suite_science}

Symphony's new LMC suite captures a statistical sample of LMC-mass systems at high resolution and in a narrow host mass range. Thus, this suite allows several basic questions to be addressed for LMC-mass hosts using simulations with comparable resolution to typical Milky Way--mass zoom-ins. For example, we have quantified how LMC-mass hosts' higher concentrations (Figure \ref{fig:Mhost_chost}, left panel), earlier formation times (Figure \ref{fig:Mhost_chost}, right panel), and earlier subhalo infall times (Appendix \ref{infall_times}) shape their surviving and disrupted subhalo populations relative to the subhalo populations of, e.g., Milky Way--mass systems.

Our LMC-mass suite is timely given recent observational progress in our understanding of the dark matter content and substructure of the LMC and galaxies of similar luminosities. For example, estimates of the LMC's total mass have recently been improved based on its dynamical impact on stellar streams (e.g., \citealt{Erkal181208192,Shipp210713004}), and ultra-faint satellites of the LMC have been identified using new, precise astrometric data (e.g., \citealt{Kallivayalil180501448,Patel200101746}). Meanwhile, brighter satellites of LMC-mass hosts throughout the Local Volume are now commonly being discovered (e.g., see \citealt{Carlsten220300014} for a recent census). Precise theoretical predictions for the subhalo and satellite populations of LMC-mass hosts in general will be important in order to interpret the LMC's satellite population and observations of satellites around nearby LMC-mass galaxies in a cosmological context. We expect that the combination of Symphony's LMC-mass suite and the population of LMC analogs falling into Milky Way--mass hosts presented in upcoming zoom-in simulations (D.\ Buch et al.\ 2023, in preparation) will be particularly helpful for interpreting observations of the actual LMC's luminous substructure. However, we caution that such analyses must account for our LMC suite's selection criteria, which yield LMC hosts in significantly underdense environments relative to typical LMC-mass halos.

Extensions to Symphony's LMC suite may include resimulations with: (1) analytic galaxy potentials or hydrodynamic physics to study the impact of LMC-mass galaxies on their subhalo populations, and (2) nonstandard dark matter physics to study its impact on extremely low-mass halos and subhalos, including those that are not expected to host galaxies. These efforts would build on previous zoom-ins that focus on or include LMC-mass systems (e.g., \citealt{Jahn190702979,Nadler210912120,Schwabe211009145}) by leveraging Symphony's large sample of LMC-mass halos to quantify the halo-to-halo scatter and environmental dependence of such baryonic and dark matter physics. We emphasize that our LMC suite robustly resolves the population statistics of halos and subhalos below the galaxy formation threshold (i.e., $M_{\mathrm{peak,sub}}\lesssim 10^8~\msun$; \citealt{Nadler191203303,Munshi210105822}), and will therefore help facilitate searches for novel dark matter physics using gravitational probes of low-mass halos, including strong gravitational lensing and stellar stream perturbations.

\subsection{Milky Way--mass Suite}
\label{sec:mw_suite_science}

The original version of Symphony's Milky Way--mass suite, from \cite{Mao150302637}, has already been used to investigate: (1) the dependence of subhalo abundance on host concentration \citep{Mao150302637}, (2) the disrupted progenitors of the Milky Way halo \citep{Deason160107905}, (3) the Doppler effect on dark matter annihilation signals \citep{Powell161102714}; (4) semianalytic models of galaxy formation and feedback physics using the Milky Way satellite population \citep{Lu160502075,Lu170307467}, (5) the impact of baryonic physics on subhalo populations \citep{Nadler171204467}, (6) the galaxy--halo connection for classical and ultra-faint Milky Way satellite galaxies \citep{Nadler180905542,Nadler191203303}, (7) constraints on dark matter physics beyond the CDM paradigm from Milky Way satellites \citep{Nadler190410000,Nadler200108754,Nadler200800022,Nadler210107810,Nadler210912120,Mau220111740}, (8) the dependence of subhalo populations on secondary host halo properties \citep{Fielder180705180}, (9) the impact of subhalos on dark matter density profiles \citep{Fielder200702964}, and (10) the connection between dark matter accretion and SFRs in zoom-in simulations \citep{Wang210211876}.

The studies listed above illustrate the breadth of science cases Symphony's Milky Way--mass suite enables; in many cases, performing similar studies at different host halo mass scales (e.g., using the LMC and/or Group suites) may yield additional insight. Here, we discuss a few specific examples of new directions for future work given observational and theoretical advances following the publication of the original Milky Way suite in \cite{Mao150302637}. Observationally, hundreds of new satellites around Milky Way--mass galaxies have been discovered in recent years, e.g., by the SAGA \citep{Geha170506743,Mao200812783} and ELVES \citep{Carlsten220300014} surveys. With appropriate extensions (e.g., by performing resimulations with analytic galaxy potentials) Symphony's Milky Way--mass suite can be used to provide precise predictions for the dependence of subhalo and satellite populations on environment in a narrow host halo mass range, thereby reducing theoretical uncertainties relevant for interpreting these observed satellite systems.

Constrained realizations of Milky Way--mass halos that satisfy additional ``Milky Way--like'' constraints, including a quiescent merger history following a \emph{Gaia}--Enceladus-like major merger at $z\approx 2$ (e.g., see \citealt{Helmi200204340} for a review) and the infall of an LMC analog at late times, are now being developed using zoom-in simulations (D.\ Buch et al.\ 2023, in preparation) and have recently been identified in large-volume cosmological simulations (e.g., \citealt{Evans200504969}). Comparing such constrained realizations of Milky Way--like systems to Symphony's Milky Way--mass hosts, which generally do not satisfy specific observational constraints related to the formation history of the Milky Way, will help place the Milky Way's dark matter structure and substructure in a cosmological context. These pursuits are timely given significant ongoing advances in our understanding of the Milky Way's formation history (e.g., \citealt{Helmi180606038,Naidu200608625}) and satellite population (e.g., \citealt{Drlica-Wagner191203302,Nadler191203303}).

\subsection{Group Suite}
\label{sec:group_suite_science}

Like the LMC-mass suite, Symphony's Group suite of strong lens analogs contains a large sample of high-resolution zoom-in simulations in a host halo mass regime that is relatively under-explored (however, see, e.g., \citealt{Fiacconi160203526,Fiacconi160909499,Despali181102569,Richings200514495}). Accurate predictions for the substructure of $\approx 10^{13}~\msun$ hosts are increasingly important due to recent observations that probe low-mass subhalos of early-type galaxy strong lenses, either statistically or on an individual basis (e.g., \citealt{Vegetti12013643,Hezaveh160101388,Hsueh190504182,Gilman190806983,Gilman190902573}). Furthermore, the sample of strong lenses available for high-resolution follow-up is expected to increase substantially in the next decade (e.g., \citealt{Oguri10012037,Weiner201015173}).

Predictions for the substructure of Group hosts used to facilitate current lensing studies often rely on semianalytic modeling (e.g., using \textsc{Galacticus}); however, the precise dependence of subhalo populations on host halo mass, accretion history, environment, etc.\ has not been characterized in this regime using a statistical sample of high-resolution simulations. We therefore anticipate that Symphony's Group suite (and extensions thereof) will inform these studies, particularly because the hosts were selected in a narrow mass range at a redshift typical of observed strong lenses ($z=0.5$). As an example, \cite{Wagner-Carena220300690} compared their lens substructure model to predictions from Symphony's Group suite in order to interpret its constraining power.

\subsection{Low-mass Cluster and Cluster Suites}
\label{sec:lcluster_suite_science}

We discuss the L-Cluster and Cluster suites together because their average host halo masses only differ by a factor of a few. Symphony's L-Cluster suite was originally presented in \cite{Bhattacharyya210608292}. These authors also presented a resimulation of the L-Cluster suite in a self-interacting dark matter model to study the impact of self-interactions on cluster subhalos, which is important to quantify given that upcoming cluster weak lensing measurements can potentially distinguish between CDM and self-interacting dark matter models (e.g., \citealt{Banerjee190612026}).

Symphony's Cluster suite was originally presented in \cite{Wu12093309,Wu12106358} and has been used to investigate: (1) galaxy cluster virial scaling relations \citep{Wu13070011}, (2) the structural properties and formation history of cluster-size halos \citep{Wu12093309}, (3) tidal stripping of cluster subhalos \citep{Wu12106358}, (4) the connection between hot gas and galaxy mass \citep{Wu150303924}, (5) baryonic growth and metal enrichment in clusters \citep{Martizzi151000718}, (6) cool cores in galaxy clusters \citep{Hahn150904289}, (7) the dark matter velocity distribution in clusters \citep{Mao12102721}, (8) assembly bias for cluster halos \citep{Mao170503888}, and (9) the impact of subhalos on hosts' measured dark matter density profiles \citep{Fielder200702964}.

As in our discussion of the Milky Way suite, the studies listed above simply illustrate the breadth of potential science cases for Symphony's L-Cluster and Cluster suites. Similar studies that explicitly compare the L-Cluster and Cluster suites may yield additional insights (e.g., by quantifying how the results depend on cluster mass). Below, we describe a few new directions for future work using these suites enabled by recent theoretical developments and upcoming cluster substructure measurements. 

Observationally, estimates of the cluster mass--richness relation have improved over the last few years (e.g., \citealt{McClintock180500039,Murata190407524}). Interpreting these observations to constrain the galaxy--halo connection or cosmological parameters (e.g., \citealt{To191001656,To201001138}) requires accurate predictions for the dark matter content of observed clusters. We anticipate that Symphony's L-Cluster and Cluster suites, and particularly their combination with \textsc{UniverseMachine}, will enable a more complete understanding of the galaxy populations surrounding massive clusters, which directly inform these science cases. These predictions are timely in anticipation of upcoming wide-field optical surveys, including the Rubin Observatory Legacy Survey of Space and Time (\citealt{Ivezic08052366}) and the Nancy Grace Roman Space Telescope High Latitude Wide Area Survey (\citealt{Eifler200405271}), which will simultaneously and accurately measure cluster richness and gravitational lensing signal. Robustly inferring cluster masses from richness-selected samples remains a major systematic uncertainty for these upcoming cluster cosmology analyses. Our high-resolution cluster suites enable self-consistent studies of clusters' predicted galaxy content and gravitational lensing signatures, and can therefore be used to assess optical cluster selection biases, e.g., related to halo triaxiality and concentration \citep{Wu220305416,Zhang220208211}.

Meanwhile, a growing body of literature has studied the ``splashback'' boundaries of dark matter halos using simulations (e.g., \citealt{Adhikari14094482,Diemer14011216,More150405591}), and the splashback boundary has been observed and used to constrain models of galaxy evolution (e.g., \citealt{Shin181106081,Adhikari200811663}). Because halo populations are well resolved out to many times the host's virial radius, Symphony can be used to quantify the detailed nature of splashback boundaries and halo populations, including their dependence on host properties and environment. Finally, we note that recent measurements have also been used to set limits on the fraction of ``orphan'' satellites, whose dark matter subhalos are heavily stripped in the cluster environment (e.g., \citealt{Kumar220500018}). Symphony resolves the detailed tidal stripping of cluster subhalos, which can be used to inform such measurements and empirical models for the contribution of surviving and disrupted satellites to intracluster light (e.g., \citealt{Behroozi12076105,Behroozi2019}).

\subsection{Science Cases for the Entire Symphony Compilation}
\label{sec:compilation_science}

To conclude this Section, we highlight a few key questions that the Symphony compilation as a whole can be used to address. This list is not exhaustive, and instead simply demonstrates the breadth of science enabled by the entire Symphony compilation.
\begin{enumerate}
    \item How do the secondary and environmental properties of host halos above and below the redshift-dependent collapse mass, shape their subhalo populations?
    \item To what extent is the evolution of subhalo populations self-similar? How do the normalization and slope of the SHMF evolve, and how does this evolution relate to host halos' formation histories in detail?
    \item What are disrupted subhalos' population statistics as a function of host mass, secondary properties, redshift, and environment?
    \item How do splashback mass functions and halo correlation statistics in the one to two-halo transition regime depend on host halo mass and environment?
    \item What aspects of halo and subhalo populations are not currently captured by semianalytic and/or empirical models of structure and galaxy formation?
\end{enumerate}

Finally, we emphasize that Symphony data products are made publicly available at \url{http://web.stanford.edu/group/gfc/symphony}, following in the footsteps of previous cosmological simulation data releases (e.g., \citealt{Lemson0608019,Riebe11090003,Skillman14072600,Heitmann190411966,Ishiyama200714720,Villaescusa-Navarro220101300}). Furthermore, Symphony's public data and codebase is designed to enable modular and efficient analysis. For example, our publicly available merger trees are provided in binary, depth-first format, and we provide a number of tutorials that demonstrate Symphony's key functionality. These features differentiate Symphony from high-resolution simulations of large cosmological volumes, some of which contain comparable numbers of Group, L-Cluster, and/or Cluster-mass hosts. Furthermore, to our knowledge, no existing cosmological simulations contain comparable numbers of LMC and/or Milky Way--mass halos that are resolved with $\approx 10^6$ particles compared to Symphony's $39$ LMC-mass hosts and $45$ Milky Way--mass hosts.

%---------------------------------------------------------------------------------------
%	SECTION 8
%---------------------------------------------------------------------------------------

\section{Conclusions}
\label{sec:conclusion}

We have presented Symphony, a compilation of 262 cosmological zoom-in simulations spanning host halo masses from $10^{11}$--$10^{15}~\msun$. The key features of the Symphony compilation include the following: 
\begin{enumerate}
    \item Symphony contains a statistical sample of $262$ zoom-in simulations over four decades of host halo mass, from $10^{11}$--$10^{15}~\msun$;
    \item Symphony includes two new suites of LMC-mass and strong lens analog Group zoom-ins, with $39$ and $49$ hosts in narrow mass ranges around $10^{11}~\msun$ and $10^{13}~\msun$, respectively;
    \item Symphony resolves halos and subhalos at similar, high resolution across all suites;
    \item Symphony includes extensive convergence tests, including $15$ higher-resolution resimulations (five for each of the LMC, Milky Way, and Group suites), and $96$ lower-resolution resimulations (one for every Cluster zoom-in);
    \item All Symphony halo catalogs and merger trees are publicly available at \url{http://web.stanford.edu/group/gfc/symphony}.
\end{enumerate}

The key results from the first Symphony analysis presented in this paper include the following:

\begin{enumerate}
    \item Symphony host halo concentrations are consistent with cosmological simulations after environmental selection effects are accounted for (Figure \ref{fig:Mhost_chost}), and their secondary properties correlate with subhalo abundance, with correlation coefficients that peak at the Milky Way host halo mass scale (Figure \ref{fig:rhoX});
    \item Subhalo mass functions are approximately self-similar, and scale linearly with host halo mass; at fixed sub-to-host halo mass ratio, higher-mass Symphony hosts' subhalo abundance is approximately two times higher compared to lower-mass hosts (Figure \ref{fig:SHMF});
    \item The normalized radial distribution of Symphony subhalos depends weakly on host halo mass, and they are systematically less concentrated than hosts' underlying dark matter density profiles (Figure \ref{fig:radial});
    \item Semianalytic predictions for subhalo populations from \textsc{Galacticus} are consistent with Symphony subhalo mass functions and radial distributions within the $2\sigma$ Poisson error on the mean for $M_{\mathrm{sub}}/M_{\mathrm{host}}>10^{-3}$ across our full range of host halo masses; however, \textsc{Galacticus} predicts $\approx 25\%$ higher subhalo abundances at lower sub-to-host halo mass ratios as well as more concentrated radial distributions (Figure \ref{fig:SHMF_galacticus});
    \item By applying the empirical galaxy--halo connection model \textsc{UniverseMachine} to all Symphony simulations, we predict the SFHs of galaxies occupying all halos and subhalos in our zoom-in regions (Figure \ref{fig:SMHM_UM}); these predictions statistically capture the correlated dark matter and stellar growth for systems with luminosities comparable to all currently observable satellite galaxies in the universe.
\end{enumerate}

We envision Symphony as a foundation for a range of dark matter science that depends on the detailed evolution of halos and subhalos over a wide dynamic range in a cosmological context. For example, extensions of Symphony suites in alternative dark matter models and/or cosmologies, and including analytic or hydrodynamic treatments of baryonic physics, will be helpful for ongoing efforts to compare substructure observations with theoretical predictions, including in relatively under-explored regimes using Symphony's new LMC-mass and strong lens analog Group suites. More generally, extending Symphony to other mass scales and cosmic environments while maintaining its statistical and self-consistent ethos will enable a more complete understanding of structure formation.

%----------------------------------------------------------------------------------------
%	Acknowledgements
%----------------------------------------------------------------------------------------

\acknowledgments

We are grateful to Matt Becker for providing the c125-1024 and c125-2048 simulations, Benedikt Diemer for providing data from the Erebos simulations, Alexander Ji for providing the Caterpillar parent box, Ralf Kaehler for assisting with the visualizations in Figure \ref{fig:vis}, and Bryn\'{e} Hadnott for helpful discussions and contributions to the public website and notebooks. We thank Tom Abel, Simon Birrer, Deveshi Buch, Fangzhou Jiang, and Shengqi Yang for helpful discussions related to this work.

This research was supported in part by the Kavli Institute for Particle Astrophysics and Cosmology at Stanford University and SLAC National Accelerator Laboratory, and from the U.S. Department of Energy under contract No.\ DE-AC02-76SF00515 to SLAC National Accelerator Laboratory. This research also received support from the National Science Foundation under grant No.\ NSF DGE-1656518 through the NSF Graduate Research Fellowship received by E.O.N. A.B.\ and X.D.\ acknowledge support from NASA ATP grant 17-ATP17-0120. Y.-Y.M.\ was supported by NASA through the NASA Hubble Fellowship grant No.\ HST-HF2-51441.001 awarded by the Space Telescope Science Institute, which is operated by the Association of Universities for Research in Astronomy, Inc., under NASA contract NAS5-26555.
H.-Y.W.\ is supported by DOE grant DE-SC0021916 and NASA grant 15-WFIRST15-0008. Part of this work was performed at the Aspen Center for Physics, which is supported by National Science Foundation grant PHY-1607611.

This research made use of computational resources at SLAC National Accelerator Laboratory, a U.S.\ Department of Energy Office, and the Sherlock cluster at the Stanford Research Computing Center (SRCC); the authors are thankful for the support of the SLAC and SRCC computing teams. Additional computing resources used in this work were made available by a generous grant from the Ahmanson Foundation.

This research used \url{https://arXiv.org} and NASA's Astrophysics Data System for bibliographic information.

\software{
Colossus \citep{Diemer171204512},
{\sc consistent-trees} \citep{Behroozi11104370},
\textsc{Gadget-2} \citep{Springel0505010},
Helpers (\http{bitbucket.org/yymao/helpers/src/master/}),
Jupyter (\http{jupyter.org}),
Matplotlib \citep{matplotlib},
NumPy \citep{numpy},
Pynbody \citep{Pontzen1305002},
{\sc Rockstar} \citep{Behroozi11104372}
SciPy \citep{scipy}, 
Seaborn (\https{seaborn.pydata.org}).
}

\bibliographystyle{yahapj2}
\bibliography{references,software}

%---------------------------------------------------------------------------------------
%	Appendices
%---------------------------------------------------------------------------------------

\appendix

\section{Figure 1 References}
\label{sec:fig1_sims}

Apart from Symphony, Figure \ref{fig:simulations} shows the following cosmological zoom-in simulations: NIHAO \citep{Wang150304818,Dutton151200453}, ELVIS \citep{Garrison-Kimmel13106746,Garrison-Kimmel180604143}, Artemis \citep{Poole-McKenzie200615159}, Caterpillar \citep{Griffen150901255}, Via Lactea \citep{Diemand08051244}, Aquarius \citep{Springel08090898}, Latte \citep{Wetzel160205957,Samuel190411508}, APOSTLE \citep{Sawala151101098}, Ponos \citep{Fiacconi160203526,Fiacconi160909499}, Phoenix \citep{Gao12011940}, the Three Hundred Project \citep{Cui180904622}, C-EAGLE \citep{Barnes170310907}, and simulations from \cite{Despali181102569} and \cite{Richings181112437}. This collection is not comprehensive, and does not include: resimulation suites (e.g., PhatELVIS; \citealt{Kelley181112413}), zoom-ins of cosmological volumes that do not target specific hosts (e.g., Copernicus Complexio; \citealt{Hellwing150506436}), hydrodynamic simulations of lower-mass galaxies (e.g., from FIRE; \citealt{Hopkins13112073}), or simulations only presented in a hydrodynamic context (e.g., ERIS; \citealt{Guedes11036030}).

\section{Convergence Tests}
\label{sec:convergence}

This Appendix describes Symphony resimulation suites used to test for numerical convergence and the resulting convergence properties. Appendix \ref{sec:resims} describes our suite of zoom-in resimulations run with varying resolution, Appendix~\ref{sec:convergence_res} describes their convergence properties, Appendix~\ref{sec:convergence_eps} describes resimulations with varying softening, and Appendix~\ref{sec:convergence_eta} describes resimulations with varying time stepping.

\subsection{Resimulations with Varying Resolution}
\label{sec:resims}

We perform high-resolution (HR) resimulations for five zoom-ins from each of our LMC, Milky Way, and Group suites, for a total of 15 resimulations. These resimulations use one additional \textsc{MUSIC} refinement level relative to each suite's fiducial initial conditions; thus, the HR LMC, Milky Way, and Group resimulations use six, five, and four refinement regions, yielding an equivalent of $32,768$, $16,384$, and $8192$ particles per side in the most refined regions, respectively. The corresponding particle mass in the highest-resolution regions of these resimulations is a factor of 8 lower than the fiducial value for each suite, yielding $m_{\mathrm{part,HR}}/M_\odot=6.3\times 10^3$, $5.0\times 10^4$, and $4.1\times 10^5$ in the highest-resolution regions for the LMC, Milky Way, and Group HR runs, respectively. The softening lengths in the highest-resolution regions are reduced by a factor of 2 relative to the fiducial values, yielding values of $\epsilon_{\mathrm{HR}}/(\mathrm{pc}\ h^{-1})=40$, $80$, and $170$, respectively. All other simulation parameters are fixed at their fiducial values.

Furthermore, the entire Cluster suite was originally resimulated at one resolution level lower than the fiducial Cluster simulations we present \citep{Wu12093309,Wu12106358}. These low-resolution (LR) resimulations were performed with a particle mass of $m_{\mathrm{part,LR}}=1.4\times 10^9~\msun$ (eight times higher than the fiducial value) and a softening of $\epsilon_{\mathrm{LR}}=6687~ \mathrm{pc}\ h^{-1}$ ($\approx$ two times higher than the fiducial value). Note that, because our fiducial Cluster suite has relatively high mass resolution compared to our other suites (see Figure \ref{fig:simulations}), host halo particle counts in these ``low-resolution'' runs only differ from those in the fiducial versions of our other suites by factors of a few.

Finally, we note that \cite{Bhattacharyya210608292} resimulated one host from the L-Cluster suite at higher resolution, using a particle mass of $m_{\mathrm{part,HR}}=2.8\times 10^7~ \msun$ (eight times lower than the fiducial value) and a softening of $\epsilon_{\mathrm{HR}}=600~ \mathrm{pc}\ h^{-1}$ (two times lower than the fiducial value). These authors find that the $V_{\rm peak,sub}$ function of surviving subhalos is enhanced at low $V_{\rm peak,sub}$ in the high-resolution resimulation, although with only a single zoom-in, it is difficult to draw strong conclusions from this test. As shown in Appendix \ref{sec:convergence_res}, the agreement between the convergence properties of our LMC, Milky Way, and Group resimulations compared to our Cluster convergence tests suggests that there is no significant mass trend in our zoom-ins' convergence properties. We therefore assume that the convergence properties of the L-Cluster suite are similar to our other suites, deferring additional resimulations to future work.

\subsection{Convergence Properties with Varying Resolution}
\label{sec:convergence_res}

\begin{figure*}[t!]
\centering
\includegraphics[width=0.49\textwidth]{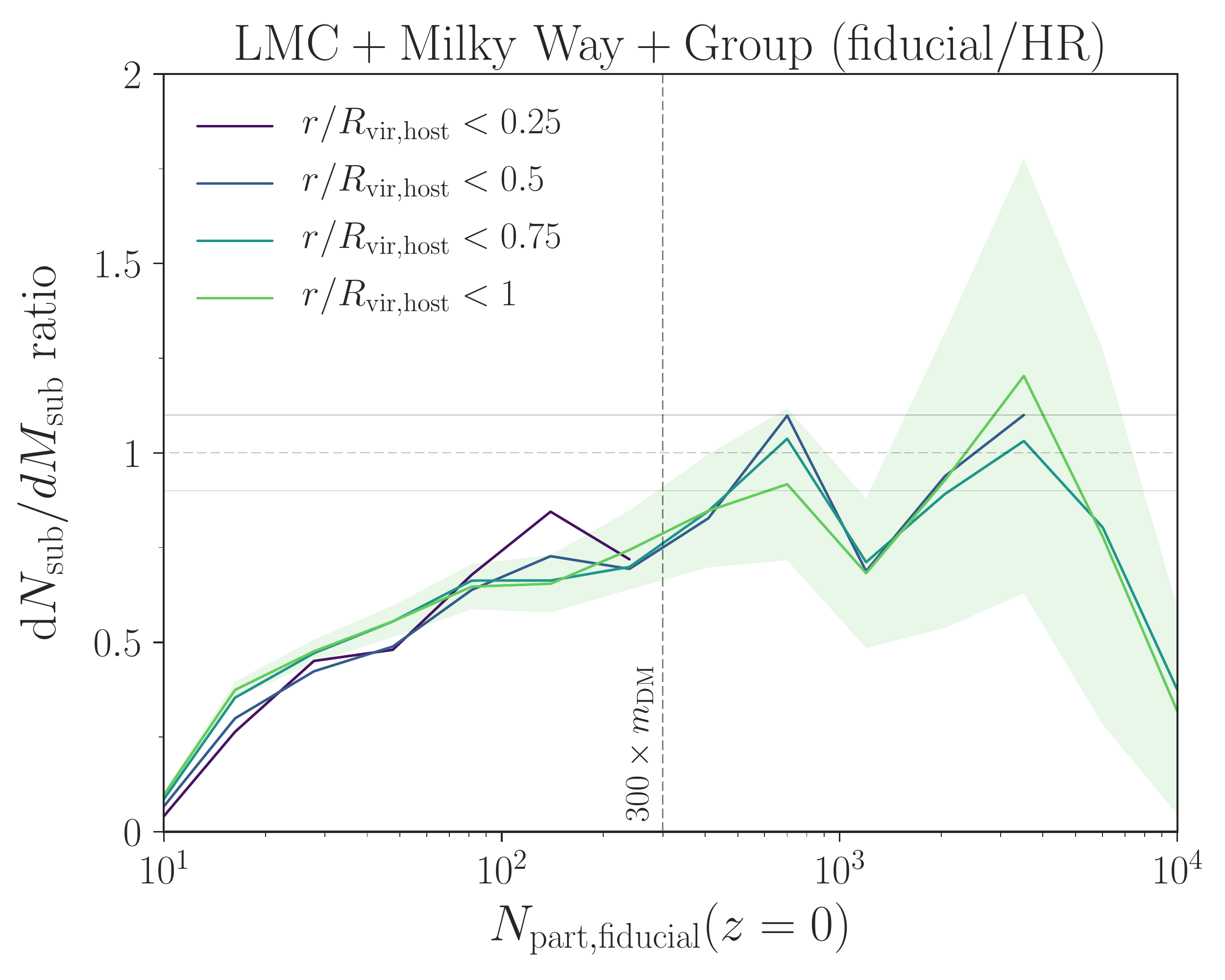}
\includegraphics[width=0.49\textwidth]{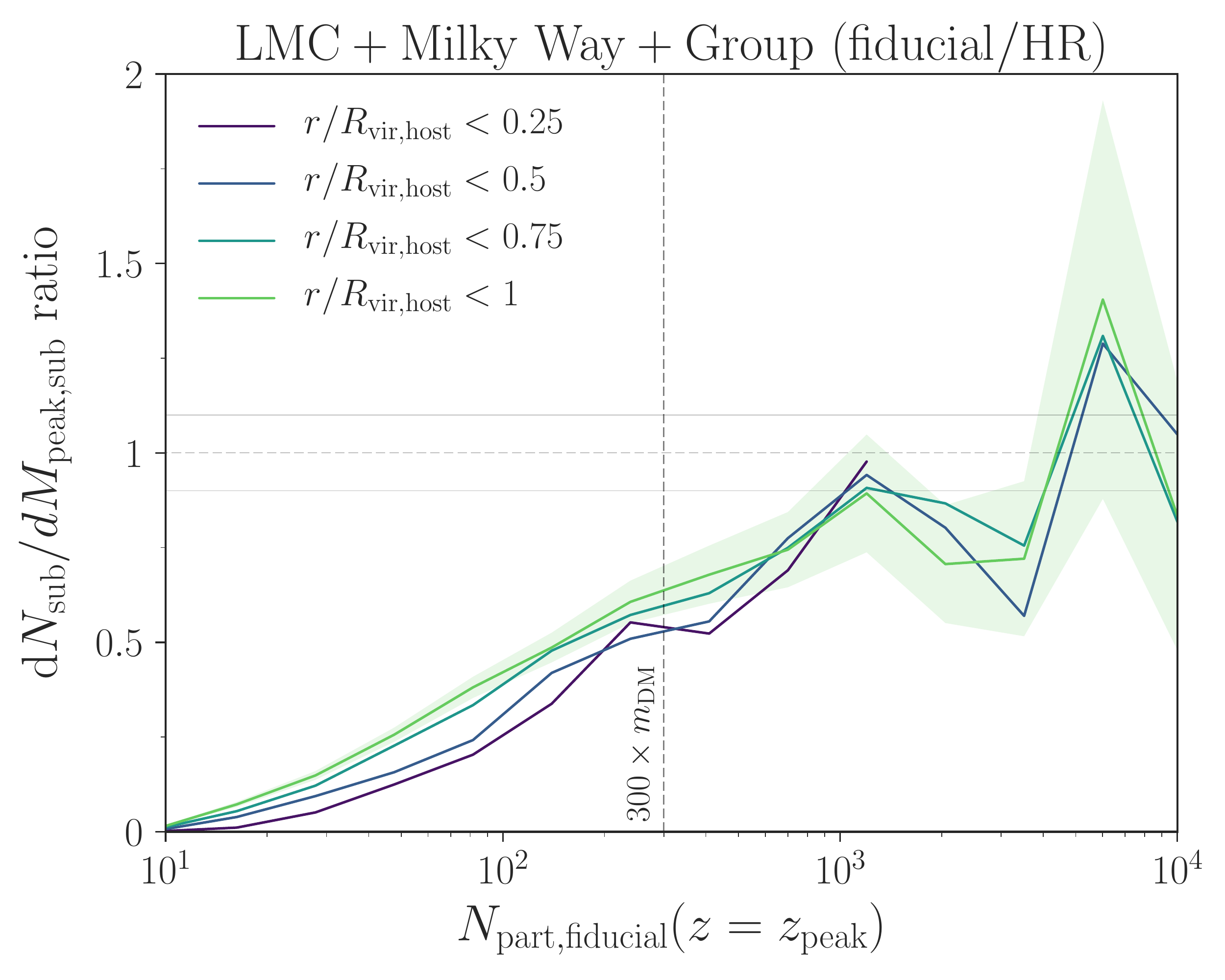}\\
\includegraphics[width=0.49\textwidth]{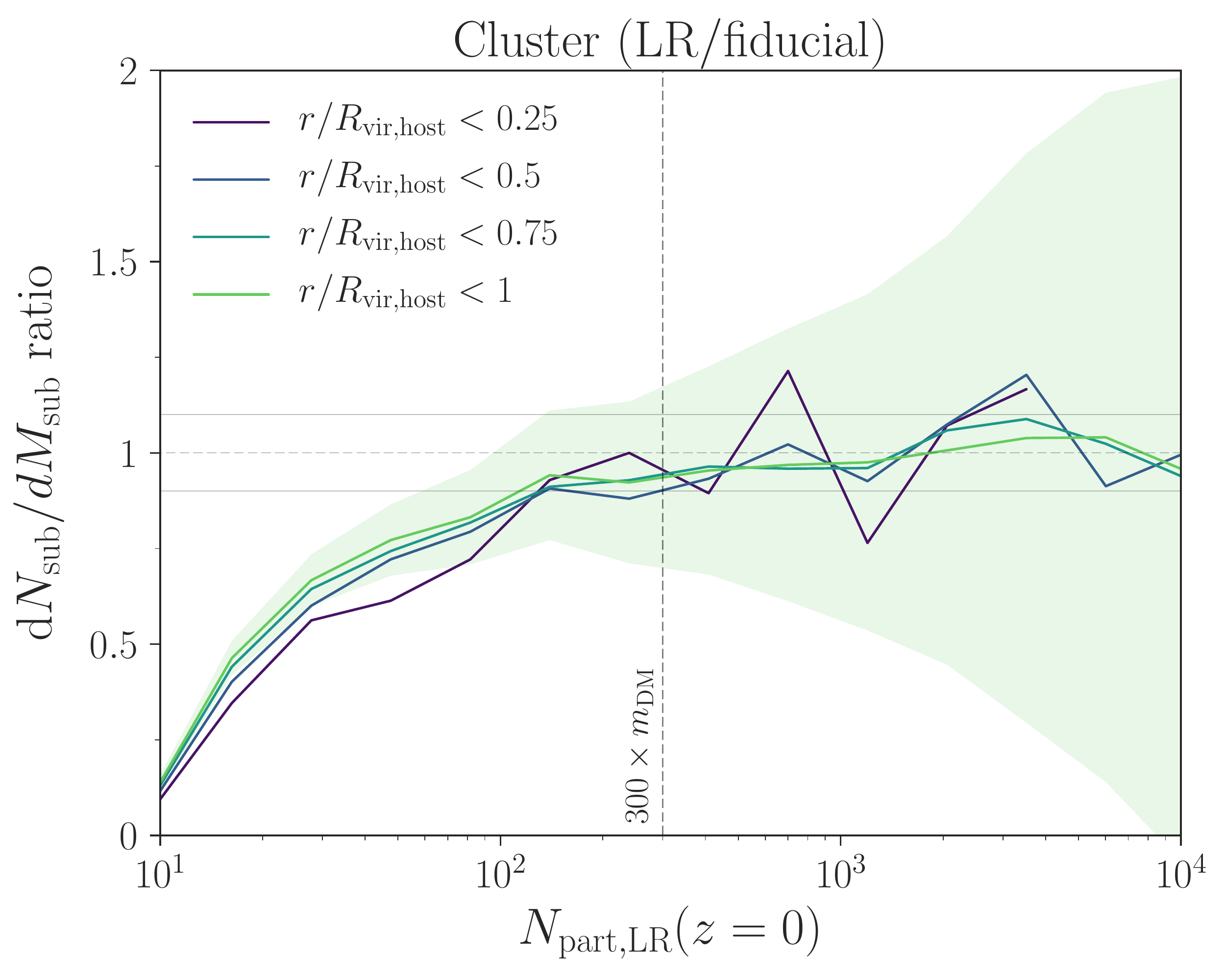}
\includegraphics[width=0.49\textwidth]{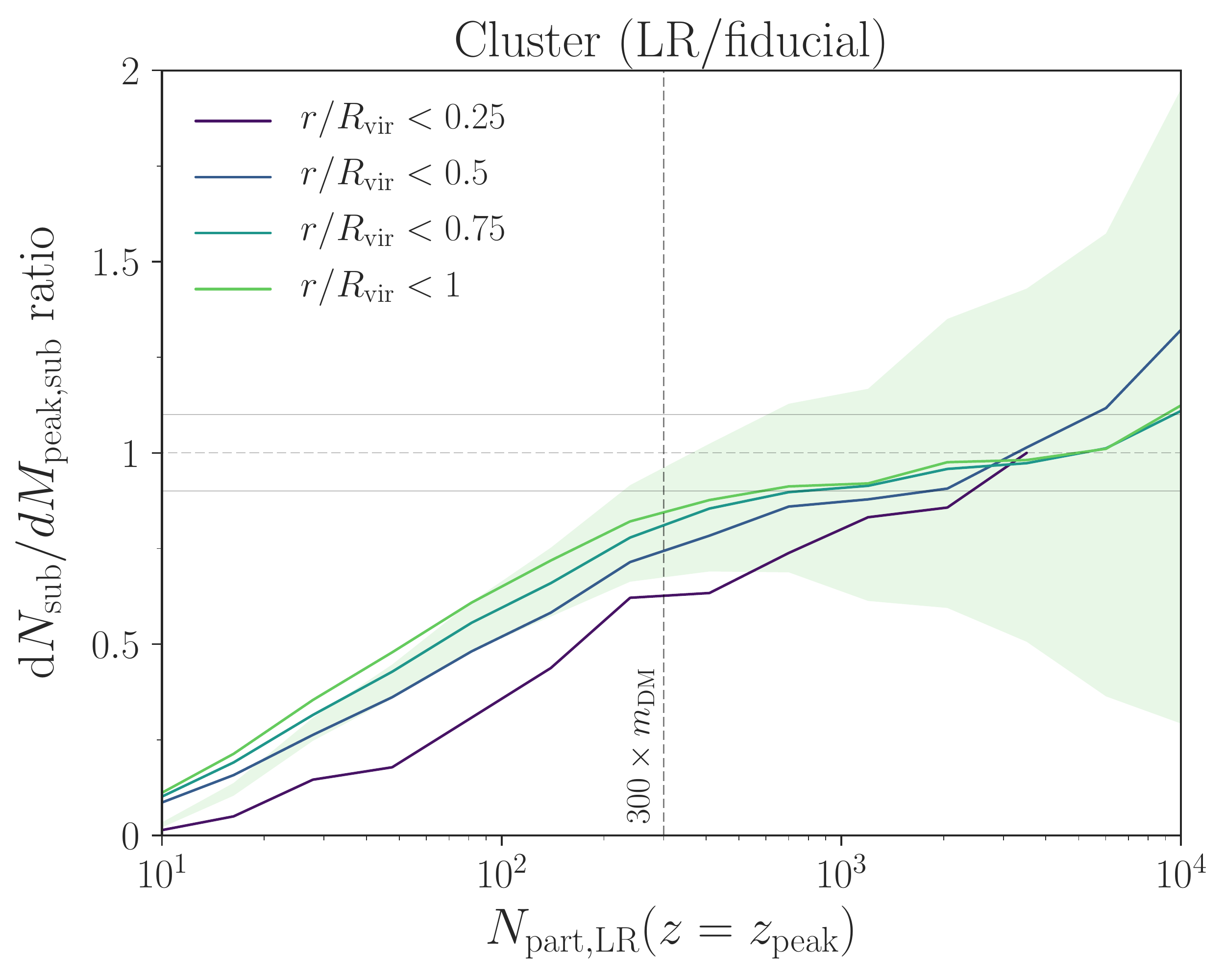}
\caption{Top-left panel: ratio of the mean differential subhalo mass function, evaluated using subhalo virial mass at $z=0$, in 15 fiducial-resolution simulations (five from each of the LMC, Milky Way, and Group suites), normalized to that in high-resolution (HR) resimulations of these halos. Results are plotted in terms of equivalent particle number in the fiducial-resolution runs, i.e., $N_{\mathrm{part,fiducial}}(z=0)=M_{\mathrm{sub}}/m_{\mathrm{part,fiducial}}$. Ratios are shown for subhalos within $0.25$ (purple), $0.5$ (blue), $0.75$ (blue-green), and $1.0$ (green) times the host halo virial radius in each fiducial or HR simulation. The green band indicates $1\sigma$ Poisson error on the mean for the result using all subhalos within the virial radius. Thin solid horizontal lines indicate ratios of $\pm 10\%$, and curves are truncated when the error on the mean exceeds $100\%$. The vertical dashed line corresponds to our fiducial convergence limit of $300 m_{\mathrm{part}}$. Top-right panel: same as the top-left panel, but for differential \emph{peak} subhalo mass functions. Note that a more stringent particle number threshold of $N_{\mathrm{part,LR}}(z=z_{\mathrm{peak}})\gtrsim 1000$ is required to achieve a similar level of convergence in terms of $M_{\mathrm{peak,sub}}$ for subhalos at large distances from the host center. Furthermore, the convergence properties of the $M_{\mathrm{peak,sub}}$ function depend strongly on distance (see Appendix \ref{sec:convergence_res} for details).
Bottom-left panel: same as the top-left panel, but using subhalo mass functions from low-resolution (LR) resimulations of all $96$ Cluster halos, normalized to those in our fiducial-resolution Cluster simulations and stacked over all hosts. Results are plotted in terms of equivalent particle number in the LR runs, i.e., $N_{\mathrm{part,LR}}(z=0)=M_{\mathrm{sub}}/m_{\mathrm{part,LR}}$. Bottom-right panel: same as the bottom-left panel, but for differential peak subhalo mass functions.}
\label{fig:cluster_convergence}
\end{figure*}

We compare simulations at varying resolution against one another in Figure \ref{fig:cluster_convergence}. The top-left panel shows the ratio of differential SHMFs for our 15 HR resimulations compared against the SHMFs for those halos at fiducial resolutions. The ratio of the SHMFs is computed separately for each suite and then averaged into a combined set of ratios.
The top-right panel shows the same for differential $M_{\mathrm{peak,sub}}$ functions, where $M_{\mathrm{peak,sub}}$ is the maximum virial mass that a subhalo ever achieved along its main branch; this quantity is often used in analyses where one wants to minimize the impact of subhalo disruption (e.g., see \citealt{Campbell170506347}). The bottom panels show the same quantities measured for our LR vs.\ fiducial Cluster simulations. Figure \ref{fig:cluster_convergence} also shows the dependence of convergence behavior on subhalo distance from the host center, $r/R_{\rm vir,host}$. Previous convergence tests have shown that convergence properties of subhalos depend on distance in idealized configurations (e.g., \citealt{VandenBosch171105276,VandenBosch180105427}) and cosmological settings (e.g., \citealt{MansfieldAvestruz2021}). This is because tidal forces and subhalo mass-loss rates increase at smaller distances, leading to more demanding numerical requirements.

SHMFs are converged at the $10\%$ level for~$N_{\rm part}=M_{\mathrm{sub}}/m_{\mathrm{part}} > 300$ at $z=0$. Trends with subhalo distance from the host center are weak and only reach statistical significance for low-particle count subhalos that are below the convergence limit. The trend with $r/R_{\rm vir,host}$ is not well characterized in the three low-mass suites due to the relatively small number of hosts tested, but the Cluster test suggests that convergence behavior at distances down to at least $\approx 0.25R_{\mathrm{vir,host}}$ is consistent with that for more distant subhalos.

SHMFs evaluated using peak virial mass show weaker convergence trends, requiring between $\approx 700$ and $1000$ particles for convergence at larger distances from the host center. Unlike SHMFs evaluated using $z=0$ masses, the convergence behavior is strongly dependent on $r/R_{\rm vir,host}$, even at high particle counts, such that subhalos closer to the host center require more particles for convergence than distant subhalos. We note that the apparent convergence of SHMFs within $r/R_{\rm vir,host}<0.25$ is consistent with being a statistical fluctuation, and it is possible that a larger suite of multiresolution halos would measure stricter convergence criteria.

While entirely consistent with existing convergence tests, this radial dependence has important implications for studies of satellite radial profiles that rely on abundance matching using subhalos' peak properties (e.g., \citealt{Reddick12072160}) or efforts to use radial profiles to constrain the galaxy--halo connection (e.g., \citealt{Graus180803654}). The effect of this lack of convergence will be to preferentially suppress subhalo number density profiles in the inner regions of host halos. We note that the true level of bias in subhalo profiles may be larger than indicated by Figure \ref{fig:cluster_convergence}, which shows the ratio of lower-to-higher-resolution runs. This is possible because it is possible that---until strict convergence is achieved---inner radial profiles in the higher-resolution runs are also biased low.

At first it may seem counterintuitive that SHMFs evaluated using peak mass are more poorly converged than those using $z=0$ mass, given that $M_{\mathrm{peak,sub}}$ is unchanged by the exact mass-loss rate of a subhalo after infall as long as that subhalo can still be tracked by the halo finder. However, the $M_{\mathrm{peak,sub}}$ function is effectively a convolution of the $M_{\mathrm{sub}}$ function with the $M_{\mathrm{peak,sub}}/M_{\mathrm{sub}}$ distribution (see Appendix \ref{stripping_distributions}), and is therefore sensitive to subhalos below the $M_{\mathrm{sub}}$ convergence limit, even for subhalos with large peak particle counts. Furthermore, the radial dependence of convergence properties---which are only significant for low-particle-count halos when using $z=0$ masses---influence the $M_{\mathrm{peak,sub}}$ function at particle counts where mass functions appear converged for all subhalos within $R_{\rm vir,host}$. The radial dependence of $M_{\mathrm{peak,sub}}$ convergence is unlikely to be a problem specific to our zoom-in suites; this behavior has been noted in earlier convergence studies \citep{VandenBosch171105276,VandenBosch180105427,MansfieldAvestruz2021}, and, as explained above, is expected from first principles.

\subsection{Resimulations with Varying Softening}
\label{sec:convergence_eps}

\begin{figure*}[t!]
\centering
\includegraphics[width=0.49\textwidth]{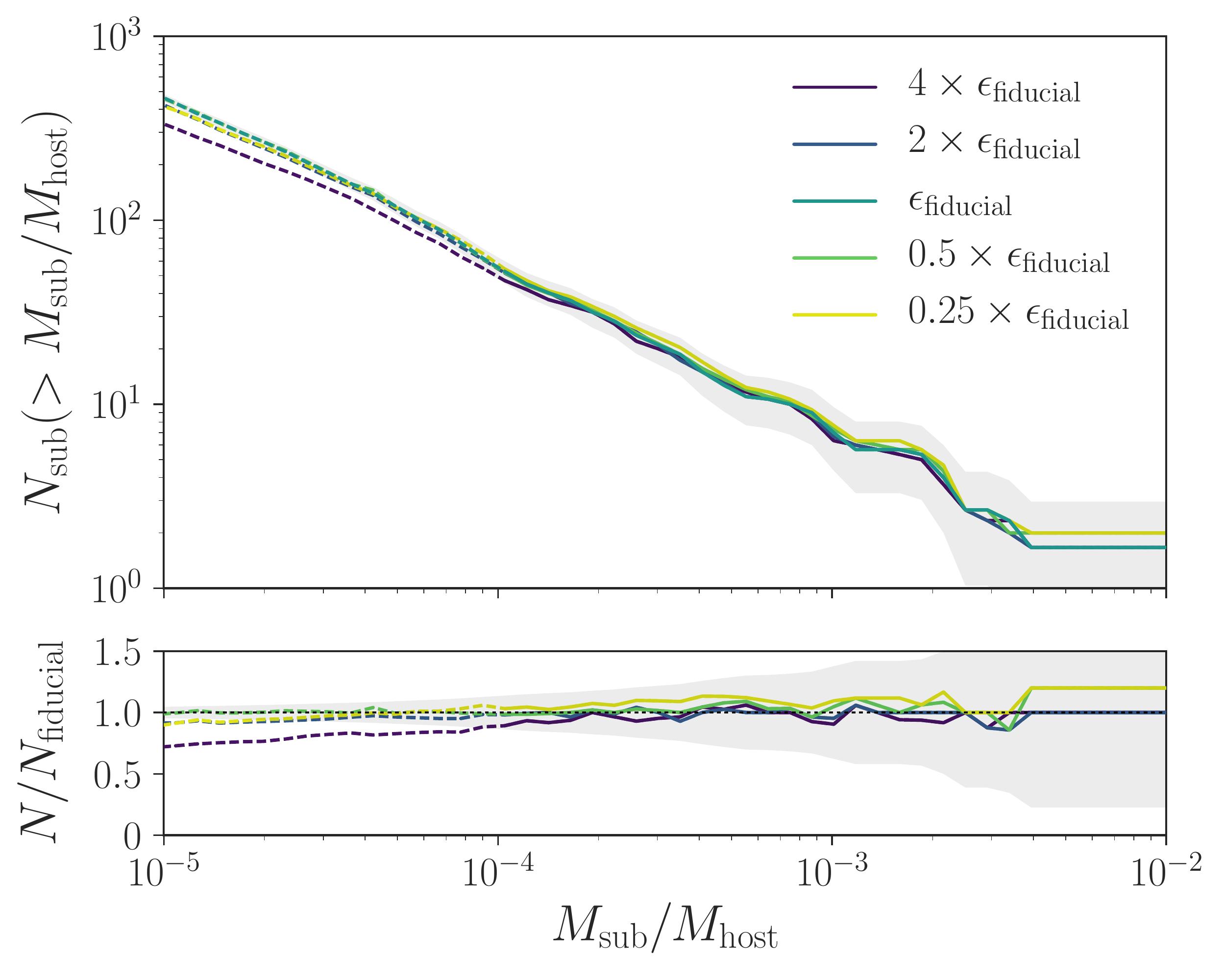}
\includegraphics[width=0.49\textwidth]{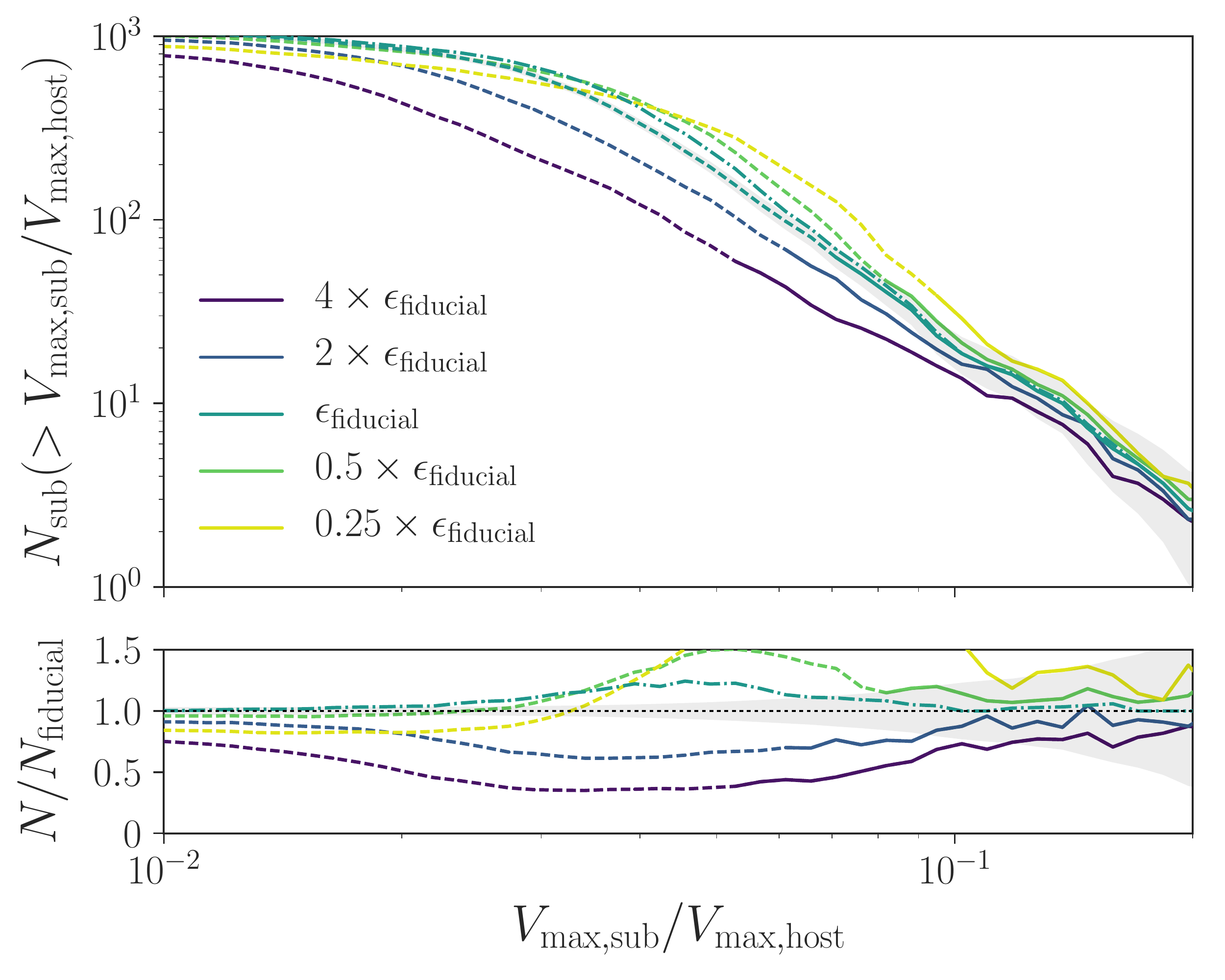}\\
\includegraphics[width=0.49\textwidth]{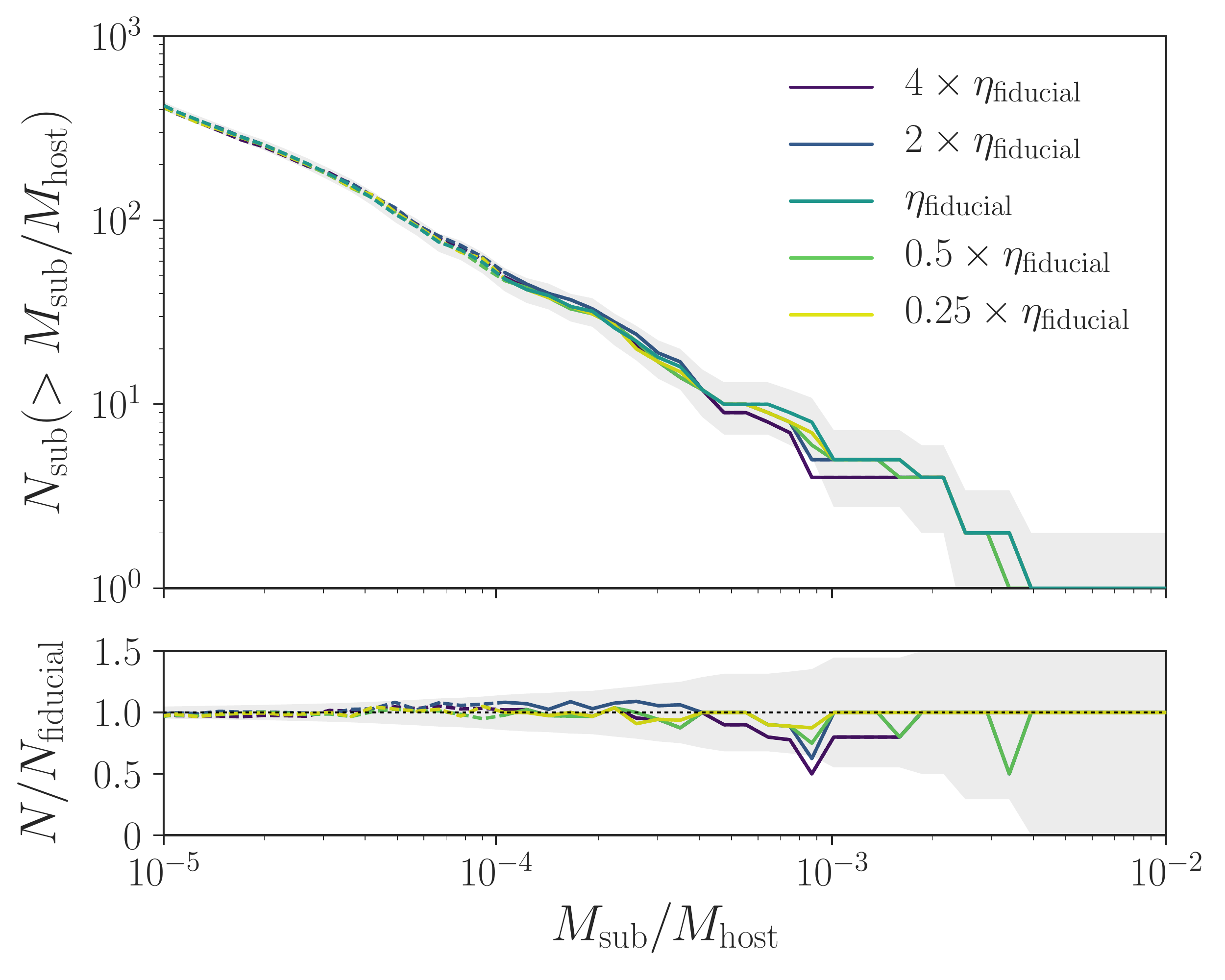}
\includegraphics[width=0.49\textwidth]{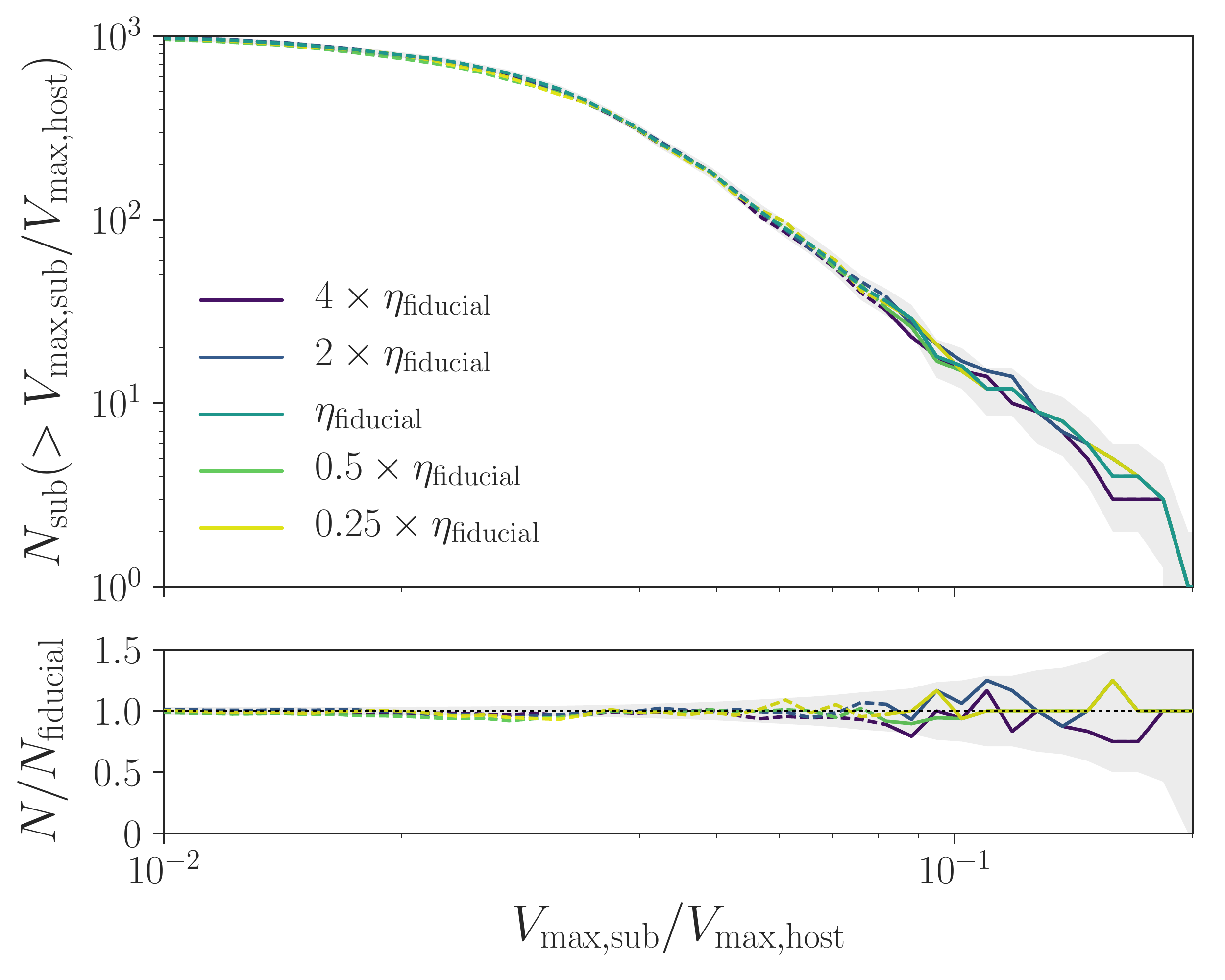}
\caption{Top-left panel: subhalo mass functions evaluated using sub-to-host halo virial mass ratio at $z=0$, stacked over three hosts from the Milky Way suite resimulated with Plummer-equivalent gravitational softening lengths of $4.0$ (purple), $2.0$ (blue), $0.5$ (green), and $0.25$ (yellow) times the fiducial value of $\epsilon_{\mathrm{fiducial}}=170~ \mathrm{pc\ h}^{-1}$ (blue-green). Solid lines show the mean cumulative mass function stacked over these three hosts, and transition to dashed at the largest value of $M_{\mathrm{sub}}/M_{\mathrm{host}}$ among these hosts set by our convergence limit of $300 m_{\mathrm{part}}$. The dotted--dashed line shows the fiducial $V_{\mathrm{max,sub}}$ function, corrected for low-$\epsilon$ suppression of centripetal forces using the \cite{MansfieldAvestruz2021} model. The shaded region shows the $1\sigma$ Poisson error on the mean, and the bottom panel shows the ratio of the runs with varying softening length relative to the fiducial setting. These subhalo mass functions are consistent with our fiducial results at the $\approx10\%$ level down to the convergence limit. Top-right panel: same as the top-left panel, but for normalized maximum circular velocity functions at $z=0$. Note that the resimulations transition from solid to dashed at slightly different values of $V_{\mathrm{max,sub}}/V_{\mathrm{max,host}}$ because there is not a one-to-one relationship between particle count and $V_{\mathrm{max,sub}}$. The systematic variations in $V_{\mathrm{max,sub}}$ as a function of softening length are discussed in Appendix \ref{sec:convergence_eps}. Bottom panels: same as the top panels, but for varying time stepping.}
\label{fig:eps_tests}
\end{figure*}

We perform additional resimulations for three host halos in the Milky Way suite using the fiducial initial conditions and particle mass but varying the softening length. In particular, these resimulations adopt $0.25$, $0.5$, $2$, and $4$ times the fiducial softening length of $\epsilon_{\mathrm{fiducial}}=170~ \mathrm{pc}\ h^{-1}$ while holding all other parameters fixed. The top-left panel of Figure~\ref{fig:eps_tests} demonstrates that SHMFs evaluated using $z=0$ subhalo mass in these runs are consistent with our fiducial results at the $\approx10\%$ level down to the convergence limit of $300m_{\mathrm{part}}$.

These results are broadly consistent with similar tests performed in \citet{MansfieldAvestruz2021} that studied the impact of force softening on the 50 largest clusters in c125-1024. The average mass of this sample was $\approx 10^{14} M_\odot$, leading to $N_{\rm part} \approx 7\times 10^5$ and resolved sub-to-host halo mass ratios of $M_{\mathrm{sub}}/M_{\rm host} \approx4\times 10^{-4}$, on average. Thus, our tests have better mass resolution but much less statistical power. They also probe force softening scales that are four times smaller relative to the mean interparticle spacing, $L/N$, than the tests in \citet{MansfieldAvestruz2021}. These authors showed that variations in $\epsilon$ only begin to suppress SHMFs significantly at very large $\epsilon \approx 0.12\times L/N,$ similar to softenings that lead to the historical ``overmerging'' problem in $N$-body simulations (see discussion in, e.g., \citealt{Klypin9708191}). Our suites' softening values are well below this value, with $0.01 < \epsilon_{\mathrm {fiducial}} /(L/N) < 0.012$ for the LMC, Milky Way, Group, and L-Cluster suites, and $\epsilon_{\rm fiducial} = 0.027\times L/N$ for the Cluster suite (see Section \ref{sec:symphony_overview}). The lack of impact on our SHMFs due to this effect is therefore expected.

In the top-right panel of Figure \ref{fig:eps_tests}, we study the impact of force softening on the subhalo maximum circular velocity functions. We find that subhalos contract and that the $V_{\rm max,sub}$ function increases with decreasing $\epsilon$. We do not find that the subhalo $V_{\rm max,sub}$ function converges with decreasing $\epsilon$. This raises the question of whether convergence in the $V_{\rm max,sub}$ function requires even smaller force softening scales than tested here. This would have serious implications, because virtually all cosmological and zoom-in simulations use force softening scales larger than our simulations' $0.25\epsilon_{\rm fiducial}$ when expressed in terms of the mean interparticle spacing.

One critical question is whether the lack of convergence in the small- and large-$\epsilon$ regimes is caused by the same effect. \citet{MansfieldAvestruz2021} argued that, in the large epsilon regime ($\epsilon \gtrsim \epsilon_{\rm fiducial}$), contraction with decreasing $\epsilon$ is due to the reduction of centripetal forces. This suppression can occur far beyond the formal force softening scale because lower centripetal forces lead to lower densities, which in turn cause lower (physical) centripetal forces. \citet{MansfieldAvestruz2021} showed that a model based on this effect predicts convergence limits across a wide range of cosmological boxes in the low-$\epsilon$ regime and quantitatively predicts the exact biases in $V_{\rm max,sub}$ for controlled test simulations. However, this model predicts that the effect would be weak at $\epsilon_{\rm fiducial}$ and nonexistent at $0.5 \epsilon_{\rm fiducial}$, suggesting that the lack of convergence at small $\epsilon$ in Figure \ref{fig:eps_tests} has a separate cause.

To illustrate this point, we apply the $\epsilon$-debiasing model from \cite{MansfieldAvestruz2021} to our fiducial run and show it as the dotted--dashed curve in the top-right panel of Figure \ref{fig:eps_tests}. This model was shown to exactly recreate the $\epsilon$-dependent $V_{\rm max}$ bias in resimulations of c125-1024, the parent simulation of these zoom-ins. The debiasing model predicts that $V_{\rm max}$ suppression is negligible at fiducial force softening scales. We note that we assume an NFW profile for subhalos when applying the \citet{MansfieldAvestruz2021}, which will overestimate the density of subhalos in their outskirts \citep[e.g.][]{Ogiya2019,Errani2022}, thus leading to larger $R_{\rm max}/R_{\rm vir}$ at a fixed value of $V_{\rm max}/V_{\rm vir}.$ The level of bias in the \citet{MansfieldAvestruz2021} model decreases with increasing $R_{\rm max}/\epsilon,$ meaning that the use of NFW profiles causes us to {\em overestimate} the level of biasing. Thus, the dotted--dashed line is a conservative upper limit.

The analysis in \citet{Ludlow2019} offers an alternative explanation for the small-$\epsilon$ behavior. The authors tested the impact of a very wide range of force softening scales on subhalo profiles and found that small, isolated halos experience substantial contraction when force softening scales are smaller than $0.0018$ to $0.003 \times (L/N)$ (see their Figure 2, and note that its upper x-axis is in units of $10^{-2} \epsilon/(L/N)$), which corresponds to $\epsilon \lesssim 0.25\epsilon_{\rm fiducial}$ for Symphony suites. This effect only occurs for simulations run using the default \textsc{Gadget-2} time-stepping criterion (\texttt{ErrTolIntAccuracy}), $\eta=0.025$, which is similar to the value of $\eta=0.01$ used in our simulations. For finer time stepping, no contraction is found in the \citet{Ludlow2019} tests. This suggests that the nonconvergence shown in the top-right panels of Figure \ref{fig:eps_tests} in the small-$\epsilon$ limit is caused by time-stepping errors.

\subsection{Resimulations with Varying Time Stepping}
\label{sec:convergence_eta}

We resimulate one host halo in the Milky Way suite for several values of the \textsc{Gadget-2} time stepping criterion, $\eta$ (\texttt{ErrTolIntAccuracy}). Specifically, these resimulations use $0.25$, $0.5$, $2$, and $4$ times our simulations' fiducial setting of $\eta=0.01$; the resulting SHMFs and maximum circular velocity functions compared to our fiducial run are shown in the bottom panels of Figure \ref{fig:eps_tests}.

\begin{figure*}[t!]
\centering
\includegraphics[width=\textwidth]{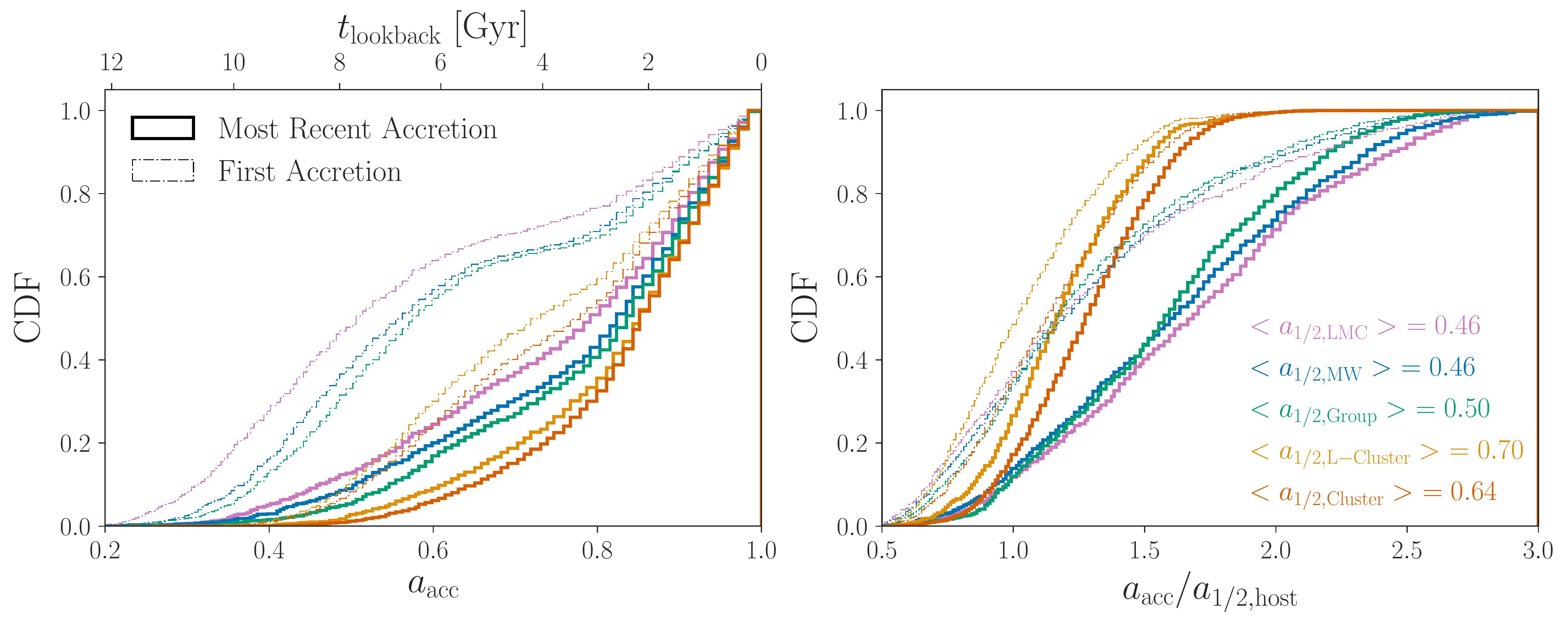}
\caption{Left panel: cumulative distributions of Symphony subhalos' most recent (solid, thick lines) and first (thin, dot-dashed lines) accretion scale factors into \emph{any} larger halo, stacked over all subhalos from each suite with $M_{\mathrm{sub}}/M_{\mathrm{host}}>2.7\times 10^{-4}$. Top ticks indicate lookback time assuming the cosmological parameters for our LMC, Milky Way, and Group suites. Right panel: same as the left panel, but with accretion scale factors normalized by the half-mass scale factor of each Symphony host. The legend indicates the mean half-mass scale factors for the LMC (pink), Milky Way (blue), Group (green), L-Cluster (gold), and Cluster (red) suites (see Table \ref{tab:sims}).}
\label{fig:acc_cdf}
\end{figure*}

We find that SHMFs at $z=0$ are robust to changes in the time-stepping parameter at the $\approx 1\%$ level down to our convergence limit. Maximum circular velocity functions are similarly well converged. Moreover, these deviations do not systematically increase as a function of the time-stepping criterion, suggesting that they are partly due to noise associated with the resimulations of the particular halo we selected. Thus, we expect that including resimulations of additional Symphony zoom-ins with varying time stepping would further reduce this systematic.

In summary, in Appendix \ref{sec:convergence_eps} we showed that large-$\epsilon$ suppression does not impact our $V_{\rm max,sub}$ functions at our fiducial $\epsilon,$ but that our subhalos continue to contract as $\epsilon$ is decreased below the fiducial level. \citet{Ludlow2019} showed that, as $\epsilon$ is decreased below our fiducial level, insufficient time stepping can also lead to subhalo contraction. However, in this section we showed that such time-stepping-dependent effects are not important {\em at} our fiducial $\epsilon.$ Therefore, our $\epsilon_{\mathrm{fiducial}}$ sits in an island of stability, in which it is not large enough to suppress rotation curves out to $V_{\rm max}$ for resolved subhalos, but not small enough to lead to time-stepping errors.

\section{Auxiliary Subhalo Population Statistics}
\label{subhalo_appendix}

This Appendix studies characteristics of Symphony subhalo populations beyond those discussed in the main text, including subhalo infall time distributions (Appendix \ref{infall_times}) and the amount of tidal stripping subhalos above our resolution limit experience (Appendix \ref{stripping_distributions}).

\subsection{Subhalo Infall Time Distributions}
\label{infall_times}

First, Figure \ref{fig:acc_cdf} shows cumulative infall time distributions for subhalos above our conservative resolution limit of $M_{\mathrm{sub}}/M_{\mathrm{host}}>2.7\times 10^{-4}$, stacked over all hosts within each Symphony suite. Infall times are measured using the \texttt{Acc\_Scale} (solid, thick distributions) and \texttt{First\_Acc\_Scale} (thin, dashed distributions) output by \textsc{consistent-trees}. Note that these variables record the scale factor at which a subhalo most recently or first accreted into \emph{any} larger halo, which may not be the Symphony host halo it resides in at $z=0$. Thus, these distributions include ``pre-processing'' effects, in which subhalos first orbited within another halo before falling into the $z=0$ host. Furthermore, even for objects that have not been subhalos of any halo other than the Symphony target host, the first and most recent accretion scale factors differ for splashback subhalos with apocenters outside of the host's virial radius.

\begin{figure*}[t!]
\centering
\includegraphics[width=\textwidth]{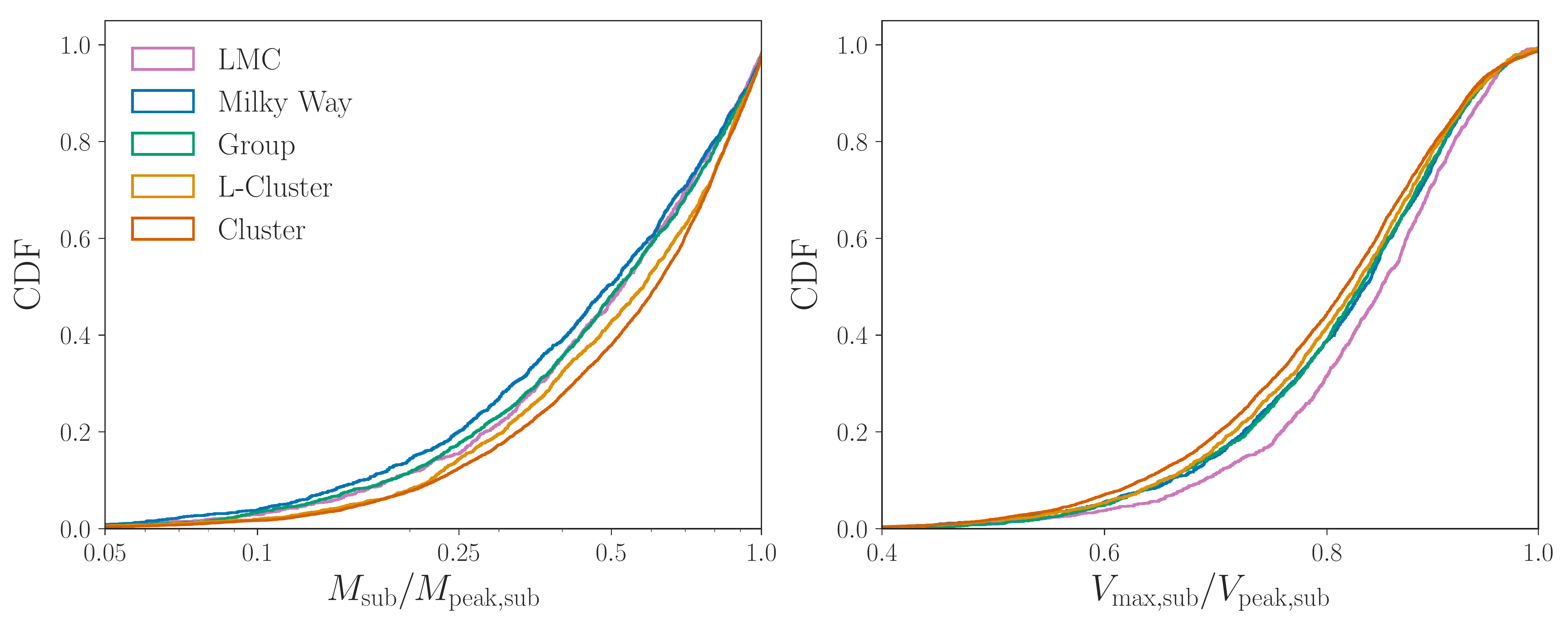}
\caption{Left panel: cumulative distributions of Symphony subhalos' $z=0$ virial mass, $M_{\mathrm{sub}}$, divided by the largest virial mass achieved along each subhalo's main branch, $M_{\mathrm{peak,sub}}$, stacked over all subhalos from each suite with $M_{\mathrm{sub}}/M_{\mathrm{host}}>2.7\times 10^{-4}$. Right panel: same as the left panel, for the ratio of $z=0$ maximum circular velocity, $V_{\mathrm{max,sub}}$, to the largest $V_{\mathrm{max,sub}}$ achieved along each subhalo's main branch, $V_{\mathrm{peak,sub}}$.}
\label{fig:mvir_mpeak_vmax_vpeak}
\end{figure*}

As expected, subhalos of lower-mass hosts accrete systematically earlier. Interestingly, the difference between first and most recent infall time distributions grows with decreasing host mass. Because the fraction of pre-processed subhalos is expected to be fairly small at all host halo mass scales we consider (e.g., \citealt{Berrier08040426,Wetzel150101972}), we mainly attribute this trend to the varying fraction of splashback subhalos as a function of host mass. In particular, the fraction of splashback subhalos is larger for lower-mass hosts because they are older and accrete more slowly than higher-mass hosts. Thus, we expect a larger difference between the first and most recent accretion scale factors for subhalos of lower-mass hosts, which is consistent with both panels of Figure \ref{fig:acc_cdf}. For the LMC, Milky Way, and Group suites, the flattening of the first accretion scale factor cumulative distributions in the left panel of Figure \ref{fig:acc_cdf} starts near $t_{\mathrm{lookback}}\approx 2~ \mathrm{Gyr}$, roughly one orbital timescale ago, and extends to $t_{\mathrm{lookback}}\approx 5~ \mathrm{Gyr}$. This is consistent with the expectation that many subhalos of lower-mass hosts accreted within the last few orbital timescales are splashback objects at $z=0$ (e.g., \citealt{Barber13100466}).

As shown in the right panel of Figure \ref{fig:acc_cdf}, the dependence of infall time on host mass is reversed when measured in units of the host's formation time. In particular, relative to their hosts' half-mass scale factors, subhalos of higher-mass hosts accrete \emph{earlier} than subhalos of lower-mass hosts. This results from a combination of two effects: (1) $a_{1/2,\mathrm{host}}$ increases for higher-mass hosts, and (2) there is a ``survivor bias'' due to the $M_{\mathrm{sub}}/M_{\mathrm{host}}>2.7\times 10^{-4}$ resolution cut imposed when calculating the cumulative distributions. Specifically, subhalos with earlier infall times, comparable to the half-mass scale factors of the LMC, Milky Way, and Group hosts, are likely to be tidally stripped below this resolution limit by $z=0$. Exploring the detailed dependence of subhalo population statistics on accretion time (e.g., following \citealt{Green210301227}) is an interesting area for future work.

\subsection{Subhalo Tidal Stripping}
\label{stripping_distributions}

Figure \ref{fig:mvir_mpeak_vmax_vpeak} shows cumulative distributions of $M_{\mathrm{sub}}/M_{\mathrm{peak,sub}}$ and $V_{\mathrm{max,sub}}/V_{\mathrm{peak,sub}}$ for subhalos above our conservative sub-to-host halo mass resolution limit of $M_{\mathrm{sub}}/M_{\mathrm{host}}>2.7\times 10^{-4}$, stacked over all hosts within each Symphony suite. Subhalos above this limit typically retain $\approx 50\%$ of their peak mass and $\approx 80\%$ of their peak maximum circular velocity. Very few subhalos with $M_{\mathrm{sub}}/M_{\mathrm{peak,sub}}\lesssim 0.05$ or $V_{\mathrm{max,sub}}/V_{\mathrm{peak,sub}}\lesssim 0.4$ remain above the mass resolution limit, regardless of whether an $M_{\mathrm{sub}}>300m_{\mathrm{part}}$ cut is applied.

Mass loss is slightly enhanced for subhalos of lower-mass hosts, likely due to a combination of their earlier infall times (Figure \ref{fig:acc_cdf}) and their hosts' higher concentrations (Figure \ref{fig:Mhost_chost}). Meanwhile, subhalos of lower-mass hosts retain slightly \emph{more} of their peak maximum circular velocity than subhalos of higher-mass hosts. This may result from the higher concentrations of lower-mass subhalos (e.g., \citealt{Moline160304057,Moline211002097}), which makes their inner regions probed by $V_{\mathrm{max,sub}}$ more resilient to stripping despite earlier infall times.

\section{Additional \textsc{Galacticus} Comparisons}
\label{sec:galacticus_comparisons_additional}

\begin{figure*}[t!]
\includegraphics[width=\textwidth]{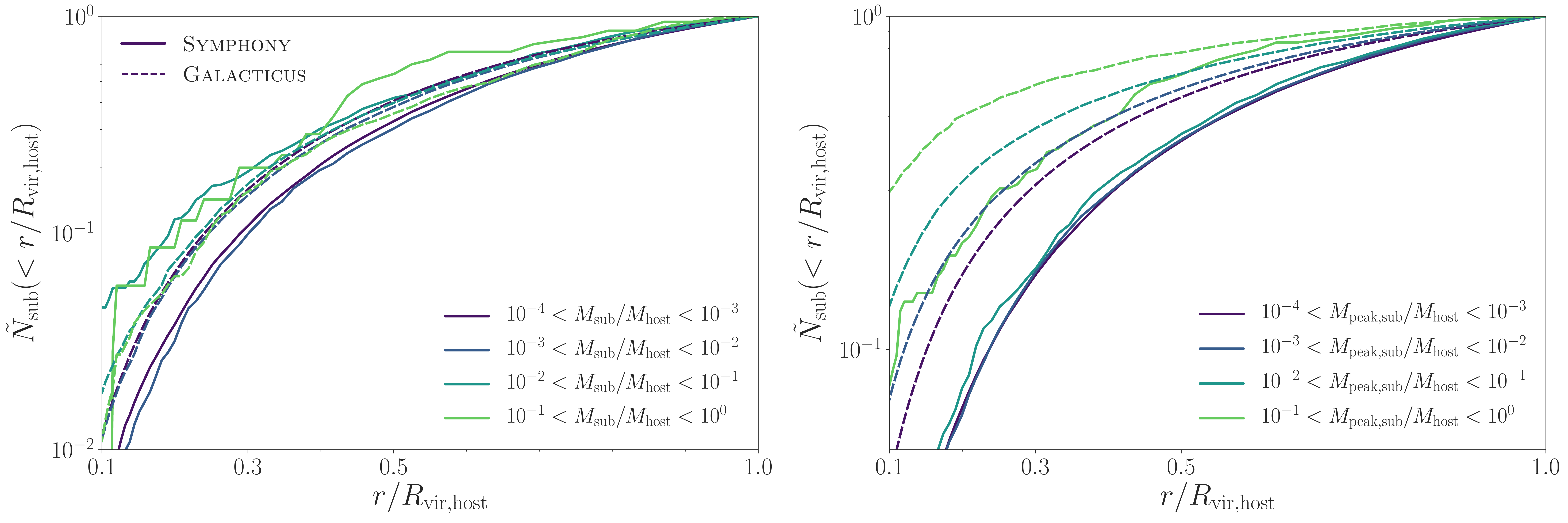}
\caption{Left panel: normalized cumulative subhalo radial distributions, stacked over all suites, in bins of $z=0$ sub-to-host halo mass ratio for Symphony (solid) and \textsc{Galacticus} (dashed). Right panel: same as the left panel for bins of peak sub-to-host halo mass ratio.}
\label{fig:galacticus_radial_peak}
\end{figure*}

\begin{figure*}[t!]
\includegraphics[width=0.5\textwidth]{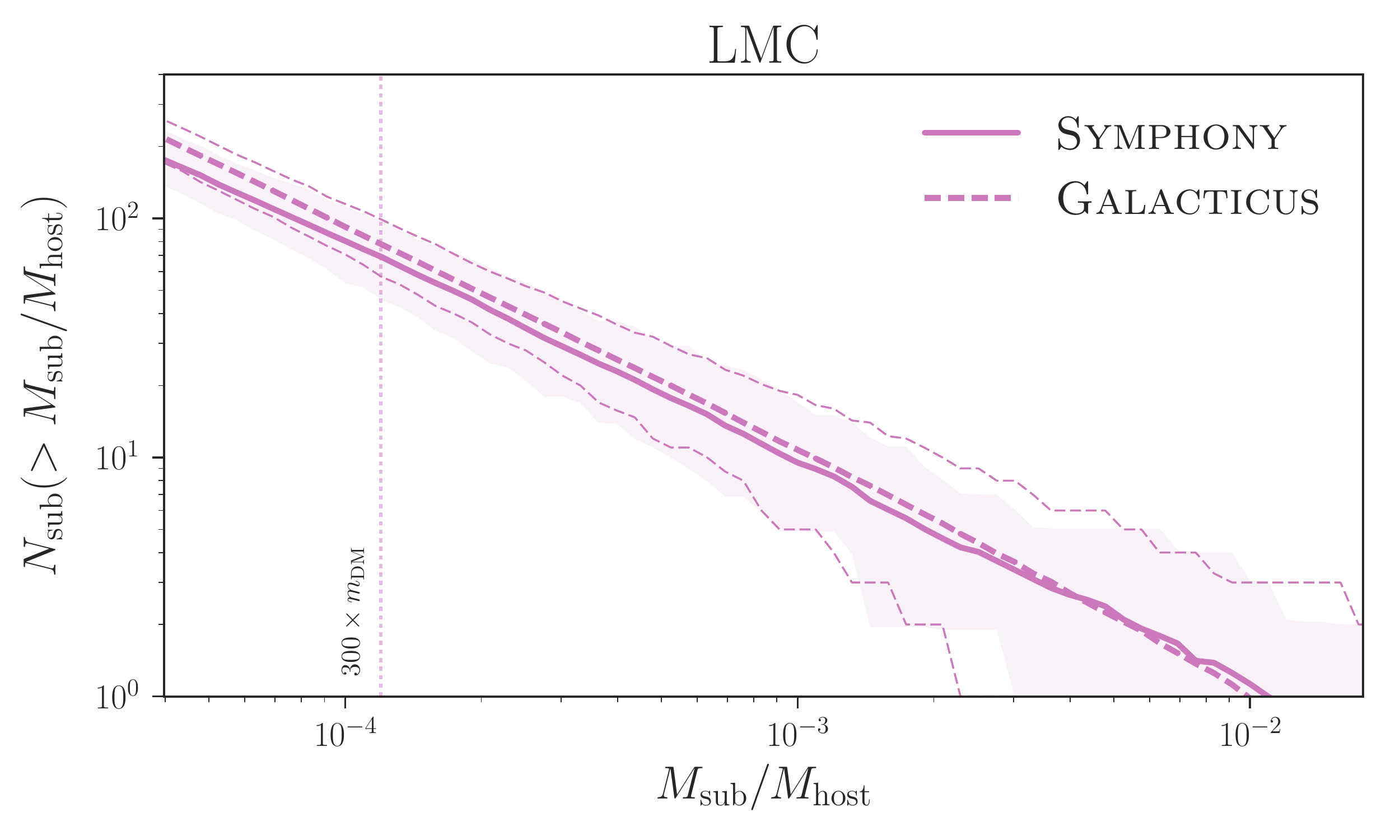}
\includegraphics[width=0.5\textwidth]{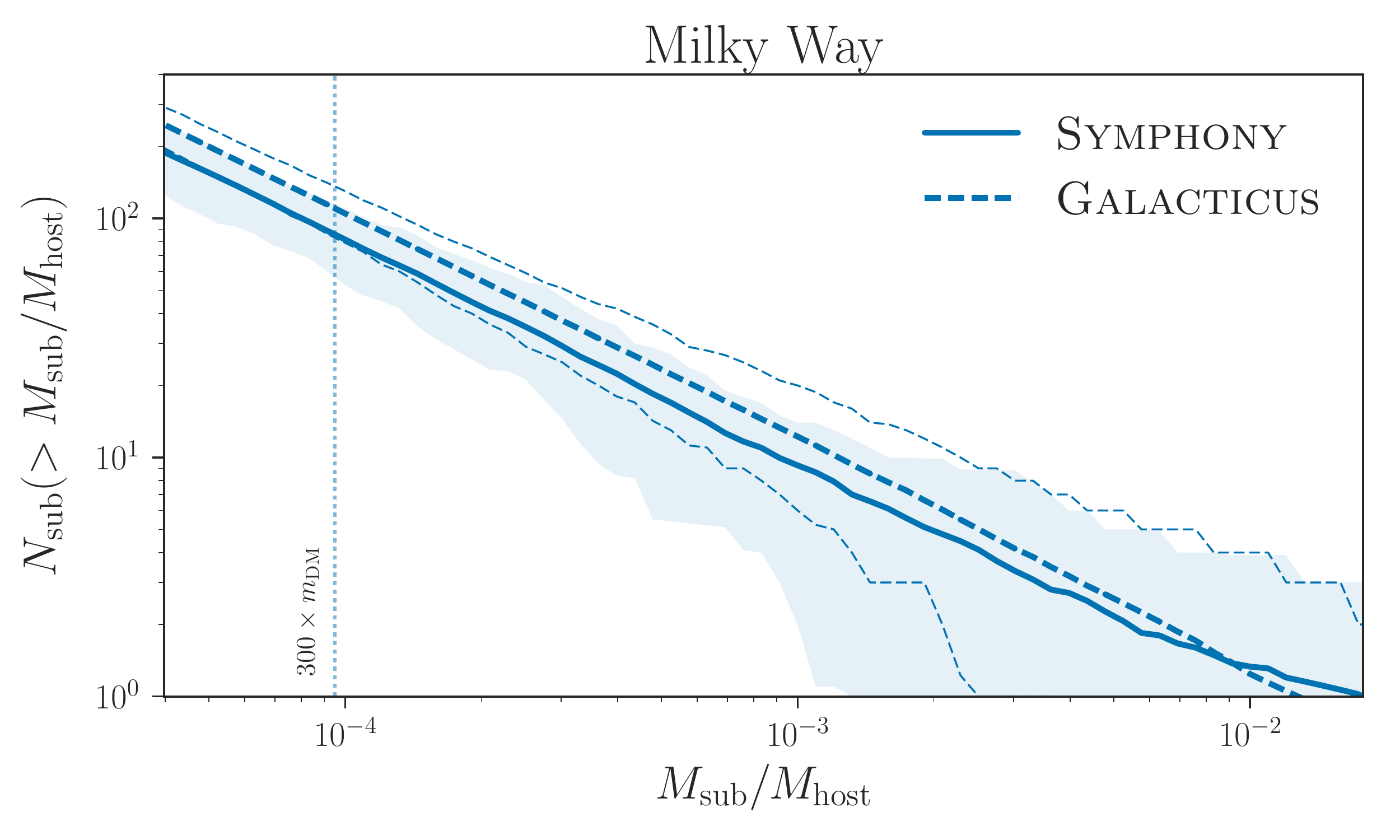}\\
\includegraphics[width=0.5\textwidth]{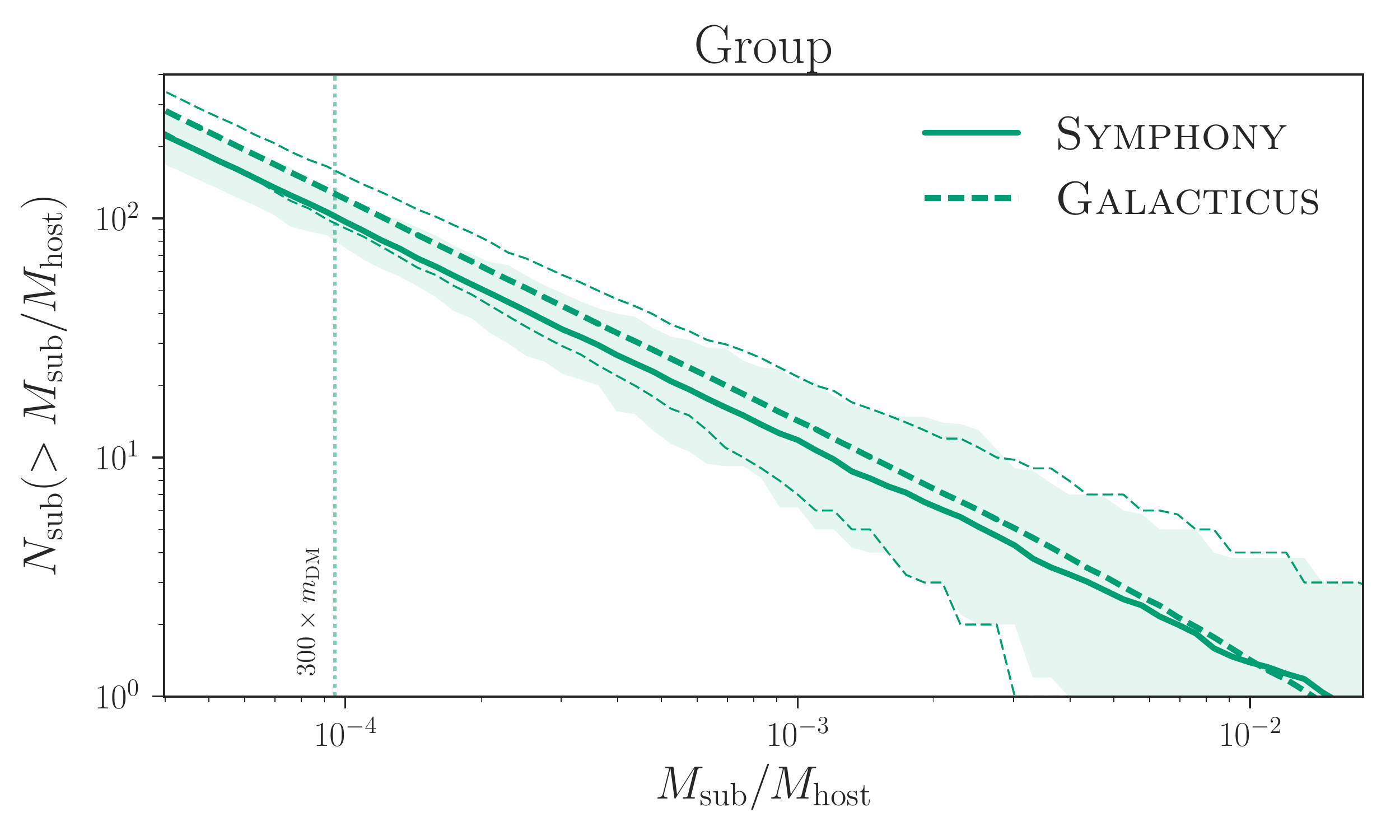}
\includegraphics[width=0.5\textwidth]{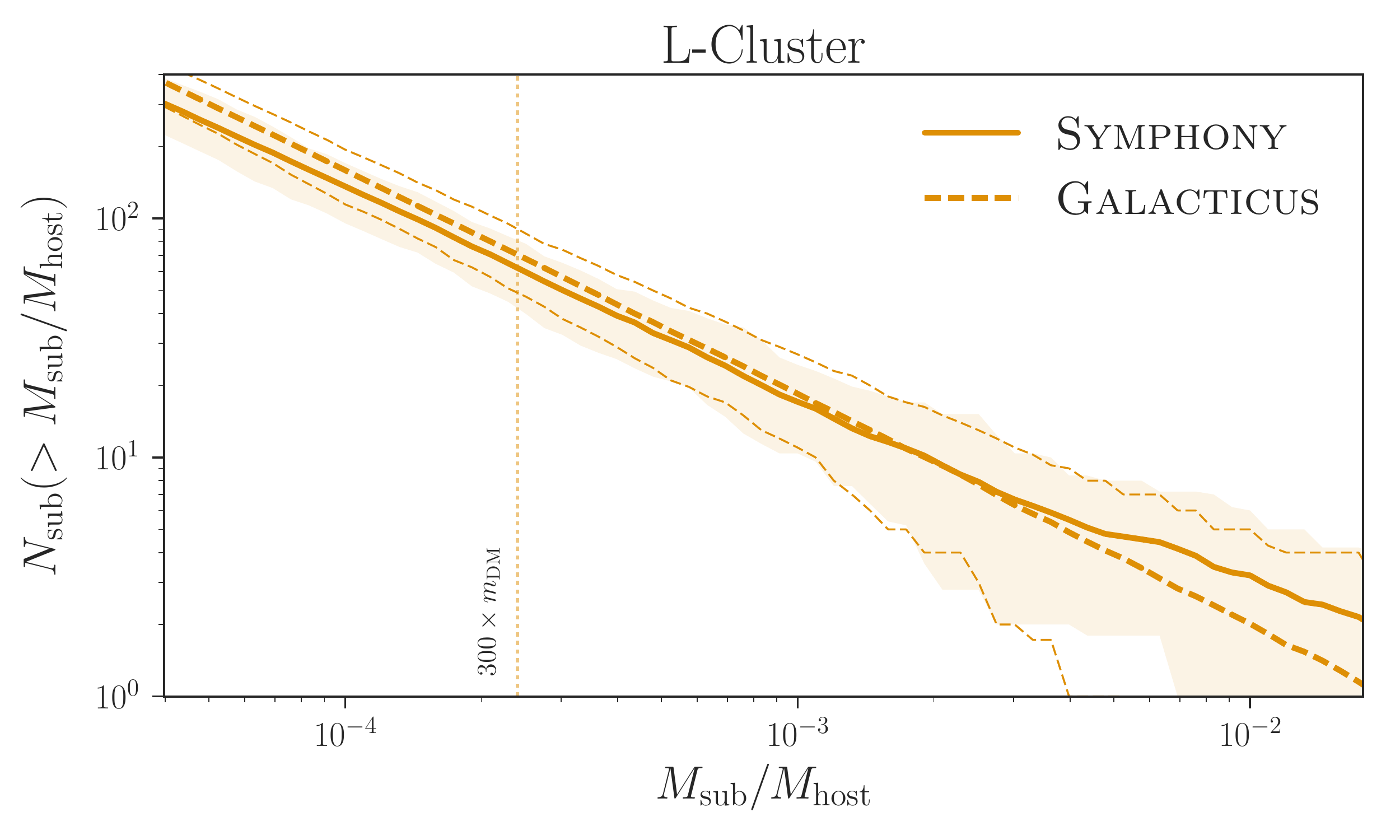}\\
\includegraphics[width=0.5\textwidth]{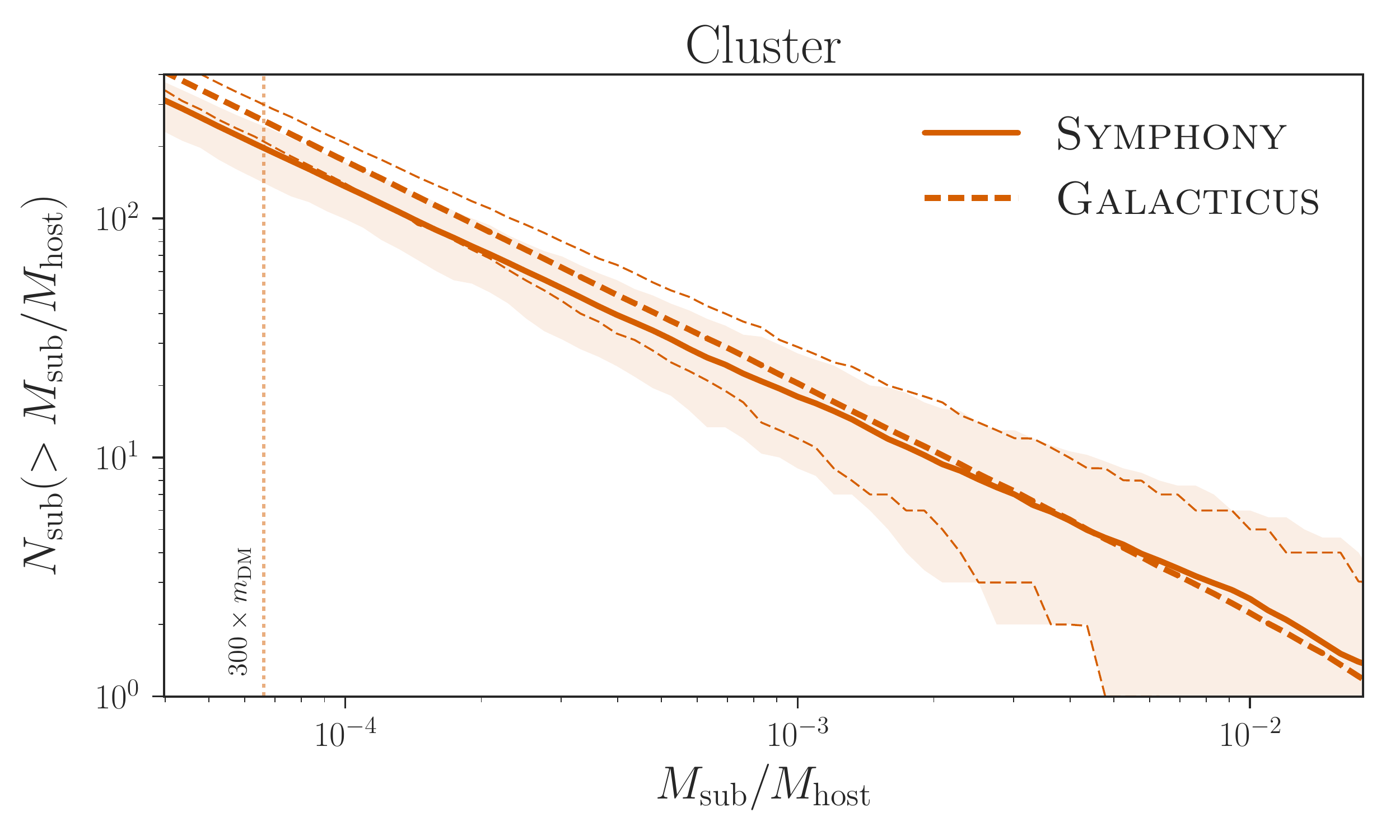}
\caption{Comparison of subhalo mass functions in our five zoom-in simulation suites with \textsc{Galacticus} predictions. In each panel, the solid line shows the mean cumulative Symphony SHMF stacked over all simulations in each suite, the shaded band shows the $2.5$th--$97.5$th percentile range, and dashed lines show the corresponding \textsc{Galacticus} predictions and percentiles. Vertical dashed lines show the conservative subhalo mass resolution limit for each suite, corresponding to $300$ times the highest-resolution dark matter particle mass.}
\label{fig:galacticus_individual_SHMFs}
\end{figure*}

This Appendix provides comparisons between Symphony and \textsc{Galacticus} predictions beyond those discussed in the main text. First, Figure \ref{fig:galacticus_radial_peak} compares Symphony and \textsc{Galacticus} normalized radial distributions stacked over all suites, binned by $M_{\mathrm{sub}}/M_{\mathrm{host}}$ (left panel) and $M_{\mathrm{peak,sub}}/M_{\mathrm{host}}$. As for Figure \ref{fig:radial}, note that the lowest-mass bin extends slightly below our fiducial $z=0$ sub-to-host mass ratio convergence limit (Equation \ref{eq:convergence_ratio}), while the radial distributions binned by peak mass are not formally converged at small radii given our convergence tests (see Appendix \ref{sec:convergence_res}), and should be interpreted with caution. Interestingly, \textsc{Galacticus} radial profiles are largely insensitive to $M_{\mathrm{sub}}/M_{\mathrm{host}}$ and become systematically more concentrated with increasing $M_{\mathrm{peak,sub}}/M_{\mathrm{host}}$. On the other hand, Symphony radial profiles are nearly identical for all $M_{\mathrm{peak,sub}}/M_{\mathrm{host}}\lesssim 10^{-2}$ and $M_{\mathrm{peak,sub}}/M_{\mathrm{host}}\lesssim 10^{-1}$, while the subhalos with the largest sub-to-host halo mass ratios in our simulations (in terms of both $z=0$ and peak mass) are more centrally concentrated due to dynamical friction.

\begin{figure*}[t!]
\includegraphics[width=0.5\textwidth]{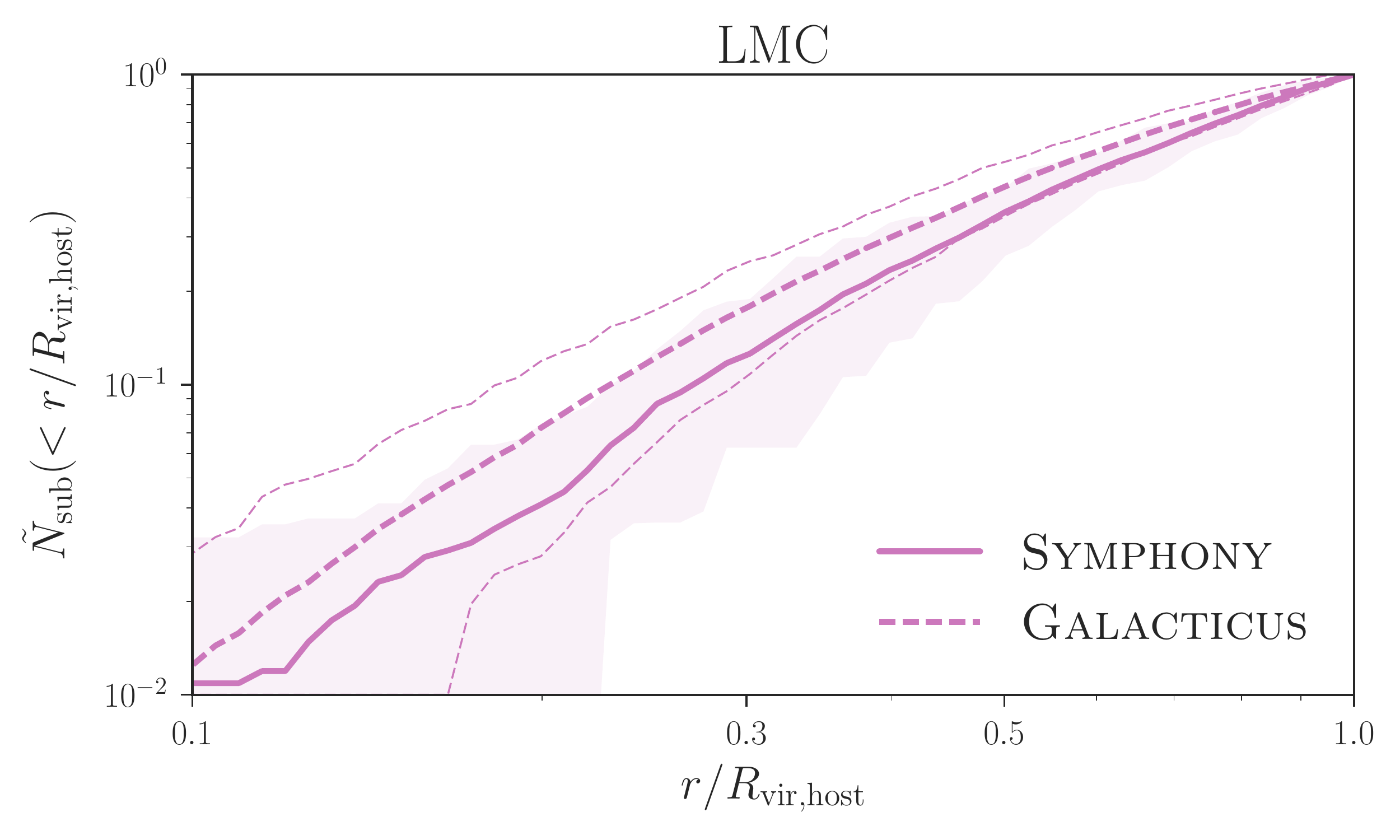}
\includegraphics[width=0.5\textwidth]{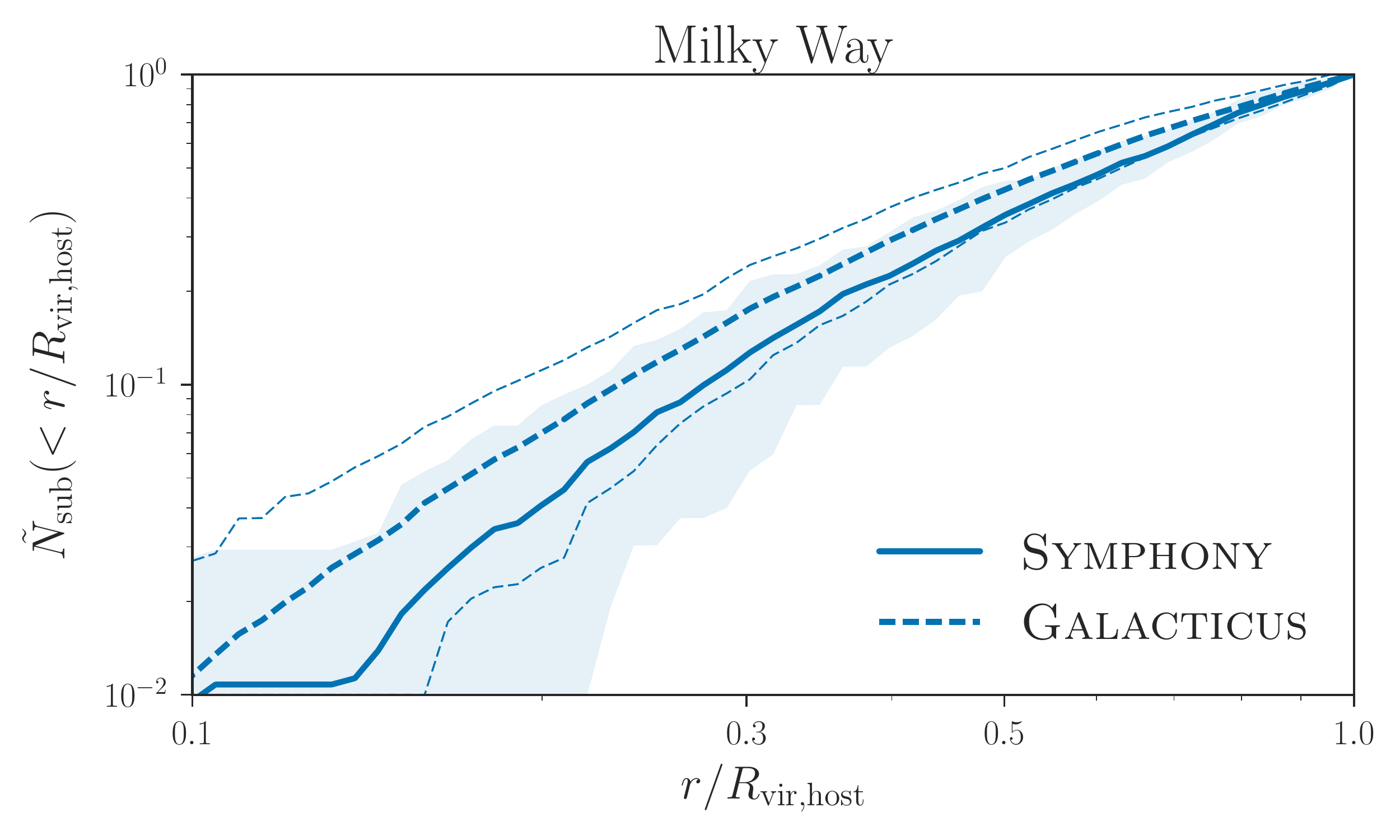}\\
\includegraphics[width=0.5\textwidth]{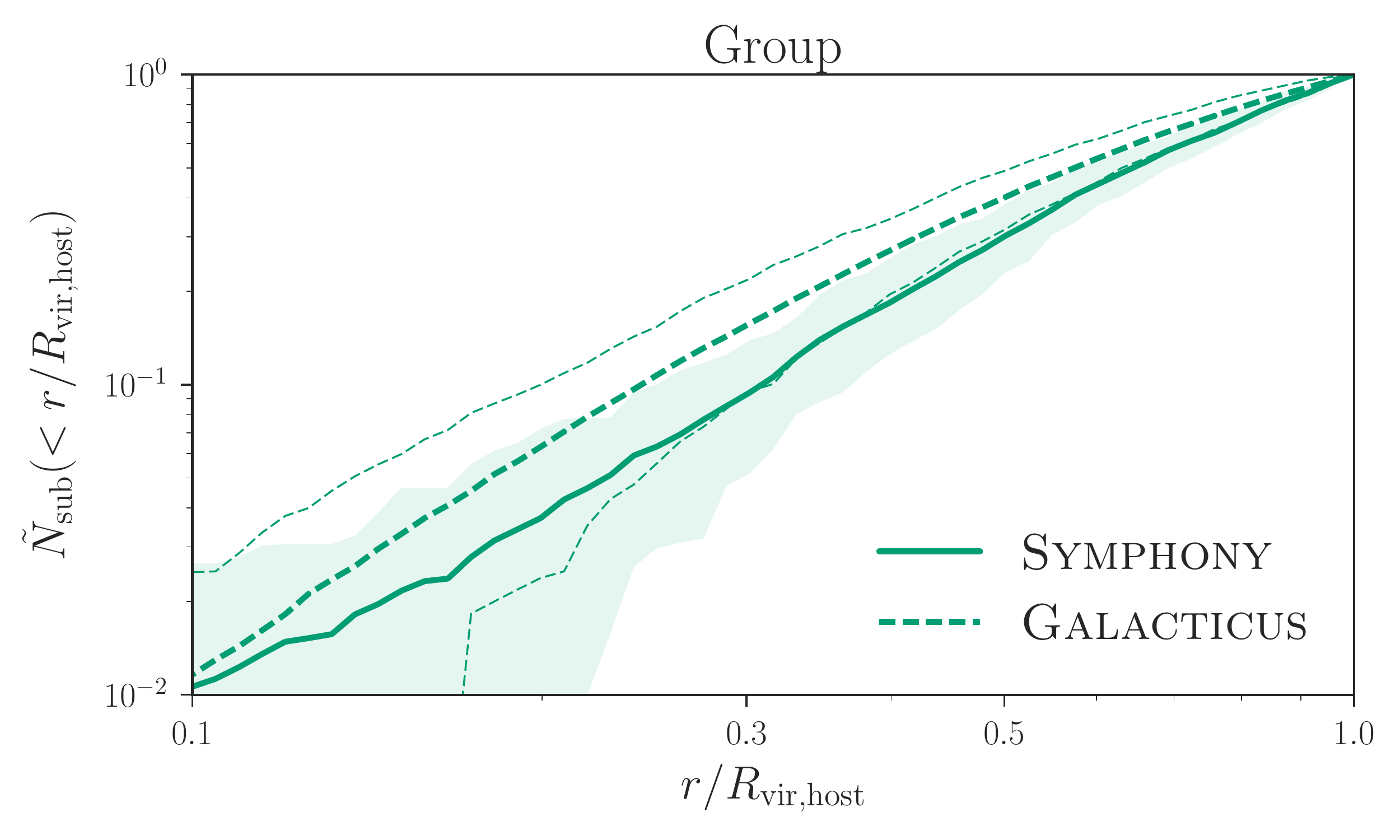}
\includegraphics[width=0.5\textwidth]{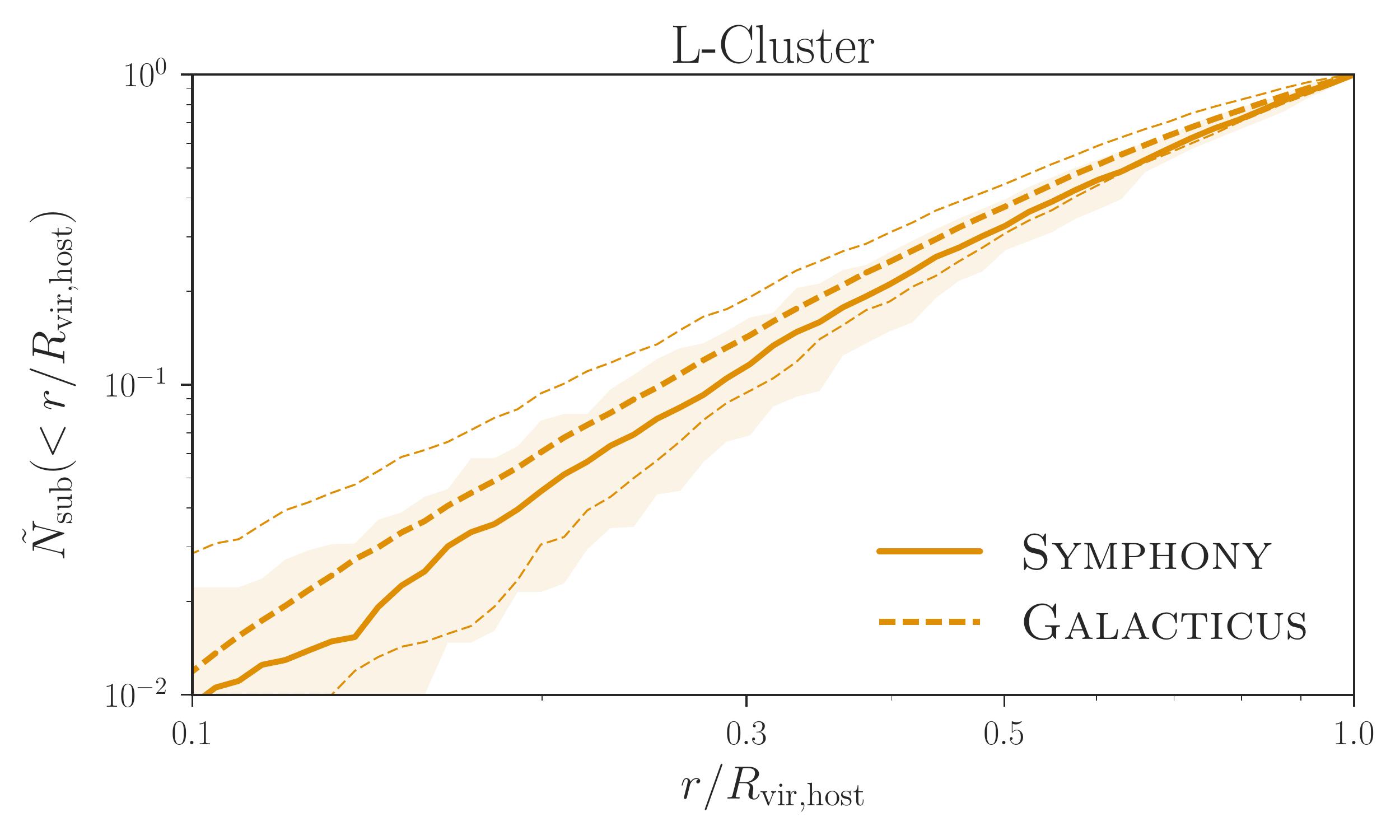}\\
\includegraphics[width=0.5\textwidth]{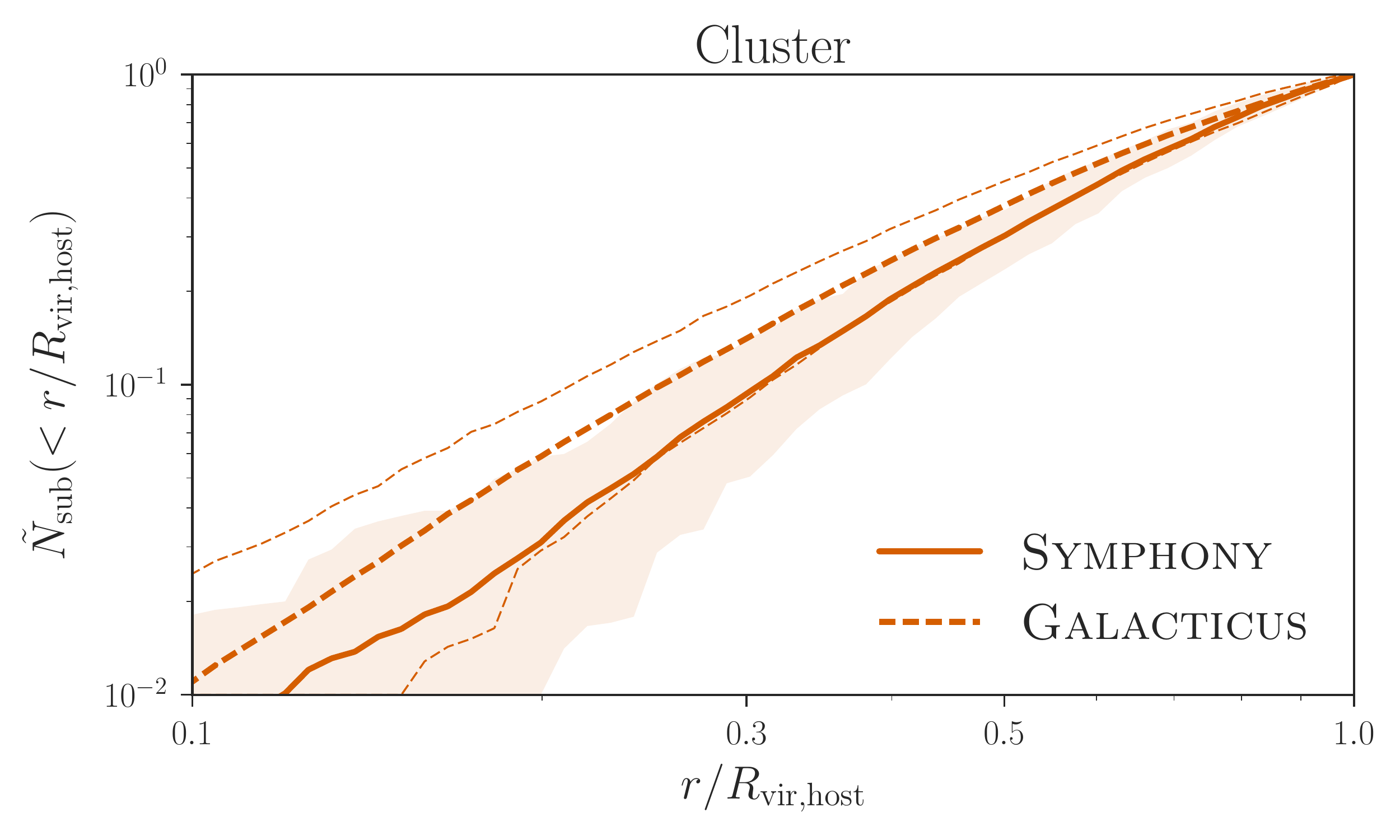}
\caption{Comparison of normalized radial distributions in our five zoom-in simulation suites with \textsc{Galacticus} predictions. In each panel, the solid line shows the mean cumulative normalized radial profile, stacked over all simulations in each Symphony suite, and the shaded band shows the corresponding $2.5$th--$97.5$th percentile range. Dashed lines show the corresponding \textsc{Galacticus} predictions and percentiles. Only subhalos above our conservative sub-to-host halo resolution threshold of $M_{\mathrm{sub}}/M_{\mathrm{host}}>2.7\times 10^{-4}$ are included.}
\label{fig:galacticus_individual_radial}
\end{figure*}

Next, Figure \ref{fig:galacticus_individual_SHMFs} shows cumulative SHMF comparisons between Symphony and \textsc{Galacticus} for each suite individually. These plots demonstrate that Symphony and \textsc{Galacticus} SHMFs are consistent within the host-to-host scatter, although \textsc{Galacticus} displays slightly but systematically steeper SHMF slopes. Finally, Figure \ref{fig:galacticus_individual_radial} shows the corresponding comparisons between Symphony and \textsc{Galacticus} cumulative normalized radial distributions. Again, these plots demonstrate that Symphony and \textsc{Galacticus} radial profiles are consistent within the host-to-host scatter, although \textsc{Galacticus} displays slightly more concentrated radial profiles in the very inner regions ($r/R_{\mathrm{vir,host}}\lesssim 0.2$) down to a fixed sub-to-host halo mass ratio. In turn, the \textsc{Galacticus} predictions slightly exceed Symphony at larger radii because the radial profiles we plot are normalized to unity at~$r/R_{\mathrm{vir,host}}=1$.

\section{Comparisons between Symphony Milky Way, Caterpillar, and Parent Boxes}
\label{sec:caterpillar_comparison}

In this Appendix, we compare the Symphony Milky Way suite SHMF to (1) its parent simulation, c125-1024 ($m_{\mathrm{part}}=2.1\times 10^8~\msun$) and a higher-resolution version of the same box, c125-2048 ($m_{\mathrm{part}}=2.6\times 10^7~\msun$), and (2) the LX14 resolution level of the Caterpillar Milky Way--mass zoom-in simulations (\citealt{Griffen150901255}; $m_{\mathrm{part}}=3.0\times 10^4~\msun$). We also compare Caterpillar SHMFs to those measured in its parent simulation, ``CaterpillarParent,'' (described in \citealt{Griffen150901255}; $m_{\mathrm{part}}=1.2\times 10^8~\msun$), and we compare both the Symphony and Caterpillar parent box SHMFs to the cosmological simulation VSMDPL (\citealt{Klypin14114001}; $m_{\mathrm{part}}=8.8\times 10^6~\msun$) as an independent check.

For the following comparisons, all SHMFs are measured using the ratio of $z=0$ subhalo virial mass to host halo mass, $M_{\mathrm{sub}}/M_{\mathrm{host}}$. For plotting purposes, we assume that SHMFs are converged for $M_{\mathrm{sub}}>300m_{\mathrm{part}}$ in each zoom-in or cosmological simulation, following Symphony convergence results in Appendix \ref{sec:convergence}; this assumption does not affect our conclusions. Furthermore, based on the consistency among the cosmological box SHMFs described below, we assume that differences in cosmological parameters between these simulations result in subdominant differences between their SHMFs compared to the selection effects that we discuss. We refer the reader to the respective references for additional details about the numerical and cosmological parameters for each simulation.

\subsection{Subhalo Mass Function Measurements}

Figure \ref{fig:symphony_caterpillar} shows the mean SHMFs, in units of $z=0$ sub-to-host halo mass ratios, for the $45$ Symphony Milky Way hosts, the $35$ Caterpillar zoom-ins hosts, and samples of hosts from c125-1024, c125-2048, VSMDPL, and the Caterpillar parent box that are chosen to either satisfy combinations of (1) the ``primary'' Symphony Milky Way mass and environmental cuts described in Section \ref{sec:mw_suite}, (2) the exact Symphony Milky Way selection, i.e., the primary mass and environmental cuts plus the ``secondary'' selection of which halos to zoom in on, and (3) the corresponding primary and secondary Caterpillar selection cuts.\footnote{The first $24$ Caterpillar hosts were presented in \cite{Griffen150901255}; we use the $35$ available at \url{https://www.caterpillarproject.org/}.} Each SHMF extends down to a sub-to-host halo mass ratio of $300$ particles for the lowest-mass host in each host sample from the corresponding zoom-in or cosmological simulations. When interpreting the differences between cosmological simulations, note that the Caterpillar parent box resolution is slightly better than c125-1024, while VSMDPL's resolution is roughly a factor of 3 better than c125-2048.

\subsection{Comparing Symphony to Its Parent Box}

First, we study the SHMF of Symphony Milky Way hosts in the parent boxes at different resolution levels relative to all hosts in c125-2048 that pass Symphony's primary mass and environmental cuts. Specifically, the solid black line in Figure \ref{fig:symphony_caterpillar} shows the SHMF for $450$ hosts that pass the Symphony Milky Way environmental cuts described in Section \ref{sec:mw_suite}, where $10$ hosts with masses closest to each Symphony Milky Way host's $z=0$ virial mass are selected from all halos that pass the environmental cuts in order to match the host mass distributions. Meanwhile, the dashed black and gray lines show SHMFs for the $45$ Symphony target halos in c125-2048 and c125-1024, respectively. The identities of these hosts in c125-1024 are known, because they are used to generate the Symphony initial conditions, and we find their matches in c125-2048 by minimizing a metric based on differences in $z=0$ mass and phase-space distance. We have verified that all of the c125-1024 and c125-2048 matches' MAHs match each Symphony zoom-in at the percent level or better.

As described in Section \ref{sec:galacticus_sub}, comparing these SHMFs reveals that, in both the zoom-ins and parent boxes, Symphony hosts' mean SHMF is $\approx 30\%$ higher than the mass and environmentally selected sample's mean SHMF at high sub-to-host halo mass ratios. Because this effect appears in both the parent and zoom-in simulations, we attribute it to either a random fluctuation or the effects of secondary selection cuts---i.e., a consequence of choices among which host halos to zoom in on after the primary mass and environmental cuts are applied---rather than an artifact (or feature) of our zoom-in simulation procedure or analysis.  Meanwhile, at sub-to-host halo mass ratios near c125-2048's resolution limit of $M_{\mathrm{sub}}/M_{\mathrm{host}}\approx 10^{-2}$, the mean SHMFs of both the Symphony zoom-in suite and the matched sample in c125-2048 are consistent with the mass and environmentally selected host sample's SHMF in both c125-2048 and VSMDPL.

We therefore conclude that:
\begin{enumerate}
    \item For $M_{\mathrm{sub}}/M_{\mathrm{host}}\gtrsim 10^{-2}$, the mean Symphony Milky Way SHMF is $\approx 30\%$ high relative to all Milky Way--mass hosts in c125-1024 and c125-2048 that satisfy the Symphony mass and environmental cuts described in Section \ref{sec:mw_suite}; the SHMF of the $45$ Symphony hosts in the fiducial and high-resolution versions of the parent box displays the same offset, which is therefore caused by the selection of the $45$ Symphony hosts, and may either be a random fluctuation or a selection effect related to these hosts' properties.
    \item For $M_{\mathrm{sub}}/M_{\mathrm{host}}\lesssim 10^{-2}$, the mean Symphony Milky Way SHMF is consistent with the SHMF of a corresponding mass and environmentally selected sample in both c125-2048 and VSMDPL.
\end{enumerate}

We note that, for computational efficiency, Symphony Milky Way hosts were typically selected to have small Lagrangian volumes relative to all hosts in the parent box that satisfy the relevant mass and environmental cuts \citep{Mao150302637}. However, we only observe a weak correlation (with significant scatter) between Lagrangian volume and the amplitude of the normalized SHMF at any sub-to-host halo mass ratio, so Lagrangian volume bias does not necessarily explain the SHMF bias at high sub-to-host halo mass ratios.

\subsection{Comparing Caterpillar to its Parent Box}

\begin{figure*}[t!]
\hspace{13mm}
\includegraphics[width=0.8\textwidth]{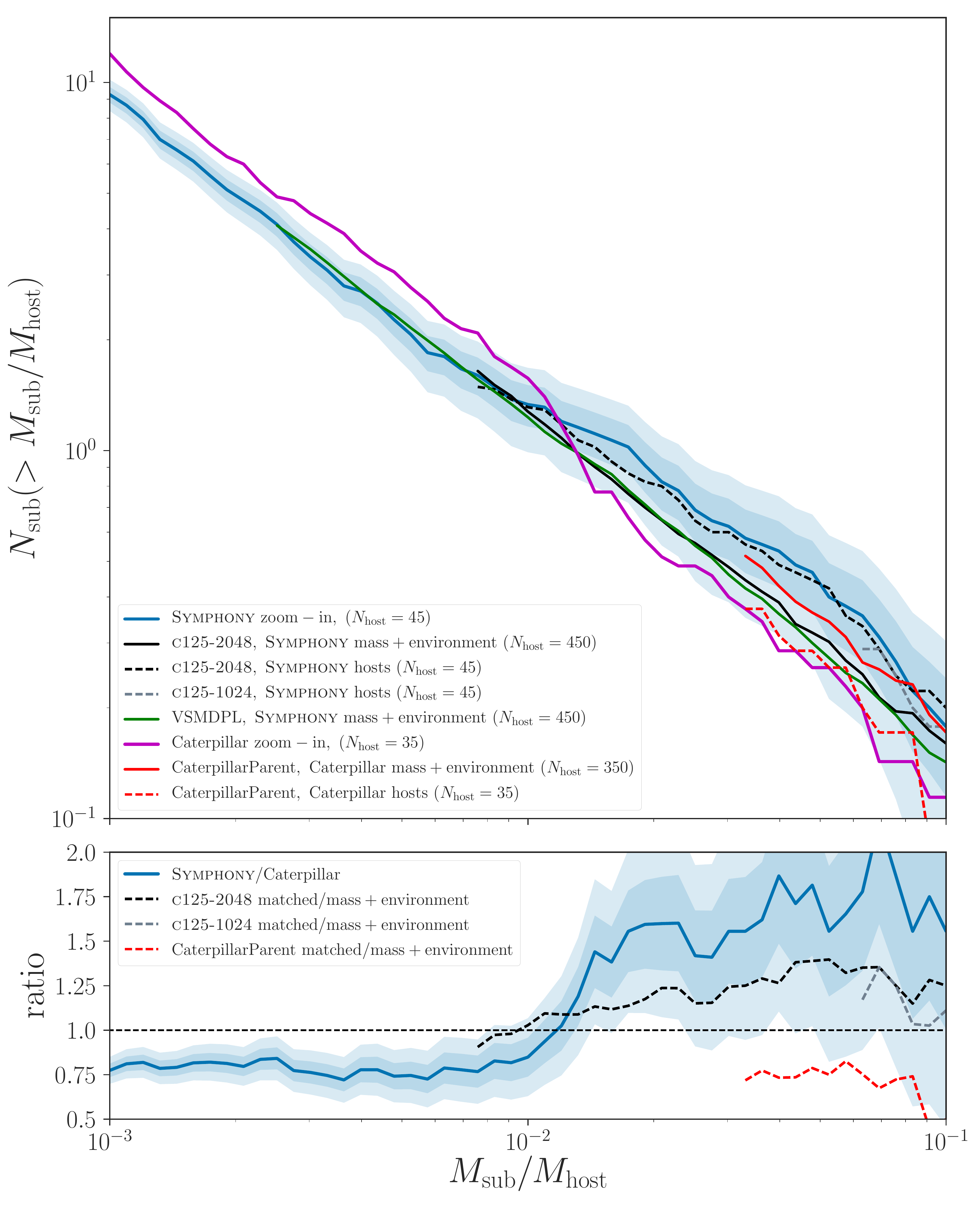}
\caption{Top panel: mean SHMFs, in units of $z=0$ sub-to-host halo mass ratio, in the Symphony Milky Way suite (blue), the c125-2048 box (a higher-resolution version of Symphony Milky Way's parent simulation) using (1) a mass-matched sample of hosts that pass the Symphony Milky Way mass and environmental cuts (solid black), and (2) the sample of Symphony target halos (dashed black), the c125-1024 box (Symphony Milky Way's parent simulation) using the sample of $45$ Symphony target hosts (dashed gray), the VSMDPL box using Symphony Milky Way's mass and environmental cuts (green), the Caterpillar zoom-in suite (magenta), and the Caterpillar parent box (``CaterpillarParent) using a mass-matched sample of hosts that pass the Caterpillar mass and environmental cuts (solid red) and the $35$ Caterpillar target host halos (dashed red). Dark (light) blue bands indicate $1\sigma$ ($2\sigma$) Poisson uncertainty on the mean Symphony zoom-in SHMF; these uncertainties are representative of other SHMFs shown that use the same (or a similar) number of hosts. Each SHMF extends down to a sub-to-host halo mass ratio of $300$ particles for the lowest-mass host in each sample. Both the Symphony and Caterpillar zoom-in resolution limits are significantly below $M_{\mathrm{sub}}/M_{\mathrm{host}}=10^{-3}$. Bottom panel: same as the top panel, but for ratios of mean SHMF in Symphony relative to Caterpillar (blue), the specific Symphony host sample in c125-2048 and c125-1024 relative to the sample with only mass and environmental cuts applied (dashed black and gray, respectively), and the specific Caterpillar sample in the Caterpillar parent box relative to the sample with only mass and environmental cuts applied (dashed red).}
\label{fig:symphony_caterpillar}
\end{figure*}

Next, we compare Caterpillar zoom-in and parent simulation SHMFs. The mean SHMF of Caterpillar hosts in the Caterpillar parent box, which includes both primary mass and environmental cuts and the effects of secondary selection for the $35$ Caterpillar hosts, is shown by the dashed red line in Figure~\ref{fig:symphony_caterpillar}. At high sub-to-host halo mass ratios, the Caterpillar sample's SHMF in both the parent box and zoom-ins is $\approx 30\%$ lower than for a mass-matched sample of hosts in the Caterpillar parent box that only pass the primary Caterpillar cuts. Thus, selection of the specific Caterpillar host sample also biases its zoom-ins' SHMF at high sub-to-host halo mass ratios, but in a different direction than in Symphony.

Meanwhile, at lower sub-to-host halo mass ratios, the mean SHMF of the Caterpillar sample is $\approx 20\%$ high relative to the host samples with Symphony's mass and environmental cuts in c125-2048 and VSMDPL. We have verified that this difference persists if Caterpillar's mass and environmental cuts are applied instead. Nonetheless, it is difficult to interpret this difference without a higher-resolution resimulation of Caterpillar's parent box for comparison. In general, we expect that cosmological and zoom-in resimulations with varying resolution are needed to robustly interpret SHMF biases at intermediate sub-to-host halo mass ratios.

Thus, we conclude that:
\begin{enumerate}
     \item For $M_{\mathrm{sub}}/M_{\mathrm{host}}\gtrsim 10^{-2}$, the mean Caterpillar SHMF is $\approx 30\%$ low relative to all hosts that pass the Caterpillar mass and environmental cuts (which are similar to the Symphony Milky Way cuts, but differ in detail; see \citealt{Griffen150901255}) in the Caterpillar parent box; the SHMF of the $35$ hosts in the Caterpillar parent box that were resimulated displays the same offset, which is therefore caused by the selection of the $35$ Caterpillar hosts, and may either be a random fluctuation or a selection effect related to these hosts' properties.
    \item For $M_{\mathrm{sub}}/M_{\mathrm{host}}\lesssim 10^{-2}$, the mean Caterpillar SHMF is $\approx 25\%$ high relative to all hosts that pass the Symphony mass and environmental cuts in c125-2048 and VSMDPL. Our tests using these cosmological simulations suggest that this difference is not caused by Caterpillar's mass and environmental cuts, and is instead either a random fluctuation due to the selection of the specific sample of $35$ Caterpillar hosts (which would also appear for a matched sample of hosts in a higher-resolution version of Caterpillar's parent box), or a result of Caterpillar's zoom-in procedure or analysis (which would only appear in the zoom-in results).
\end{enumerate}

\subsection{Comparing Symphony to Caterpillar}

Finally, we compare Symphony and Caterpillar SHMFs in the context of our previous findings. Before doing so directly, we first compare their respective parent simulations' SHMFs to each other and to an independent measurement from VSMDPL. We reiterate that Caterpillar's primary mass and environmental cuts (which differ in detail from those used in Symphony's Milky Way suite) do not appear to drive the differences between Symphony and Caterpillar SHMFs. Thus, we interpret discrepancies between Symphony and Caterpillar through the comparisons to matched samples in their parent cosmological simulations described above.

A comparison between Symphony and Caterpillar SHMFs indicates that:
\begin{enumerate}
    \item At high sub-to-host halo mass ratios, $M_{\mathrm{sub}}/M_{\mathrm{host}}\gtrsim 10^{-2}$, the upward fluctuation in the mean Symphony SHMF combines with the downward fluctuation in the mean Caterpillar SHMF to yield a $\approx 60\%$ (or $\approx 2\sigma$) discrepancy between the zoom-in suites. This difference is driven by the selection of the specific $45$ Symphony hosts and $35$ Caterpillar hosts that were resimulated, and may either be a random fluctuation or a selection effect related to these hosts' properties.\footnote{Curiously, our tests on Symphony Milky Way hosts using the modified version of \textsc{Rockstar} described in \cite{Griffen150901255} and used to analyze the Caterpillar zoom-in simulations indicate that even more high-mass subhalos are recovered with the Caterpillar analysis tools; thus, the discrepancy at the high-mass end of the SHMF may be even larger when using a unified analysis pipeline.}
    \item At intermediate sub-to-host halo mass ratios, $10^{-3}\lesssim M_{\mathrm{sub}}/M_{\mathrm{host}}\lesssim 10^{-2}$, the overabundance in the mean Caterpillar SHMF relative to cosmological simulations yields a $\approx 25\%$ (or $\approx 2\sigma$) discrepancy relative to Symphony. This difference may be caused by a fluctuation due to the selection of the specific Caterpillar host sample after mass and environmental cuts are applied, or may be caused by Caterpillar's zoom-in procedure and/or analysis. Resimulations of the Caterpillar parent box at varying resolution would help to assess these potential explanations.
    \item At low sub-to-host halo mass ratios, $M_{\mathrm{sub}}/M_{\mathrm{host}}\lesssim 10^{-3}$, the Symphony and Caterpillar SHMF slopes do not significantly differ. Thus, the SHMF overabundance in Caterpillar relative to Symphony at intermediate subhalo masses propagates to very low sub-to-host halo mass ratios, resulting in the $\approx 25\%$ discrepancy at the lowest resolved masses shown in Figure \ref{fig:SHMF_galacticus}.
\end{enumerate}
We emphasize that Caterpillar's LX14 resolution is better by a factor of roughly 8 than the Symphony Milky Way suite's fiducial resolution, and that future comparisons will need to account for resolution differences carefully to draw firm conclusions about the physical or numerical origins of SHMF discrepancies at low sub-to-host halo mass ratios.

To conclude, we note that our direct comparison between Symphony and Caterpillar SHMFs is consistent with the comparison between Symphony's Milky Way suite and \textsc{Galacticus} predictions discussed in Section~\ref{sec:galacticus_sub}. This is expected, because \cite{Yang200310646} calibrated the \textsc{Galacticus} model we use to reproduce Caterpillar SHMFs, and is confirmed by our tests.

\end{document}